\documentclass[12pt,onecolumn,draftcls]{IEEEtran}
\usepackage{stackrel}
\usepackage{pifont}
\newcommand\floor[1]{\lfloor#1\rfloor}
\newcommand\ceil[1]{\lceil#1\rceil}
\usepackage{verbatim}
\usepackage{graphicx}
\usepackage{amsmath,amssymb,amsfonts,bm}
\usepackage{subfigure}
\usepackage{psfrag}
\usepackage[dvips]{epsfig}
\usepackage{amsthm}
\usepackage{color}
\usepackage[usenames,dvipsnames]{pstricks}
\usepackage[dvips]{epsfig}
\usepackage{pst-grad} 
\usepackage{pst-plot} 
\usepackage{textcomp}
\usepackage{tikz,pgf}
\usetikzlibrary{decorations.text}
\usetikzlibrary{decorations.pathreplacing}
\usepackage{amsmath,graphicx}
\usepackage{amsmath,graphicx}
\usepackage{enumerate}
\usepackage{amsbsy}
\usepackage{amssymb}
\usepackage{amsthm}
\usepackage{amscd}
\usepackage{subfigure}
\usepackage{color}
\usepackage{cite}
\usepackage{listings}

\usepackage{tikz}
\usepackage{pgf}

\usepackage{MnSymbol}
\usepackage{stackrel}

\newtheorem{lemm}{Lemma}

\newtheorem{rema}{Remark}

\def\bx{{\bf x}}

\def\bX{{\bf X}}

\def\bx{{\bf x}}

\begin{document}

\title{A Learning-Inspired Strategy to Design Binary Sequences with Good Correlation Properties: SISO and MIMO Radar Systems}

\author{Omid Rezaei, Mahdi Ahmadi, Mohammad Mahdi Naghsh\IEEEauthorrefmark{1}, \emph{Senior~Member, IEEE}, Augusto Aubry, \emph{Senior~Member, IEEE}, Mohammad Mahdi Nayebi,  \emph{Senior~Member, IEEE}, and Antonio De Maio, \emph{Fellow, IEEE}
	
	\thanks{ 
		O. Rezaei and M. M. Nayebi are with the Department of Electrical Engineering, Sharif University of Technology,
		Tehran, 11155-4363, Iran.
		M. Ahmadi and M. M. Naghsh are with the Department of Electrical and Computer Engineering, Isfahan University of Technology, Isfahan, 84156-83111, Iran. 
		A. Aubry and A. De Maio are with the Università degli Studi di Napoli ``Federico II", Dipartimento di Ingegneria Elettrica e delle Tecnologie dell’Informazione, Via Claudio 21, I-80125 Napoli, Italy. \IEEEauthorrefmark{1} Please address all the correspondence to M. M. Naghsh, Email: mm\_naghsh@cc.iut.ac.ir} }

\maketitle
\begin{abstract}
In this paper, the design of binary sequences exhibiting low values of aperiodic/periodic correlation functions, in terms of  Integrated Sidelobe Level (ISL), is pursued via a learning-inspired method. Specifically, the synthesis of either a single or a burst of codes is addressed, with reference to both Single-Input Single-Output (SISO) and Multiple-Input Multiple-Output (MIMO) radar systems. Two optimization machines, referred to as two-layer and single-layer Binary Sequence Correlation Network (BiSCorN), able to learn actions to design binary sequences with small ISL/Complementary ISL (CISL) for SISO and MIMO systems are proposed. These two networks differ in terms of the capability to synthesize Low-Correlation-Zone (LCZ) sequences and computational cost.
Numerical experiments show that proposed techniques can outperform state-of-the-art algorithms for the design of binary
sequences and Complementary Sets of Sequences (CSS) in terms of ISL and, interestingly, of Peak Sidelobe Level (PSL). 
\end{abstract}
\pagebreak

\section{Introduction} \label{intro}
An active sensing device such as a radar system can determine useful information on the prospective targets by transmitting waveforms toward an intended region and then analyzing the received signal. Transmit waveform is thus a critical factor that can improve the performance of active sensing systems. Indeed, an appropriate selection of the probing signals may lead to a better target detectability as well as a more accurate parameter estimation process \cite{stoicabook,gini2012waveform,aubry2015optimizing}.
Designing sequences with good correlation properties, i.e., small Integrated Sidelobe Level (ISL)/Peak Sidelobe Level (PSL), has attracted a lot of attention in radar signal processing community since such sequences rule the quality of the pulse compression process. 
Notably, in modern active sensing and radar systems a special interest is deserved to the design of binary codes due to their implementation simplicity.

The design of binary sequences has been addressed in several works. For example,
M-sequences as well as Gold and Kasami codes have good periodic auto-correlation properties \cite{scol} ensuring optimal performance under certain conditions, e.g., M-sequences are only available for specific length $2^{N-1}$. Another important instance is represented by the Barker codes that are unfortunately limited to a finite length of 13.  
The authors in \cite{bark16,bark17,bark18} extended the Barker sequences by some heuristic algorithms to propose generalized poly-phase Barker sequences. The mentioned works are also limited to length 77 \cite{bark19}.
\cite{CAN} proposed a cyclic algorithm to design in a computationally efficient way constant modulus sequence with small ISL (see \cite{lin2019efficient} for a related paper to design binary codes). In \cite{palom} the ISL minimization problem was addressed by means of Majorization-Minimization (MM) technique.
This MM-based algorithm can be faster than procedure proposed in \cite{CAN} under some conditions.
Moreover, the authors of \cite{CD} consider the problem of sequence design with good correlation properties, in terms of ISL and PSL, for sequences with either continuous or discrete phases. 
The optimization problem in the aforementioned reference was tackled by an iterative algorithm based on Coordinate-Descent (CD) optimization.

Another important line of studies focuses on the Multiple-Input Multiple-Output (MIMO) radar systems which can offer improved target detection and recognition performance.
MIMO radars often rely on almost orthogonal sets of sequences with good auto- and cross-correlation properties. Therefore, the design of waveforms with small ISL of the auto- and cross- correlations for MIMO radar systems has been pursued
in a plethora of works. Among them, it is worth mentioning Multi-CAN \cite{he2009designing}, MM-Corr \cite{song2016sequence}, ISLNew \cite{li2016design,li2017fast}, Iterative Direct Search \cite{cui2017constant}, Doppler resilient complete complementary codes \cite{tang2014construction}, and consensus-ADMM/PDMM \cite{wang2021designing} algorithms.
All of the aforementioned procedures almost meet the lower bound on ISL for sets of sequences with continuous phases \cite{welch}. The BiST algorithm \cite{alaee2019designing} extends the previous works to design sets of binary sequences with good correlation properties for MIMO radar systems.

An idea to improve the correlation properties of a sequence is increasing the degrees of freedom at the design stage. To achieve this goal, some researchers consider Complementary Sets of Sequences (CSS) where multiple codes are transmitted instead of a single one.
Most of the studies concerning CSS have been focused on the generation in closed-form of CSS and involve restrictions on the sequence length and set cardinality. However, an efficient computational method was introduced in \cite{soltanalian2012computational}. Among others, some low complexity procedures were devised e.g., CANARY \cite{soltanalian2013fast}, MM-CSS \cite{song2016sequence}, QOZCP \cite{wang2021quasi}, and
Bare MM-CSS, MM-CSS-SQUARE, MM-CSS-SD algorithms \cite{wu2019fast}.

Recently, machine learning approaches have been employed to address various engineering problems like computer vision \cite{chai2021deep}, watermarking \cite{Redmark}, low-resolution receiver design \cite{khobahi2020model}, and UAV navigation \cite{zeng2021simultaneous}. Particularly, machine learning is used for waveform design and power allocation in communication systems \cite{sharma2019distributed,d2019uplink}. This emerging paradigm has provided successful solution to the aforementioned problems. Therefore, we consider learning approaches for radar sequence design, a promising research area deserving special attention.


In light of above discussions, in this paper we devise two learning-inspired frameworks to deal with the synthesis of binary sequences with good correlation properties. The proposed design methodologies, referred
to as, two-layer and single-layer \textbf{Bi}nary \textbf{S}equence \textbf{Cor}relation \textbf{N}etwork (BiSCorN), exploit neural network tools to tackle sequence design problems. The design problems are formulated in terms of Weighted ISL (WISL) minimization for various radar system setups/models, including Single-Input Single-Output (SISO), MIMO, CSS, MIMO-CSS, and subject to a constraint imposing binary elements. All the resulting code optimization problems are handled under a unified mathematical umbrella offered by BiSCorN. Specially, single-layer and two-layer BiSCorN are employed to deal with the design problems; these two methods are different in terms of the capability to design Low-Correlation-Zone (LCZ) sequences and computational complexity (to be discussed shortly). Note that the proposed method is not a conventional learning method as it is common in the literature; however, it is a learning method, since discover how probing the environment by a training phase. Therefore, the main contributions of the work can be summarized as: (i) developing the theoretical background for learning-inspired sequence design and (ii) devising a network architecture as well as a network feeding approach for the exploitation of the ADAM algorithm\footnote{Note that ADAM is a powerful optimizer common in learning literature.}. At the analysis stage, numerical experiments are reported to show the effectiveness of the proposed BiSCorN also in comparison with the state-of-the-art methods. Note that a limited part of this work has been presented in \cite{rezaei2020learning}; more precisely, the aforementioned conference article addressed only SISO systems without mathematical backgrounds.

The rest of this paper is organized as follows. In Section \ref{pro} the design problems related to the different system setups are formulated and then, two-layer BiSCorN is developed to deal with the aforementioned design problems. In Section \ref{lcz}, the single-layer BiSCorN is proposed to design LCZ sequences for all considered system setups. The implementation process as well as computational complexity are discussed in Section~\ref{ext}.
The performance of the proposed techniques is assessed in Section \ref{per}. Finally, the conclusions are drawn in Section \ref{con}, followed by possible future research lines.

\emph{Notation:} Bold lowercase letters and bold uppercase letters are used for vectors and matrices respectively. The $l_2$-norm of a vector $\bx$ and Frobenius norm of a matrix $\bX$ are denoted by $\|\bx\|_2$ and $\|\bX\|_F$, respectively. $\mathbb{R}^{N\times N}$ ($\mathbb{R}^N$) indicates the set of ${N \times N}$ (${N \times 1}$) real matrices (column vectors).
$\mathbb{R}^{N}_{+}$ represents set of (${N \times 1}$) vectors with non-negative real entries. The absolute value of the scalar $x$ is denoted by $\vert x \vert$, whereas ${(\cdot)^{{T}}}$ indicates the transpose operator. $\mathbb{E}_{\bx} [\cdot ]$ stands for the statistical expectation with respect to (w.r.t.) $\bx$. $\floor{x}$ is the greatest integer less than or equal to $x$ and $\ceil{x}$ is the least integer greater than or equal to $x$. $\mathcal{N}(\boldsymbol{\omega},\mathbf{\Sigma})$ denotes a Gaussian distribution with mean $\boldsymbol{\omega}$ and covariance matrix $\mathbf{\Sigma}$. ${x}_{\textrm{\tiny{mod}}\hspace{1pt}y}$ is equal to $x- \floor{\frac{x}{y}} y$ for $x \in \mathbb{R}$ and $y \in \mathbb{R}_{+}$. The notation ${\nabla}_{{\mathbf{x}}} f(\cdot)$ indicates the gradient of a function w.r.t. $\mathbf{x}$. $\textrm{Diag}(\mathbf{x})$ denotes the diagonal matrix formed by the entries of the vector argument
$\mathbf{x}$. $\mathbf{I}_N \in \mathbb{R}^{N\times N}$ represents the identity matrix and $\mathbf{e}_n \in \mathbb{R}^{N}$ is the $n^{th}$ standard vector. Bold numbers $\mathbf{1}$ and $\mathbf{0}$ are for vectors whose elements are all one and all zero, respectively.
\section{Problem Formulation and the Proposed BiSCorN} \label{pro}
In this section, first, a SISO system is considered and a learning-based network, referred to as two-layer BiSCorN, is proposed for designing binary sequences with small periodic/aperiodic WISL. Then, the proposed approach is extended to MIMO systems. Finally, BiSCorN is modified to design binary complementary sequences with small Weighted Complementary ISL (WCISL) in (either SISO or MIMO) systems.  
\subsection{SISO}\label{SISOpro}
Let us define $\mathbf{x}=[x(1), x(2), ... , x(N)]^{T} \in \mathbb{R}^N$ as a transmit fast-time SISO radar sequence with length $N$. The aperiodic and periodic auto-correlation functions of $\mathbf{x}$ can be defined as
\begin{equation} \label{first}
	r_{AP} (k)= \sum_{n=1}^{N-k} x(n) \hspace{1pt} x(n+k),~~~k=0, ..., N-1,
\end{equation}
and
\begin{equation} \label{second}
	r_{P} (k)= \sum_{n=1}^{N}  x(n) \hspace{1pt} x{({(n+k)}_{\textrm{\tiny{mod}}\hspace{1pt}N})},~~~k=0, ..., N-1,
\end{equation} 
respectively. Note that in the definitions above, $r_{AP} (0)$ and $r_P (0)$ coincide with the energy of the sequence $\mathbf{x}$, while all the other aperiodic/periodic auto-correlation values, i.e., $\lbrace r_{AP} (k) \vert^{N-1}_{k=1} \rbrace$ and $\lbrace r_P (k)\vert^{N-1}_{k=1} \rbrace$, are referred to as sidelobes \cite{stoicabook}. 

In the following, WISL is considered as the design metric to measure the quality of sequences in terms of correlation properties. Precisely, letting $\mathbf{w}_{\mathcal{S}}=[w_{\mathcal{S},1},w_{\mathcal{S},2}, ... ,w_{\mathcal{S},N-1}]^T$ with $w_{\mathcal{S},k} \geq 0, \hspace{2pt} k=1,...,N-1,$ the WISL for the SISO case can be defined as\footnote{In some references $\textrm{WISL}= \displaystyle \sum_{k=1 }^{N-1} w_{\mathcal{S},k} \vert r (k) \vert^2$. Also, the subscript $\mathcal{S}$ is an abbreviation for \textit{SISO}.}:
\begin{equation}
	\textrm{WISL}= 2\sum_{k=1}^{N-1} w_{\mathcal{S},k} \vert r (k) \vert^2,
\end{equation}
where, with a slight abuse of notation, $r(k)$ refers to either $r_{AP} (k)$ or $r_P (k)$, depending on the specific design instance. Note that the special case of ISL can be obtained by selecting $\mathbf{w}_{\mathcal{S}}= \mathbf{1}$.
Hence, the synthesis of binary sequences for SISO systems can be formulated as
\begin{eqnarray}\label{maxmin21}
	\min_{\mathbf{x} } & \displaystyle \sum_{k=1}^{N-1} w_{\mathcal{S},k} \vert r (k) \vert^2 \\ \nonumber
	\mbox{s. t.}\;\;& x(n) \in \lbrace -1,+1 \rbrace,~~n=1,...,N,
\end{eqnarray}
which is a non-convex NP-hard optimization problem \cite{CAN}.

In the sequel, a learning-based network is proposed to deal with the minimization problem in \eqref{maxmin21}.
We start with a relaxation of the binary constraint on the code elements in \eqref{maxmin21}. Precisely, we allow the elements of the code vary within $[-1,1]$. Since $\mathbf{x}=\mathbf{0}$ is the trivial solution to the relaxed problem, the term $w_{\mathcal{S},b} \vert r(0)-N \vert^2$ is added to the objective function. Therefore, the solution present a value of $r(0)$ closer and closer to $N$ as ${w}_{\mathcal{S},b}$ approaches one avoiding the trivial solution. This trick leads to the following reformulated version of \eqref{maxmin21}:
\begin{eqnarray}\label{maxmin23}
	\min_{\mathbf{x} } &  \widetilde{\mathbf{w}}^T_{\mathcal{S}} \hspace{1pt}\hspace{1pt}\widetilde{\mathbf{r}}_\mathcal{S}   \\ \nonumber
	\mbox{s. t.}\;\;& \vert x(n) \vert \leq 1,~~n=1,...,N,
\end{eqnarray}
where $\widetilde{\mathbf{w}}_{\mathcal{S}} =[w_{\mathcal{S},b}, w_{\mathcal{S},s} \times \mathbf{w}^T_{\mathcal{S}}]^T \in \mathbb{R}^N_{+}$ with  $w_{\mathcal{S},b}+w_{\mathcal{S},s}=1, \hspace{2pt} w_{\mathcal{S},b} \geq 0, \hspace{2pt} w_{\mathcal{S},s} \geq 0$ and
\begin{equation} \label{key}
	\widetilde{{r}}_{\mathcal{S}} (k)=
	\begin{cases}
		\vert r(0)-N \vert^2 ,~~k=1,
		\\
		\vert r(k) \vert^2,~~~~~~~k=2,...,N.
	\end{cases}
\end{equation}
This trick also paves the way to partially manage the synthesis loss, viz. the loss/degradation in the objective function of \eqref{maxmin21} required to obtain feasible binary elements for the codes from the solution to the reformulated version (see Remark~1 for details). Specifically, being $ \vert x(n) \vert \leq 1$ it follows that $r(0)= \sum_{n=1}^{N} \vert x(n) \vert^2 \leq N$. Thus, decreasing the term $\vert r(0)-N \vert^2$ leads to $\vert x(n) \vert$ approaching one for any $n$, i.e., near binary solutions for the real-valued variables.
In light of above discussion, one can conclude that the parameters $w_{\mathcal{S},b}$ and $w_{\mathcal{S},s}$ can be used to manage binarization and sidelobe level: increasing $w_{\mathcal{S},b}$ makes the sequence closer to the binary solution and increasing $w_{\mathcal{S},s}$ gives more emphasis on the sidelobe terms.
The following lemma sheds light on the theoretical connections between Problems \eqref{maxmin21} and \eqref{maxmin23}.
\begin{lemm} \label{fmn}
	Let $\mathcal{P}^{(i)}$, $i\geq 1$, be the optimization problem in (\ref{maxmin23}) with $w_{\mathcal{S},b}^{(i)}=\frac{\bar{w}_{\mathcal{S},b}^{(i)}}{\bar{w}_{\mathcal{S},s}+\bar{w}_{\mathcal{S},b}^{(i)}}$  and $w_{\mathcal{S},s}^{(i)}=\frac{\bar{w}_{\mathcal{S},s}}{\bar{w}_{\mathcal{S},s}+\bar{w}_{\mathcal{S},b}^{(i)}}$, where ${\bar{w}_{\mathcal{S},b}^{{(i)}}}>0$, $i\geq 1$, while $\bar{w}_{\mathcal{S},s}>0$ is fixed. Now, denoting by $\mathbf{x}^{(i)}$ an optimal solution to Problem $\mathcal{P}^{(i)}$, $i\geq 1$, if $\bar{w}_{\mathcal{S},b}^{(i)}\rightarrow \infty$ as $i \,\to\, + \infty$, it follows that 
	\begin{itemize}
		\item for any $n=1,...,N$, $\left \vert x^{(i)} (n) \right \vert \,\to\, 1$ as $i \,\to\, + \infty$;
		\item any cluster point to the generated sequence $\mathbf{x}^{(i)}$ is an optimal solution to the original problem in \eqref{maxmin21}.
	\end{itemize}	
\end{lemm}
\textit{Proof:} See Appendix \ref{app4}.  $\hspace{325pt} \blacksquare$
%
%
%
%
%
%
%
According to Lemma 1, we can deal with the optimization problem in \eqref{maxmin21} by considering the modified design in \eqref{maxmin23}; in particular, it is possible by either supposing a sequence of design problems $\mathcal{P}^{(i)}$ e.g., with $\bar{w}_{\mathcal{S},b}^{(i)}$ proportional to $2^{(i-1)}$, $i\geq 1$, or employing a practically fixed large enough $\bar{w}_{\mathcal{S},b}^{(i)}$. Indeed, by solving Problem \eqref{maxmin23} for increasing values of ${w}_{\mathcal{S},b}$, an optimal solution to \eqref{maxmin21} can be asymptotically achieved\footnote{Note that Lemma~1 is derived for SISO system model; however, it can be straightforwardly applied to design problems associated with more general system architectures including MIMO, CSS, and MIMO-CSS.}.

\begin{figure}[!t]
	\centering
	\begin{tikzpicture}[even odd rule,rounded corners=2pt,x=11.8pt,y=12pt,scale=.8]
		\draw[thick] (0-3,.5) rectangle ++(3,3) node[midway]{ $\mathbf{X}_{\mathcal{S}}$ };
		\draw[thick] (5.5-3,.5) rectangle ++(3,3) node[midway]{$\mathbf{X}^{T}_{\mathcal{S}}$ };
		
		\draw[->,line width=1,black!100] (-4.7-3,2)--+(4.5,0);
		\draw[->,line width=1,black!100] (3-3,2)--+(2.5,0);
		\draw[->,line width=1,black!100] (8.5-3,2)--+(5.2,0);
		
		\node [] at (-2.4-3,2.7) {\footnotesize Input ($\mathbf{s}_{\mathcal{S}}$)};
		\node [] at (11.3-3,2.8) {\footnotesize Output ($\mathbf{y}_{\mathcal{S}}$)};
		
		\node [] at (11.3-14.3,-11) {\footnotesize a)~Network architecture};
		\node [] at (11.3+7,-11) {\footnotesize b)~Network weights};


		\draw[thick,rounded corners=0pt,fill=blue!10] (16+15,-7.9) rectangle ++(2.5,2) node[midway]{\footnotesize 0};
		\draw[thick,rounded corners=0pt] (16+15,-7.9+2) rectangle ++(2.5,3*2) node[midway]{\footnotesize $\vdots$};	
		\draw[thick,rounded corners=0pt,fill=blue!10] (16+15,-7.9+4*2) rectangle ++(2.5,2) node[midway]{\footnotesize 0};
		\draw[thick,rounded corners=0pt,fill=green!20] (16+15,-7.9+5*2) rectangle ++(2.5,2) node[midway]{\tiny $x(N)$};
		\draw[thick,rounded corners=0pt] (16+15,-7.9+6*2) rectangle ++(2.5,2*2) node[midway]{\footnotesize $\vdots$};
		\draw[thick,rounded corners=0pt,fill=green!20] (16+15,-7.9+8*2) rectangle ++(2.5,2) node[midway]{\tiny $x(2)$};
		\draw[thick,rounded corners=0pt,fill=green!20] (16+15,-7.9+9*2) rectangle ++(2.5,2) node[midway]{\tiny $x(1)$};

		\draw[thick,rounded corners=0pt,fill=blue!10] (16+15+2.5,-7.9) rectangle ++(2.5,2) node[midway]{\footnotesize 0};
		\draw[thick,rounded corners=0pt] (16+15+2.5,-7.9+2) rectangle ++(2.5,2*2) node[midway]{\footnotesize $\vdots$};
		\draw[thick,rounded corners=0pt,fill=blue!10] (16+15+2.5,-7.9+3*2) rectangle ++(2.5,2) node[midway]{\footnotesize 0};
		\draw[thick,rounded corners=0pt,fill=green!20] (16+15+2.5,-7.9+4*2) rectangle ++(2.5,2) node[midway]{\tiny $x(N)$};
		\draw[thick,rounded corners=0pt] (16+15+2.5,-7.9+5*2) rectangle ++(2.5,2*2) node[midway]{\footnotesize $\vdots$};	
		\draw[thick,rounded corners=0pt,fill=green!20] (16+15+2.5,-7.9+7*2) rectangle ++(2.5,2) node[midway]{\tiny $x(2)$};	
		\draw[thick,rounded corners=0pt,fill=green!20] (16+15+2.5,-7.9+8*2) rectangle ++(2.5,2) node[midway]{\tiny $x(1)$};
		\draw[thick,rounded corners=0pt,fill=blue!10] (16+15+2.5,-7.9+9*2) rectangle ++(2.5,2) node[midway]{\footnotesize 0};

		\draw[thick,rounded corners=0pt,fill=white!10] (16+15+2*2.5,-7.9) rectangle ++(2*2.5,10*2) node[midway]{\footnotesize $\hdots$};

		\draw[thick,rounded corners=0pt,fill=green!20] (16+15+4*2.5,-7.9) rectangle ++(2.5,2) node[midway]{\tiny $x(N)$};
		\draw[thick,rounded corners=0pt] (16+15+4*2.5,-7.9+2) rectangle ++(2.5,3*2) node[midway]{\footnotesize $\vdots$};	
		\draw[thick,rounded corners=0pt,fill=green!20] (16+15+4*2.5,-7.9+4*2) rectangle ++(2.5,2) node[midway]{\tiny $x(2)$};	
		\draw[thick,rounded corners=0pt,fill=green!20] (16+15+4*2.5,-7.9+5*2) rectangle ++(2.5,2) node[midway]{\tiny $x(1)$};
		\draw[thick,rounded corners=0pt,fill=blue!10] (16+15+4*2.5,-7.9+6*2) rectangle ++(2.5,2) node[midway]{\footnotesize 0};	
		\draw[thick,rounded corners=0pt] (16+15+4*2.5,-7.9+7*2) rectangle ++(2.5,2*2) node[midway]{\footnotesize $\vdots$};
		\draw[thick,rounded corners=0pt,fill=blue!10] (16+15+4*2.5,-7.9+9*2) rectangle ++(2.5,2) node[midway]{\footnotesize 0};

		\draw [decorate,decoration={brace,amplitude=6.2pt},xshift=-2pt,yshift=0pt]
		(16+15,-7.8) -- (16+15,2.10) node [black,midway,xshift=-0.55cm] 
		{\tiny $N-1$};
		
		\draw [decorate,decoration={brace,amplitude=6.2pt,mirror,raise=4pt},xshift=-2pt,yshift=0pt]
		(16.05+4+16,-7.75) -- (16.05+4+16,0.1) node [black,midway,xshift=0.68cm] 
		{\tiny $N-2$};
		
		\draw [decorate,decoration={brace,amplitude=6.2pt},xshift=-2pt,yshift=0pt]
		(18.15+8.8+14,-4.9+9) -- (18.15+8.8+14,1.95+10) node [black,midway,xshift=-0.55cm] 
		{\tiny $N-1$};

		\draw[thick,rounded corners=0pt,fill=white!20] (29-13-1.5,-4.1) rectangle ++(13,12.2) ;

		\draw[thick,rounded corners=0pt,fill=green!20] (29-2-13+.5,-4.1) rectangle ++(3,2) node[midway]{\tiny $x(N)$}; 
		\draw[thick,rounded corners=0pt,fill=green!20] (29-2-13+3+.5,-4.1) rectangle ++(3,2) node[midway]{\tiny $x(1)$};
		\draw[thick,rounded corners=0pt,fill=green!20] (29-2-13+6+4+.5,-4.1) rectangle ++(3,2) node[midway]{\tiny $x(N-1)$};

		\draw[thick,rounded corners=0pt,fill=green!20] (29-2-13+3+.5,-4.1+2) rectangle ++(3,2) node[midway]{\tiny $x(N)$};
		\draw[thick,rounded corners=0pt,fill=green!20] (29-2-13+.5,-4.1+2) rectangle ++(3,2) node[midway]{\tiny $x(N-1)$};
		\draw[thick,rounded corners=0pt] (29-2-13+3+.5,-4.1+2+2) rectangle ++(3,4.2) node[midway]{$\vdots$};
		\draw[thick,rounded corners=0pt] (29-2-13+.5,-4.1+2+2) rectangle ++(3,4.2) node[midway]{$\vdots$};

		\draw[thick,rounded corners=0pt,fill=green!20] (29-2-13+.5,-4.1+2+2+4.2) rectangle ++(3,2) node[midway]{\tiny $x(2)$};
		\draw[thick,rounded corners=0pt,fill=green!20] (29-2-13+.5,-4.1+2+2+4.2+2) rectangle ++(3,2) node[midway]{\tiny $x(1)$};

		\draw[thick,rounded corners=0pt,fill=green!20] (29-2-13+3+.5,-4.1+2+2+4.2) rectangle ++(3,2) node[midway]{\tiny $x(3)$};
		\draw[thick,rounded corners=0pt,fill=green!20] (29-2-13+3+.5,-4.1+2+2+4.2+2) rectangle ++(3,2) node[midway]{\tiny $x(2)$};
		
		\draw[thick,rounded corners=0pt,fill=white!20] (29-2-13+3+3+.5,-4.1) rectangle ++(4,12.2) node[midway]{$\hdots$};

		\draw[thick,rounded corners=0pt,fill=green!20] (29-2-13+3+4+3+.5,-4.1+2+2+4.2+2) rectangle ++(3,2) node[midway]{\tiny $x(N)$};
		\draw[thick,rounded corners=0pt,fill=green!20] (29-2-13+3+4+3+.5,-4.1+2+2+4.2+2-2) rectangle ++(3,2) node[midway]{\tiny $x(1)$};
		
		\draw[thick,rounded corners=0pt,fill=green!20] (29-2-13+3+4+3+.5,-4.1+2+2+4.2+2-4.2-2-2) rectangle ++(3,2) node[midway]{\tiny $x(N-2)$};
		\draw[thick,rounded corners=0pt] (29-2-13+3+4+3+.5,-4.1+2+2+4.2+2-4.2-2) rectangle ++(3,4.2) node[midway]{$\vdots$};

		\node [] at (21.5+15+.8,-9.3) {\footnotesize Aperiodic: $\mathbf{X}_{{\mathcal{S}},AP}(\mathbf{x})$ };
		\node [] at (36.5-2-13-1+.9,-5.5) {\footnotesize Periodic: $\mathbf{X}_{{\mathcal{S}},P} (\mathbf{x})$};

	\end{tikzpicture}
	\caption{Two-layer BiSCorN to design aperiodoc/periodic binary sequences within the SISO radar context.}
	\label{h0t42}
	\centering
\end{figure}
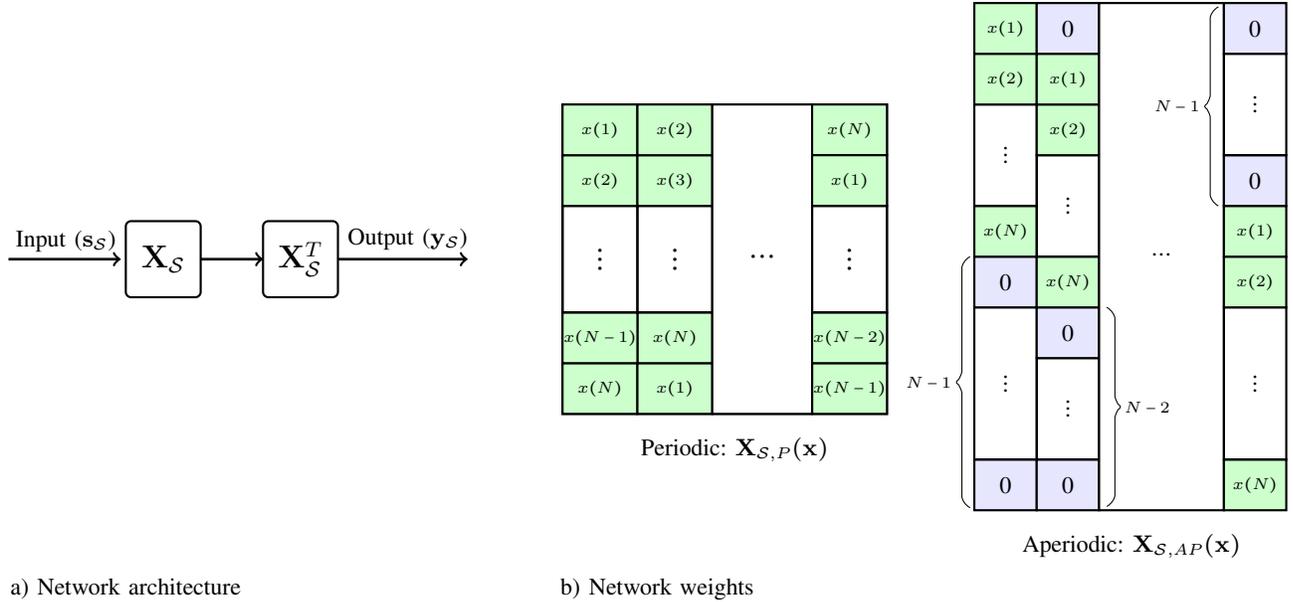

We begin the code design for SISO case via a learning-based framework. Indeed, the objective function of Problem \eqref{maxmin23} can be considered as the loss function of a neural network in which the desired solution is not the network output but the network weights, as opposed to the typical usages of neural networks in the literature \cite{chai2021deep,Redmark}. Precisely, we propose a two-layer fully connected BiSCorN with the following loss function
\begin{align} \label{key12}
	\mathcal{L}_{\mathcal{S}} & =  \left \| \mathbf{y}_{\mathcal{S}} \left( \mathbf{s}_{\mathcal{S}}, \mathbf{x}  \right) -N \mathbf{s}_{\mathcal{S}} \right \| _2^2  , 
\end{align}
where ${\mathbf{s}_{\mathcal{S}}} \in \mathbb{R}^N$ is the network input, $N\hspace{1pt} \mathbf{s}_{\mathcal{S}}$ is the desired output, and $\mathbf{y}_{\mathcal{S}}$ is the output of the network. As shown in Fig.~\ref{h0t42}.a, the output vector can be obtained as $\mathbf{y}_{\mathcal{S}} \left( \mathbf{s}_{\mathcal{S}}, \mathbf{x}  \right)=\mathbf{X}^{T}_{\mathcal{S}} \mathbf{X}_{\mathcal{S}} \mathbf{s}_{\mathcal{S}}$, where the matrix $\mathbf{X}_{\mathcal{S}}$ and its transpose can be used respectively for the first and second layer weights of the BiSCorN\footnote{This structure can be considered as an auto-encoder \cite{hinton1994autoencoders} composed of an encoder $\mathbf{X}_{\mathcal{S}}$ and a decoder $\mathbf{X}_{\mathcal{S}}^T$. Note that since the decoder is transpose of the encoder, our proposed network can be regarded as a kind of weight-tied auto-encoders \cite{li2018random}.}. For notation simplicity, $\mathbf{X}_{\mathcal{S}}$ denotes either $\mathbf{X}_{\mathcal{S},P}$ or $\mathbf{X}_{\mathcal{S},AP}$, i.e.,
\begin{equation} \label{trans}
	\mathbf{X}_{\mathcal{S}} (\mathbf{x})=
	\begin{cases}
		\mathbf{X}_{\mathcal{S},AP} (\mathbf{x}), \hspace{8pt} \textrm{Aperiodic},\\
		\mathbf{X}_{\mathcal{S},P} (\mathbf{x}), \hspace{15pt} \textrm{Periodic},
	\end{cases}
\end{equation}
where
\begin{equation}
	\mathbf{X}_{\mathcal{S},P}(\mathbf{x})=[\mathbf{x}_{P,0}(\mathbf{x}),\mathbf{x}_{P,1}(\mathbf{x}),...,\mathbf{x}_{P,N-1}(\mathbf{x})]\in \mathbb{R}^{N \times N},
\end{equation}
\begin{equation}
	\mathbf{X}_{\mathcal{S},AP}(\mathbf{x})=[\mathbf{x}_{AP,0}(\mathbf{x}),\mathbf{x}_{AP,1}(\mathbf{x}),...,\mathbf{x}_{AP,N-1}(\mathbf{x})]\in \mathbb{R}^{(2N-1) \times N},
\end{equation}
with $\mathbf{x}_{P,i}(\mathbf{x})$ and $\mathbf{x}_{AP,i}(\mathbf{x})$ are respectively the circularly shifted versions of $\mathbf{x}$ and $\mathbf{x}_{\textrm{pad}}=[\mathbf{x}^T,0,0,...,0]^T$ $\in \mathbb{R}^{2N-1}$ by $i$ elements (as shown in Fig.~\ref{h0t42}.b). Note that the constraint $\vert x (n) \vert \leq 1,~n=1,...,N-1,$ is applied to the network weights (see \cite{courbariaux2016binarized} for a similar case). 
Then, by substituting $\mathbf{X}_{\mathcal{S}}$, the $k^{th}$ entry of the network output can be written as (see Appendix~\ref{app2})
\begin{equation} \label{al}
	y_{\mathcal{S},k} = \begin{cases}
		\sum_{i=1}^{N} r(i-1) \hspace{2pt}s_{\mathcal{S},i} , \hspace{113pt} k=1,\\ 
		\sum_{i=1}^{k} r(k-i) \hspace{2pt}s_{\mathcal{S},i} +
		\sum_{i=k+1}^{N} r(i-k) \hspace{2pt}s_{\mathcal{S},i}  
		, \hspace{8pt} 2 \leq k \leq N-1, \\ 
		\sum_{i=1}^{N} r(N-i) \hspace{2pt}s_{\mathcal{S},i},\hspace{108pt} k=N.
	\end{cases}
\end{equation}
Herein, by selecting $\mathbf{s}_{\mathcal{S}}=[{a_{\mathcal{S}}},0,0,...,0]^T \in \mathbb{R}^{N}$ (for arbitrary $a_{\mathcal{S}}\neq0$) and by using \eqref{key} as well as \eqref{al}, the loss function in \eqref{key12} boils down to 
\begin{align} \label{key1}
	\mathcal{L}_{\mathcal{S}} & = a^2_{\mathcal{S}} \left( (r(0)-N )^2 + r^2(1)  + ...+ r^2(N-1)  \right) 
	=\mathbf{a}^T_{\mathcal{S}} \hspace{2pt} \widetilde{\mathbf{r}}_{\mathcal{S}},
\end{align}
where $\mathbf{a}_{\mathcal{S}}=[a^2_{\mathcal{S}},a^2_{\mathcal{S}},...,a^2_{\mathcal{S}}]^T \in \mathbb{R}^{N}_{+}$. 
Accordingly, the architecture in Fig.~\ref{h0t42} can deal with the design problem in \eqref{maxmin23} as long as equal ISL weights are considered, i.e., $\widetilde{\mathbf{w}}_{\mathcal{S}}=\mathbf{a}_{\mathcal{S}}$. Note that the equal weights do not necessarily lead to optimal binary solutions as addressed by Lemma~1; however, we numerically observed that binary sequences associated with the equal weights possess good correlation properties (see Remark~1 below and Section~\ref{per}). Furthermore, another version of BiSCorN accounting for arbitrary weights and hence those addressed in Lemma~1 will be devised in Section~\ref{lcz}.

\begin{figure}
	\centering
	\subfigure[Normalized evaluated loss function for 10 trials.]{\includegraphics[width=9cm,height=4.5cm]{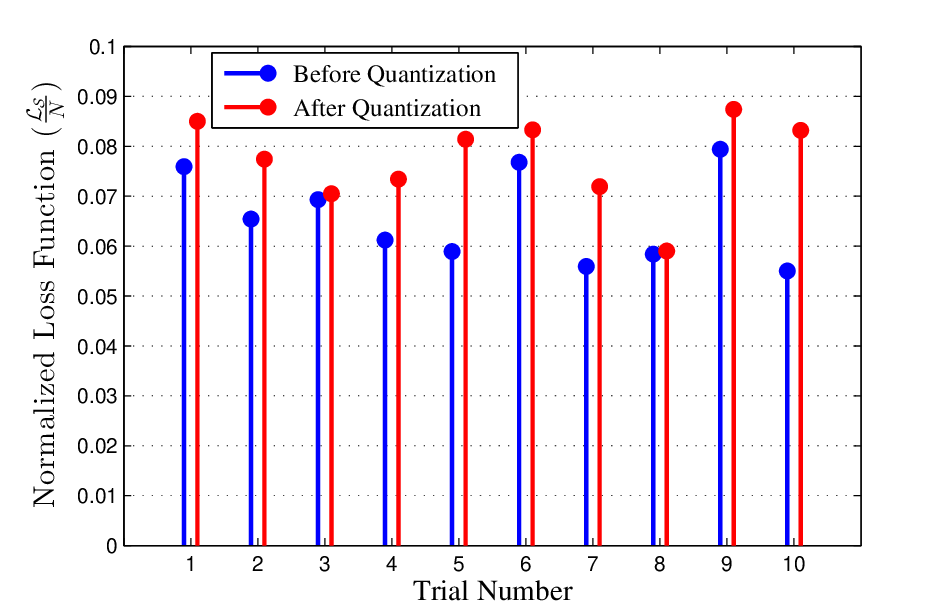}}
	\vfill
	\subfigure[The grayscale image of matrix $\mathbf{E}=\frac{1}{N} \mathbf{X}^T_{\mathcal{S}} \mathbf{X}_{\mathcal{S}}$ before quantization.]{\includegraphics[width=7cm,height=4.5cm]{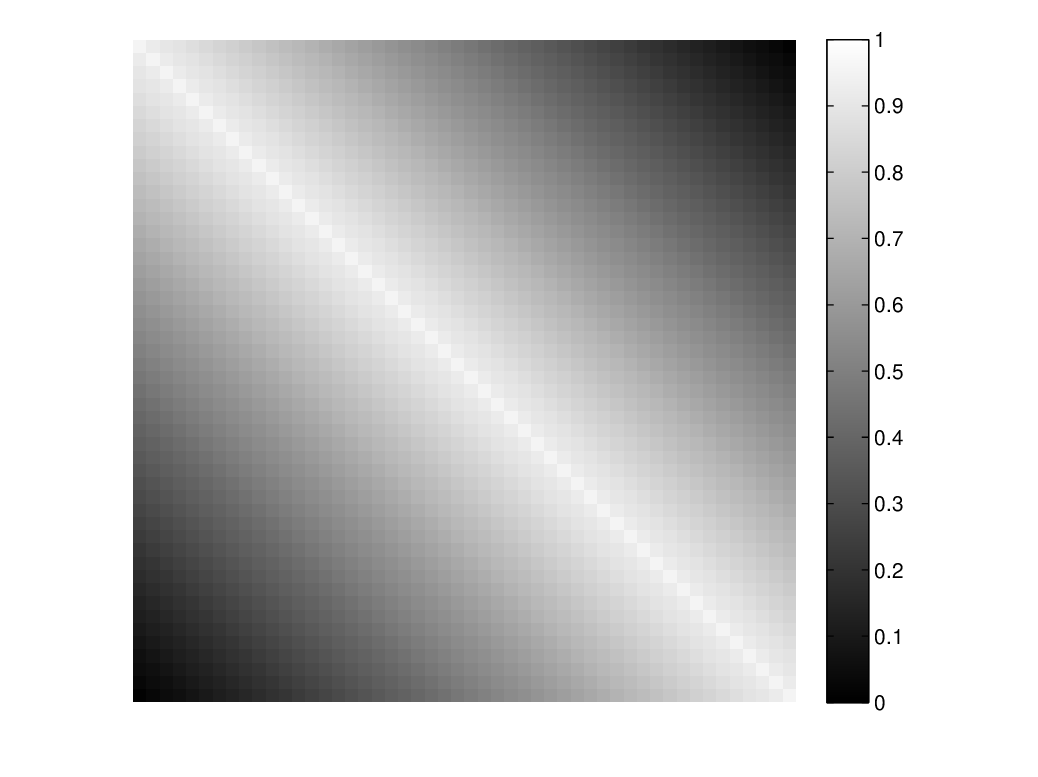}}
	\hfill
	\subfigure[The grayscale image of matrix $\mathbf{E}=\frac{1}{N} \mathbf{X}^T_{\mathcal{S}} \mathbf{X}_{\mathcal{S}}$ after quantization.]{\includegraphics[width=7cm,height=4.5cm]{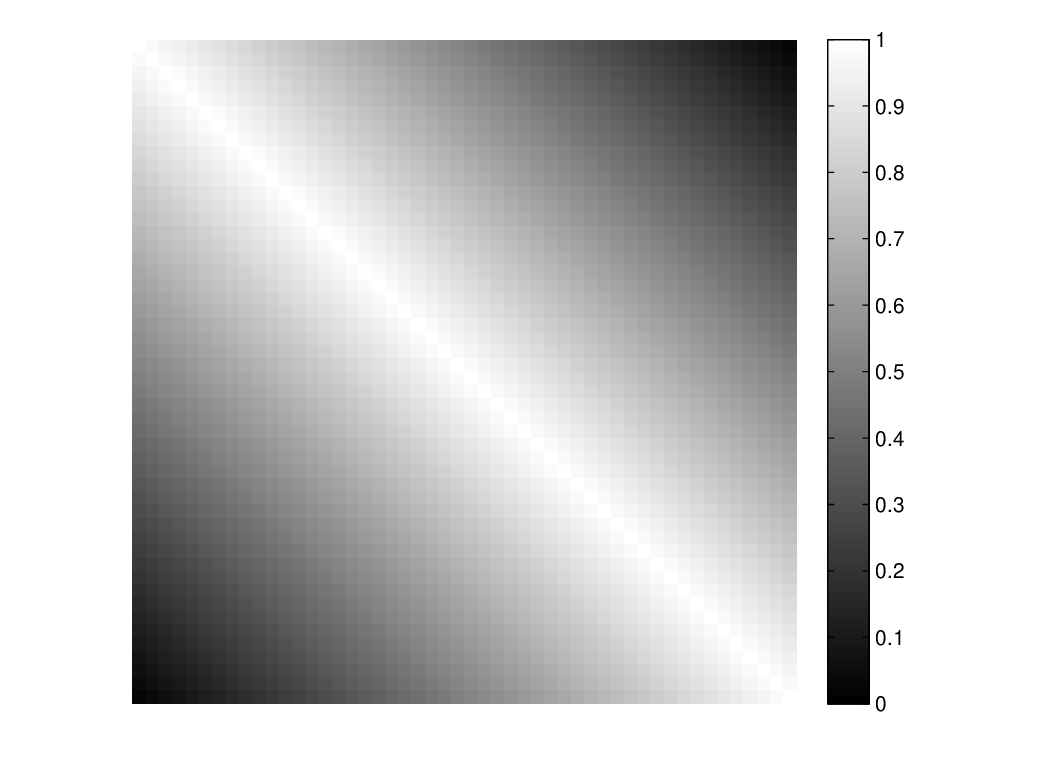}}
	\vfill
	\subfigure[Amplitude of input/output signals before quantization.]{\includegraphics[width=8.1cm,height=4.5cm]{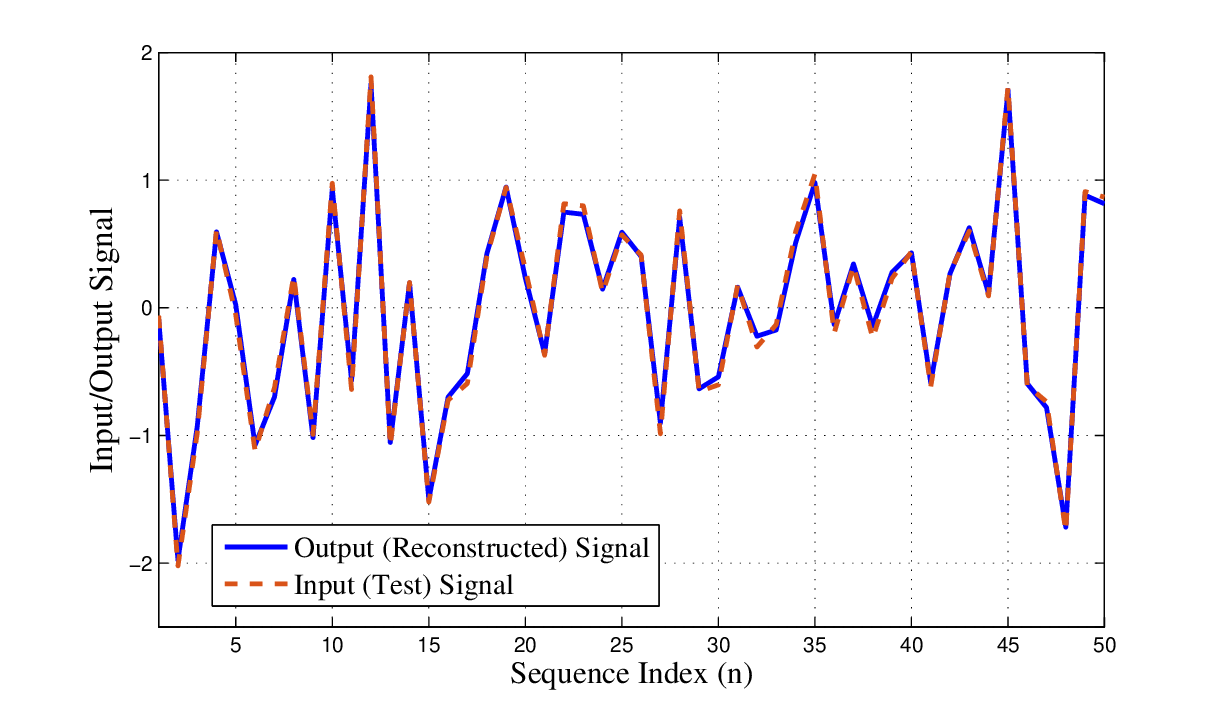}}
	\hfill
	\subfigure[Amplitude of input/output signals after quantization.]{\includegraphics[width=8.1cm,height=4.5cm]{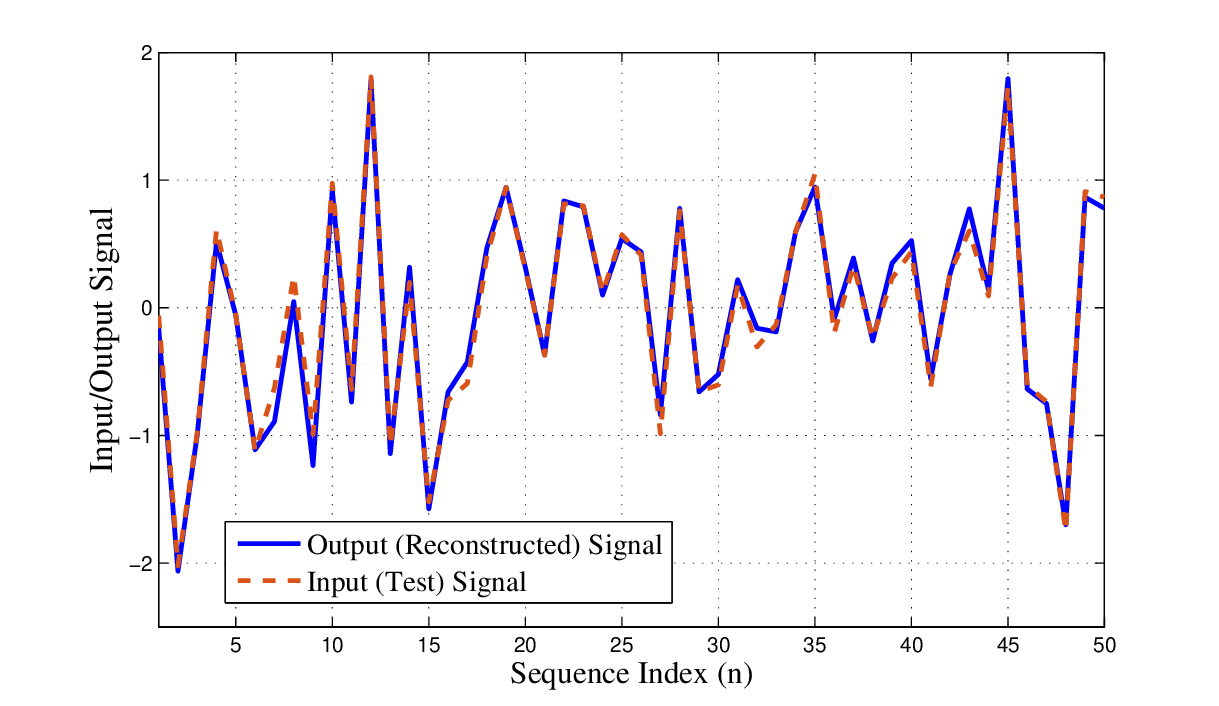}}
	\hfill
	\caption{Performance evaluation of SISO two-layer BiSCorN for (test) random input signals before/after quantization for sequence length $N=50$.}

	\label{htttt1ttt0}
\end{figure}

\begin{rema}[synthesis stage]
The feasible set of the design problem in \eqref{maxmin21} for the special case of $N=3$ consists of vertices of a cube\footnote{For $N>3$, the feasible set is a hypercube.} centered at origin with the side length equal to 2. Whereas, for the relaxed problem in \eqref{maxmin23}, the feasible set is the total volume inside the aforementioned cube. Note that the objective function in Problem \eqref{maxmin23} includes a penalty term for non-binary solutions (that is, deviation from the vertices of the cube). However, the proposed method might converge to sequences with a few non-binary elements (see Subsection \ref{bibi}). For example, Fig.~\ref{ht2} shows a solution for the relaxed problem that is located at one of the cube edges: having two binary elements and one non-binary element. Therefore, a final stage is used after convergence of the method to quantize non-binary elements to the nearest values in the set $\lbrace -1, +1 \rbrace$ and obtain a feasible sequence to the original design problem in \eqref{maxmin21}.
%
%
\end{rema}
\begin{figure}
	\centering
	\begin{tikzpicture}[even odd rule,rounded corners=2pt,x=12pt,y=12pt,scale=.35,every node/.style={scale=.8}]

		\draw[-,line width=1,dashed,black!100] (-3,-3.5)--+(0,7);
		\draw[->,line width=1,black!100] (-3,3.5)--+(0,7);
		
		\draw[-,line width=1,dashed,black!100] (-3,-3.5)--+(8.30,-.9);
		\draw[->,line width=1,black!100] (5.5,-4.4)--+(7,-.73);
		
		\draw[-,line width=1,dashed,black!100] (-3,-3.5)--+(-2.5,-2.5);
		\draw[->,line width=1,black!100] (-5.5,-6)--+(-2.5/2,-2.5/2);


		\draw[thick,fill=red!100,black] (-15,2) circle (.15cm);
		\node [] at (-16,4) {\scriptsize (1,-1,1)};

		\draw[thick,fill=black!100] (-8.2-1.8-.5-2.2-1-1.2,-1.8-1.5+4-4+3-1.5) circle (.2cm);

		\draw[-,line width=1,blue!100] (-6.5,-4.5)--+(-1.7,2.7);	
		\draw[-,line width=1,blue!100] (-8.2,-1.8)--+(-1.8,-1.5);	
		\draw[-,line width=1,blue!100] (-8.2-1.8,-1.8-1.5)--+(-.5,4);	
		\draw[-,line width=1,blue!100] (-8.2-1.8-.5,-1.8-1.5+4)--+(-2.2,-4);
		\draw[-,line width=1,blue!100] (-8.2-1.8-.5-2.2,-1.8-1.5+4-4)--+(-1,3);	
		\draw[->,line width=1,blue!100] (-8.2-1.8-.5-2.2-1,-1.8-1.5+4-4+3)--+(-1.2,-1.5);

%

		\draw[thick,fill=black!100] (-6.5,-4.5) circle (.18cm);
		\node [text=red!100] at (-3.0,-2.7) {\scriptsize initial point};

		\draw[thick,fill=black!100] (-10,-10) circle (.15cm);
		\node [] at (-7.5,-12) {\scriptsize (-1,-1,-1)};
		
		\draw[thick,fill=black!100] (4,0) circle (.15cm);
		\node [] at (6.7,-.5) {\scriptsize (1,1,1)};

		\draw[thick,fill=black!100] (9,-12) circle (.15cm);	
		\node [] at (6.5,-10.5) {\scriptsize (-1,1,-1)};
		
		\draw[thick,fill=black!100] (-15,-17) circle (.15cm);
		\node [] at (-11,-16) {\scriptsize (1,-1,-1)};
		
		\draw[thick,fill=black!100] (-10,7) circle (.15cm);
		\node [] at (-8,8.5) {\scriptsize (-1,-1,1)};
		
		\draw[thick,fill=black!100] (9,5) circle (.15cm);
		\node [] at (7.5,7) {\scriptsize (-1,1,1)};
		
		\draw[thick,fill=black!100] (4,-19) circle (.15cm);	
		\node [] at (1,-17) {\scriptsize (1,1,-1)};
		
		\draw[-,line width=1,dashed,black!100] (-10,-10)--+(19,-2);
		
		\draw[-,line width=1,dashed,black!100] (-10,-10)--+(-5,-7);
		
		\draw[-,line width=1,dashed,black!100] (-10,-10)--+(0,17);
		
		\draw[-,line width=1,dashed,black!100] (-10,7)--+(-5,-5);
		
		\draw[-,line width=1,dashed,black!100] (-15,-17)--+(0,19);
		
		\draw[-,line width=1,dashed,black!100] (-10,7)--+(19,-2);
		
		\draw[-,line width=1,dashed,black!100] (-15,2)--+(19,-2);
		
		\draw[-,line width=1,dashed,black!100] (-15,-17)--+(19,-2);
		
		\draw[-,line width=1,dashed,black!100] (9,-12)--+(-5,-7);
		
		\draw[-,line width=1,dashed,black!100] (9,-12)--+(0,17);
		
		\draw[-,line width=1,dashed,black!100] (4,-19)--+(0,19);
		
		\draw[-,line width=1,dashed,black!100] (4,0)--+(5,5);
	\end{tikzpicture}
	\caption{The feasible set of the SISO problem in \eqref{maxmin23} for $N=3$.}
	\label{ht2}
	\centering
\end{figure}
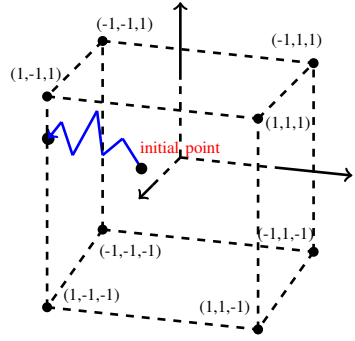
\begin{rema}[another perspective on BiSCorN]
It is known that the transmitted binary sequence $\mathbf{x}$ has ideal auto-correlation properties if $r(0)=N$ and $r(k)=0,~k=1,...,N-1$. Therefore, for an ideal sequence design scheme $\mathbf{X}^T_{\mathcal{S}} \mathbf{X}_{\mathcal{S}} =N \mathbf{I}_N$. We expect that exploiting BiSCorN provides a matrix $\mathbf{X}_{\mathcal{S}}$ such that $\mathbf{X}^T_{\mathcal{S}} \mathbf{X}_{\mathcal{S}} \approx  N \mathbf{I}_N$.
To evaluate the performance of BiSCorN, one may define $\mathbf{E} = \frac{1}{N} \mathbf{X}^T_{\mathcal{S}} \mathbf{X}_{\mathcal{S}}$; and next feed the network with a random test signal for monitoring the entries of $\mathbf{E}$ (see Section~\ref{ext} for more details). That is, the lower the energy of off-diagonal entries the better the performance. The evaluation of the normalized loss function, i.e., $\frac{\mathcal{L}_{\mathcal{S}}}{N}$, for $10$ random input signals (as test signals) before and after quantization are shown in Fig.~\ref{htttt1ttt0}.a for the case of $N=50$. Also, Fig.~\ref{htttt1ttt0}.b-c exemplify the matrix $\mathbf{E}$ in a grayscale image and its corresponding network input/output are shown for a trial in Fig.~\ref{htttt1ttt0}.d-e. Precisely, Fig.~\ref{htttt1ttt0}.d-e show that how closely the network output follows the network input. This figure illustrates the goodness of BiSCorN even after applying quantization. 
\end{rema}
\subsection{MIMO}\label{MIMOpro}	
A MIMO radar equipped with $N_T$ transmit antennas is considered, where each antenna emits a different code vector with $N$ subpulses. Let us denote by $\mathbf{x}_m=[x_m(1), x_m(2), ... , x_m(N)]^{T} \in \mathbb{R}^N,~m=1,...,N_T,$ the transmit code vector at the $m^{th}$ antenna. The aperiodic and periodic cross-correlation functions between $\mathbf{x}_m$ and $\mathbf{x}_l$ in MIMO system are defined as
\begin{equation}
	r_{AP,ml} (k)= \sum_{n=1}^{N-k} x_m(n) \hspace{1pt} x_l(n+k),~~m,l=1, ..., N_T,~~-N+1 \leq k \leq N-1,
\end{equation}
and
\begin{equation}
	r_{P,ml} (k)= \sum_{n=1}^{N}  x_m(n) \hspace{1pt} x_l{({(n+k)}_{\textrm{\tiny{mod}}\hspace{1pt}N})},~~m,l=1, ..., N_T,~~-N+1 \leq k \leq N-1,
\end{equation} 
respectively. Defining $\widetilde{\mathbf{X}}_{\mathcal{M}}=[\mathbf{x}_1,\mathbf{x}_2,...,\mathbf{x}_{N_T} ]$, the design of binary space-time radar codes can be cast as\footnote{The subscript $\mathcal{M}$ is associated with \textit{MIMO}.}
\begin{eqnarray}\label{maxmin221}
	\min_{ \widetilde{\mathbf{X}}_{\mathcal{M}} } & \hspace{-7pt} w_{\mathcal{M},0} \sum_{\substack{m,l=1\\m \neq l}}^{N_T} \vert  r_{ml} (0) \vert^2 +  \sum_{k=1 }^{N-1} w_{\mathcal{M},k} \sum_{m,l=1}^{N_T} \vert r_{ml} (k) \vert^2 + \sum_{k=-N+1 }^{-1} w_{\mathcal{M},-k} \sum_{m,l=1}^{N_T} \vert r_{ml} (k) \vert^2 \\ \nonumber
	\mbox{s. t.}\;\;& x_m (n) \in \lbrace -1,+1 \rbrace,~m=1,...,N_T,~n=1,...,N,
\end{eqnarray}
where $\mathbf{w}_{\mathcal{M}}=[w_{\mathcal{M},0},w_{\mathcal{M},1}, ... ,w_{\mathcal{M},N-1}]^T$ with $w_{\mathcal{M},k} \geq 0, \hspace{2pt} k=1,...,N-1,$ is the weighting vector associated with the correlation terms and $r_{ml}(k)$ refers to either $r_{AP,ml} (k)$ or $r_{P,ml} (k)$. Note that Problem \eqref{maxmin221} is a non-convex NP-hard problem \cite{CAN}. 

Using the symmetry property $r_{ml} (k)= r_{lm} (-k), \hspace{2pt}\forall m,l,k$, the objective of \eqref{maxmin221} can be cast as
\begin{equation}
	w_{\mathcal{M},0} \sum_{\substack{m,l=1\\m \neq l}}^{N_T} \vert  r_{ml} (0) \vert^2 +  2\sum_{k=1 }^{N-1} w_{\mathcal{M},k} \sum_{m,l=1}^{N_T} \vert r_{ml} (k) \vert^2.
\end{equation}
Next, similar to the procedures in Subsection~\ref{SISOpro}, the binary constraint on $x_m (n)$ is relaxed in $\vert x_m (n)\vert \leq 1$ and the term ${w}_{\mathcal{M},b}  \sum_{m=1}^{N_T}  \vert r_{mm}(0)-N \vert^2$ is added to the objective function to penalize non-binary solutions. Precisely, the design problem in \eqref{maxmin221} is approximated as
\begin{eqnarray}\label{maxmin22563}
	\min_{ \widetilde{\mathbf{X}}_{\mathcal{M}} } &  \widetilde{\mathbf{w}}^T_{\mathcal{M}} \hspace{1pt}\hspace{1pt}\widetilde{\mathbf{r}}_\mathcal{M}   \\ \nonumber
	\mbox{s. t.}\;\;& \vert x_m (n) \vert \leq 1,~~m=1, ..., N_T,~~n=1,...,N,
\end{eqnarray}
where
\begin{equation} \label{key245}
	\widetilde{{r}}_{\mathcal{M}} (k) =
	\begin{cases}
		\sum_{m=1}^{N_T}  \vert r_{mm}(0)-N \vert^2,\hspace{2pt}~~k=1, \\ 
		\sum_{\substack{m,l=1\\m \neq l}}^{N_T} \vert  r_{ml} (0) \vert^2,\hspace{2pt}~~~~~~~\hspace{1pt}k=2,
		\\
		2\sum_{m,l=1}^{N_T} \vert r_{ml} (k-1) \vert^2,\hspace{2pt}~k=3,...,N+1,
	\end{cases}
\end{equation}
and $\widetilde{\mathbf{w}}_{\mathcal{M}} =[{w}_{\mathcal{M},b}, {w}_{\mathcal{M},s} \times \mathbf{w}^T_{\mathcal{M}}]^T \in \mathbb{R}^{N+1}_{+}$ is the design parameter managing the sidelobe level suppression and amplitude variability of the synthesized code.

\begin{figure}[!t]
	\centering
	\begin{tikzpicture}[even odd rule,rounded corners=2pt,x=12pt,y=12pt,scale=.7,every node/.style={scale=.9}]
		
		\draw[thick] (0-6,.5) rectangle ++(3,3) node[midway]{ $\mathbf{X}_{\mathcal{M}}$ };
		\draw[thick] (5.5-6,.5) rectangle ++(3,3) node[midway]{$\mathbf{X}^{T}_{\mathcal{M}}$ };
		
		\draw[->,line width=1,black!100] (-5.5-6,2)--+(5.5,0);
		\draw[->,line width=1,black!100] (3-6,2)--+(2.5,0);
		\draw[->,line width=1,black!100] (8.5-6,2)--+(6,0);
		
		\node [] at (-2.8-6,2.7) {\footnotesize Input ($\mathbf{S}_{\mathcal{M}}$)};
		\node [] at (11.6-6,2.8) {\footnotesize Output ($\mathbf{Y}_{\mathcal{M}}$)};

		\node [] at (-6.7,-2.5) {\small a)~Network architecture};
		\node [] at (20-1,-2.5) {\small b)~Network weights};

		\draw[thick,rounded corners=0pt,fill=green!20] (19-1,0) rectangle ++(5,3.5) node[midway]{ $\widetilde{\mathbf{X}}_{\mathcal{S}}(\mathbf{x}_1)$ };
		\draw[thick,rounded corners=0pt,fill=green!20] (24-1,0) rectangle ++(5,3.5) node[midway]{$\widetilde{\mathbf{X}}_{\mathcal{S}}(\mathbf{x}_2)$ };
		\draw[thick,rounded corners=0pt] (29-1,0) rectangle ++(4,3.5) node[midway]{ $\hdots$ };
		\draw[thick,rounded corners=0pt,fill=green!20] (33-1,0) rectangle ++(5.1,3.5) node[midway]{ $\widetilde{\mathbf{X}}_{\mathcal{S}}(\mathbf{x}_{N_T})$ };

		\node [] at (17-1,2) { $\mathbf{X}_{\mathcal{M}}$: };
		
	\end{tikzpicture}
	\caption{Two-layer BiSCorN to design aperiodoc/periodic binary sequence sets within MIMO radar context.}
	\label{h0t51}
	\centering
\end{figure}
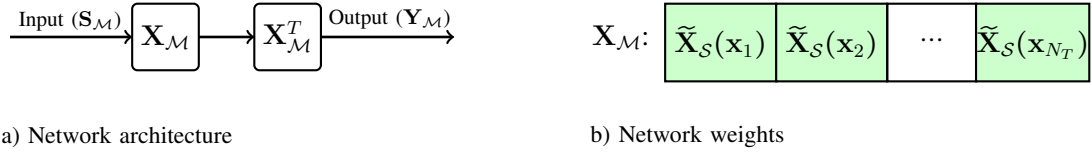

To proceed further, let
\begin{equation} \label{lj}
	\mathbf{X}_{\mathcal{M}}=
	\begin{cases}
		\mathbf{X}_{\mathcal{M},AP}= [\widetilde{\mathbf{X}}_{\mathcal{S},AP}(\mathbf{x}_1),\widetilde{\mathbf{X}}_{\mathcal{S},AP}(\mathbf{x}_2), ..., \widetilde{\mathbf{X}}_{\mathcal{S},AP}(\mathbf{x}_{N_{T}})]~ \in \mathbb{R}^{(2N-1) \times NN_T }, \hspace{8pt} \textrm{Aperiodic},
		\\
		\mathbf{X}_{\mathcal{M},P}=[\widetilde{\mathbf{X}}_{\mathcal{S},P}(\mathbf{x}_1),\widetilde{\mathbf{X}}_{\mathcal{S},P}(\mathbf{x}_2), ..., \widetilde{\mathbf{X}}_{\mathcal{S},P}(\mathbf{x}_{N_{T}})]~ \in \mathbb{R}^{N \times NN_T }, \hspace{56pt} \textrm{Periodic},
	\end{cases}
\end{equation}
where
\begin{equation}
	\widetilde{\mathbf{X}}_{\mathcal{S},AP}(\mathbf{x}_m)=[\mathbf{x}_{AP,0}(\mathbf{x}_m),\sqrt{2}\hspace{2pt}\mathbf{x}_{AP,1}(\mathbf{x}_m),...,\sqrt{2}\hspace{2pt}\mathbf{x}_{AP,N-1}(\mathbf{x}_m)]\in \mathbb{R}^{(2N-1) \times N}, 
\end{equation}
\begin{equation}
	\widetilde{\mathbf{X}}_{\mathcal{S},P}(\mathbf{x}_m)=[\mathbf{x}_{P,0}(\mathbf{x}_m),\sqrt{2}\hspace{2pt}\mathbf{x}_{P,1}(\mathbf{x}_m),...,\sqrt{2}\hspace{2pt}\mathbf{x}_{P,N-1}(\mathbf{x}_m)]\in \mathbb{R}^{N \times N},
\end{equation}
with $\mathbf{x}_{P,i}(\mathbf{x}_m)$ and $\mathbf{x}_{AP,i}(\mathbf{x}_m)$ are defined in Subsection~\ref{SISOpro}. Then, let us define $\widetilde{\mathbf{X}}_{\mathcal{S}}$ to get either $\widetilde{\mathbf{X}}_{\mathcal{S},AP}$ or $\widetilde{\mathbf{X}}_{\mathcal{S},P}$. In this case, the loss function can be defined as
\begin{align} \label{l2}
	\mathcal{L}_{\mathcal{M}} &=   \left \| \mathbf{Y}_{\mathcal{M}} \left( \mathbf{S}_{\mathcal{M}}, \widetilde{\mathbf{X}}_{\mathcal{M}} \right) - N\mathbf{S}_{\mathcal{M}} \right \|_F^2 , 
\end{align}
where $\mathbf{S}_{\mathcal{M}} \in \mathbb{R}^{NN_T \times N_T}$ is the input matrix, $N\mathbf{S}_{\mathcal{M}}$ is the desired output and ${\mathbf{Y}}_{\mathcal{M}} = \mathbf{X}_{\mathcal{M}}^T \mathbf{X}_{\mathcal{M}} \hspace{1pt} \mathbf{S}_{\mathcal{M}}$ is the output of the MIMO two-layer BiSCorN (as shown in Fig.~\ref{h0t51}). Let us select the input matrix $\mathbf{S}_{\mathcal{M}}$ as the following signal
\begin{equation}\label{keys}
	\mathbf{S}_{\mathcal{M}}={a_\mathcal{M}}\left[
	\mathbf{e}_1,  \mathbf{e}_{N+1}, ..., \mathbf{e}_{(N_T -1)N+1} \right]
	\in \mathbb{R}^{NN_T \times N_T},
\end{equation}
where $\mathbf{e}_m \in \mathbb{R}^{NN_T}$ is the $m^{th}$ standard vector and $a_\mathcal{M} \neq0$. 
By using \eqref{key245}, \eqref{lj}, and \eqref{keys}, the loss function in \eqref{l2} can be rewritten as (see Appendix~\ref{app3})
\begin{align} \label{13}
	\mathcal{L}_{\mathcal{M}} &=  a^2_\mathcal{M} \left( \sum_{m=1}^{N_T}  \vert r_{mm}(0)-N \vert^2 + \sum_{\substack{m,l=1\\m \neq l}}^{N_T} \vert  r_{ml} (0) \vert^2+ 2\sum_{k=1}^{N-1} \sum_{m,l=1}^{N_T} \vert r_{ml} (k) \vert^2 \right) = \mathbf{a}^T_{\mathcal{M}} \hspace{1pt} \widetilde{\mathbf{r}}_{\mathcal{M}},
\end{align}
where $\mathbf{a}_{\mathcal{M}}=[{a}^2_{\mathcal{M}},{a}^2_{\mathcal{M}},...,{a}^2_{\mathcal{M}}]^T \in \mathbb{R}^{N+1}_+$. Evidently, the loss function in \eqref{13} corresponds to \eqref{maxmin22563} provided that $\widetilde{\mathbf{w}}_{\mathcal{M}}=\mathbf{a}_{\mathcal{M}}$. 

\subsection{CSS}\label{CSSpro}
This subsection refers to a system where $M$ sequences, i.e., ${\mathbf{x}}_{m^{\prime}}=[{x}_{m^{\prime}} (1), {x}_{m^{\prime}}(2),...,$ ${x}_{m^{\prime}}(N)]^T \in \mathbb{R}^{N},~m^{\prime}=1,..,M$, are transmitted sequentially in time domain. Let $r_{m^{\prime}} (k)$ be the auto-correlation function of ${\mathbf{x}}_{m^{\prime}}$. At the receive side, $\sum_{m^{\prime}=1}^{M} r_{m^{\prime}} (k)$ is employed for detection purposes (see \cite{soltanalian2013fast} for more details). In this context, the set $\widetilde{\mathbf{X}}_{\mathcal{C}}=[{\mathbf{x}}_1, {\mathbf{x}}_2,...,{\mathbf{x}}_M]\in \mathbb{R}^{N \times M} $ is called \textit{complementary} iff the following condition is satisfied \cite{soltanalian2013fast}
\begin{equation}
	\sum_{m^{\prime}=1}^{M} r_{m^{\prime}} (k) =0,~~ 1 \leq \vert k \vert \leq N-1.
\end{equation}
Then, letting $\mathbf{w}_{\mathcal{C}}=[{w}_{\mathcal{C},1}, $ ${w}_{\mathcal{C},2},...,{w}_{\mathcal{C},N-1}]^T$ with ${w}_{\mathcal{C},k} \geq 0, \hspace{2pt} k=1,...,N-1,$ the WCISL given by\footnote{The subscript $\mathcal{C}$ is used for \textit{complementary}.}
\begin{equation}
	\textrm{WCISL}=2 \sum_{k=1}^{N-1} {w}_{\mathcal{C},k}  \left \vert  \sum_{m^{\prime}=1}^{M} r_{m^{\prime}} (k)  \right \vert^{2},
\end{equation}
can be used to measure the complementarity level of a set of codes. In this case, the design problem for binary sequence sets can be cast as
\begin{eqnarray}\label{maxmin1221}
	\min_{ \widetilde{\mathbf{X}}_{\mathcal{C}} } & \displaystyle \sum_{k=1}^{N-1} {w}_{\mathcal{C},k}  \left \vert  \sum_{m^{\prime}=1}^{M} r_{m^{\prime}} (k)  \right \vert^{2} \\ \nonumber
	\mbox{s. t.}\;\;& x_{m^{\prime}} (n) \in \lbrace -1,+1 \rbrace,~{m^{\prime}}=1,...,M,~n=1,...,N,
\end{eqnarray}
which is still a non-convex NP-hard optimization problem \cite{soltanalian2013fast}.

\begin{figure}[!t]
	\centering
	\begin{tikzpicture}[even odd rule,rounded corners=2pt,x=12pt,y=12pt,scale=.7,every node/.style={scale=.9}]
		
		\draw[thick] (0-14.2,.5-1.2) rectangle ++(3,3) node[midway]{ $\mathbf{X}_{\mathcal{C}}$ };
		\draw[thick] (5.5-14.2,.5-1.2) rectangle ++(3,3) node[midway]{$\mathbf{X}^{T}_{\mathcal{C}}$ };
		
		\draw[->,line width=1,black!100] (-4.8-14.2,2-1.2)--+(4.8,0);
		\draw[->,line width=1,black!100] (3-14.2,2-1.2)--+(2.5,0);
		\draw[->,line width=1,black!100] (8.5-14.2,2-1.2)--+(5.3,0);
		
		\node [] at (-2.4-14.2,2.7-1.2) {\footnotesize Input ($\mathbf{s}_{\mathcal{C}}$)};
		\node [] at (11.3-14.2,2.8-1.2) {\footnotesize Output ($\mathbf{y}_{\mathcal{C}}$)};

		\node [] at (-14,-8.5) {\small a)~Network architecture};
		\node [] at (17,-8.5) {\small b)~Network weights};

		\draw[thick,rounded corners=0pt,fill=green!20] (17.5-1.5,-6) rectangle ++(5,3.2) node[midway]{ $\mathbf{X}_{\mathcal{S}}(\mathbf{x}_M)$ };
		\draw[thick,rounded corners=0pt] (17.5-1.5,-2.8) rectangle ++(5,3.2) node[midway]{ $\vdots$ };
		\draw[thick,rounded corners=0pt,fill=green!20] (17.5-1.5,.4) rectangle ++(5,3.2) node[midway]{  $\mathbf{X}_{\mathcal{S}}(\mathbf{x}_2)$};
		\draw[thick,rounded corners=0pt,fill=green!20] (17.5-1.5,3.6) rectangle ++(5,3.2) node[midway]{ $\mathbf{X}_{\mathcal{S}}(\mathbf{x}_1)$ };

		\node [] at (15.5-1.5,.5) { $\mathbf{X}_{\mathcal{C}}$: };
		
	\end{tikzpicture}
	\caption{Two-layer BiSCorN to design aperiodoc/periodic complementary binary sequence sets within the SISO radar context.}
	\label{h0t52}
	\centering
\end{figure}
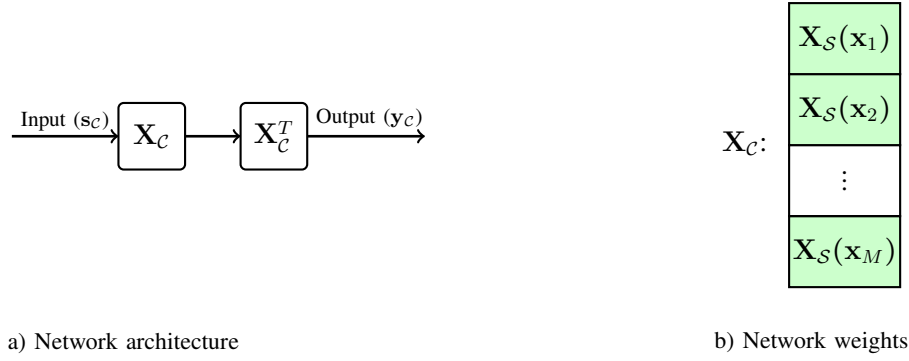

Following a line of reasoning similar to those pursued in Subsections \ref{SISOpro} and \ref{MIMOpro}, yields the following reformulated version of \eqref{maxmin1221}:
\begin{eqnarray}\label{maxmin225631}
	\min_{\lbrace \mathbf{x}_{m^{\prime}} \rbrace^{M}_{{m^{\prime}}=1} } &  \widetilde{\mathbf{w}}^T_{\mathcal{C}} \hspace{1pt}\hspace{1pt}\widetilde{\mathbf{r}}_\mathcal{C}   \\ \nonumber
	\mbox{s. t.}\;\;& \vert x_{m^{\prime}} (n) \vert \leq 1,~~{m^{\prime}}=1, ..., M,~~n=1,...,N,
\end{eqnarray}
where $\widetilde{\mathbf{w}}_{\mathcal{C}} =[{w}_{\mathcal{C},b}, {w}_{\mathcal{C},s} \times \mathbf{w}^T_{\mathcal{C}}]^T \in \mathbb{R}^{N}_{+}$ with ${w}_{\mathcal{C},b}+{w}_{\mathcal{C},s}=1,\hspace{2pt} {w}_{\mathcal{C},b} \geq 0, \hspace{2pt} {w}_{\mathcal{C},s} \geq 0 $ and
\begin{equation} \label{key2451}
	\widetilde{{r}}_{\mathcal{C}} (k) =
	\begin{cases}
		\left \vert  \sum_{m^{\prime}=1}^{M} r_{m^{\prime}}(0)- MN \right \vert^2,~~k=1, 
		\\
		\left \vert  \sum_{m^{\prime}=1}^{M}   r_{m^{\prime}}(k) \right \vert^2,~~~~~~~~~~k=2,...,N.
	\end{cases}
\end{equation}
Hence, we devise BiSCorN framework to deal with \eqref{maxmin225631}. In fact, let us denote
\begin{equation} \label{lj2}
	\mathbf{X}_{\mathcal{C}}=
	\begin{cases}
		\mathbf{X}_{\mathcal{C},AP}= [\mathbf{X}^T_{\mathcal{S},AP}(\mathbf{x}_1),\mathbf{X}^T_{\mathcal{S},AP}(\mathbf{x}_2), ..., \mathbf{X}^T_{\mathcal{S},AP}(\mathbf{x}_M) ]^T~ \in \mathbb{R}^{(2N-1)M \times N }, \hspace{8pt} \textrm{Aperiodic},
		\\
		\mathbf{X}_{\mathcal{C},P}=[\mathbf{X}^T_{\mathcal{S},P}(\mathbf{x}_1),\mathbf{X}^T_{\mathcal{S},P}(\mathbf{x}_2), ..., \mathbf{X}^T_{\mathcal{S},P}(\mathbf{x}_M) ]^T~ \in \mathbb{R}^{NM \times N }, \hspace{56pt} \textrm{Periodic},
	\end{cases}
\end{equation}
where $\mathbf{X}_{\mathcal{S},AP}$ and $\mathbf{X}_{\mathcal{S},P}$ given in Subsection~\ref{SISOpro}. Now, considering $\mathbf{s}_{\mathcal{C}}=[{a}_{\mathcal{C}},0,0,...,0]\in \mathbb{R}^{N}$ as input vector, $\mathbf{y}_{\mathcal{C}}=\mathbf{X}_{\mathcal{C}}^T \mathbf{X}_{\mathcal{C}} \hspace{1pt} \mathbf{s}_{\mathcal{C}} \in \mathbb{R}^{N}$ as output vector (as shown in Fig.~\ref{h0t52}), and $MN\mathbf{s}_{\mathcal{C}}$ as desired output, the loss function of the two-layer complementary BiSCorN is given by 
\begin{align} \label{cd}
	\mathcal{L}_{\mathcal{C}}=&    \left \| \mathbf{y}_{\mathcal{C}} \left( \mathbf{s}_{\mathcal{C}}, \widetilde{\mathbf{X}}_{\mathcal{C}}  \right) - MN\mathbf{s}_{\mathcal{C}} \right \|^2_2  =\mathbf{a}_{\mathcal{C}}^{T} \hspace{2pt} \widetilde{\mathbf{r}}_{\mathcal{C}},
\end{align}
where $\mathbf{a}_{\mathcal{C}}=[{a}^2_{\mathcal{C}},{a}^2_{\mathcal{C}},...,{a}^2_{\mathcal{C}}]^T \in \mathbb{R}^{N}_{+}$. It can be observed that the objective in \eqref{maxmin225631} boils down to \eqref{cd} with $\widetilde{\mathbf{w}}_{\mathcal{C}}=\mathbf{a}_{\mathcal{C}}$.

\subsection{MIMO-CSS}  \label{mimocsspro}
A MIMO architecture is considered where a set of CSS is assigned to each transmit antennas (see Fig.~\ref{ht}). Precisely, the $m^{th}$ transmit antenna is endowed with a set of $M$ transmit codes of size $N$, i.e., $\widetilde{\mathbf{X}}_m=[\mathbf{x}_{m,1},\mathbf{x}_{m,2},...,\mathbf{x}_{m,M}] \in \mathbb{R}^{N \times M},\hspace{2pt}m=1,...,N_T$. First, let us denote by $r_{ml,m^{\prime}} (k),\hspace{2pt}m,l=1,...,N_T, \hspace{2pt}m^{\prime}=1,...,M,\hspace{2pt}-N+1 \leq k \leq N-1,$ the aperiodic/periodic cross-correlation of $m^{th}$ and $l^{th}$ antenna sequences in time slot $m^{\prime}$, i.e., $\mathbf{x}_{m,m^{\prime}}$ and $\mathbf{x}_{l,m^{\prime}}$ at lag $k$.
Then, the WCISL for a MIMO system can be defined as
\begin{align} \label{kku1}
	\textrm{WCISL}=& w_{\mathcal{MC},0} \sum_{\substack{m,l=1\\m \neq l}}^{N_T} \left \vert  \sum_{m^{\prime}=1}^{M} r_{ml,m^{\prime}} (0) \right \vert^2 +  \sum_{k=1 }^{N-1} w_{\mathcal{MC},k} \sum_{m,l=1}^{N_T} \left \vert  \sum_{m^{\prime}=1}^{M} r_{ml,m^{\prime}} (k) \right \vert^2 \\ \nonumber & + \sum_{k=-N+1 }^{-1} w_{\mathcal{MC},-k} \sum_{m,l=1}^{N_T} \left \vert  \sum_{m^{\prime}=1}^{M} r_{ml,m^{\prime}} (k) \right \vert^2,
\end{align}
where $\mathbf{w}_{\mathcal{MC}}=[w_{\mathcal{MC},0},w_{\mathcal{MC},1}, ... ,w_{\mathcal{MC},N-1}]^T$ with $w_{\mathcal{MC},k} \geq 0, \hspace{2pt} k=0,...,N-1,$ is weight vector\footnote{The subscript $\mathcal{MC}$ is related to \textit{MIMO complementary}.}.
Next, the MIMO design problem for binary CSS can be cast as
\begin{eqnarray}\label{maxmin225631251}
	\min_{\lbrace \widetilde{\mathbf{X}}_m \rbrace^{N_T}_{m=1} } &   \textrm{WCISL}\hspace{3pt} \textrm{in}\hspace{3pt}  \eqref{kku1}
	\\ \nonumber
	\mbox{s. t.}\;\;&  x_{m,m^{\prime}} (n)  \in \lbrace -1,+1  \rbrace,~m=1, ..., N_T,~m^{\prime}=1,...,M,~n=1,...,N.
\end{eqnarray}
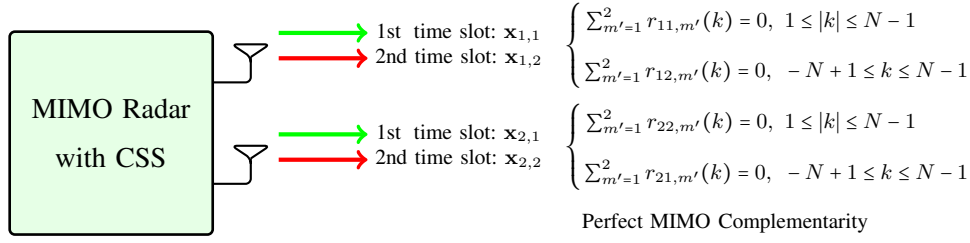
\begin{figure}[!t]
	\centering
	
	\begin{tikzpicture}[even odd rule,rounded corners=2pt,x=12pt,y=12pt,scale=.8,every node/.style={scale=.9}]
		\draw[thick,fill=green!10, line width=1pt] (-4,-2) rectangle (4,6) node[midway]{\large  \shortstack[l]{\normalsize MIMO  Radar \\ \\ \\ \\ \hspace{5pt} \normalsize with CSS}};
		\draw[thick] (4,4)--++(1.5,0)--+(0,1);
		\draw[thick] (5.5,5)--++(.75,0.5)--++(-1.5,0)--++(.75,-.5)--++(0,-.1);
		\draw[thick] (4,0)--++(1.5,0)--+(0,1);
		\draw[thick] (5.5,1)--++(.75,0.5)--++(-1.5,0)--++(.75,-.5)--++(0,-.1);
		
		\draw[->,line width=1.6,green!100] (8.6-1-1,8.45+1-3.5)--+(3.5,0);
		\node [] at (8.6+7.9+1+.7-2-1-1-.5,8.45+1-3.5) {\footnotesize \hspace{.001pt}1st \hspace{-.8pt} time slot:$~\mathbf{x}_{1,1}$};
		\draw[->,line width=1.6,red!100] (8.6-1-1,7.45+1-3.5)--+(3.5,0);
		\node [] at (8.6+7.9+1+.7-2-1-1-.5,7.45+1-3.5) {\footnotesize 2nd time slot:$~\mathbf{x}_{1,2}$};

		\node [] at (8.6+7.9+1+.7-2-1-1+11,8.45+1-3.5+.5) {\footnotesize $\sum_{m^{\prime}=1}^{2} r_{11,m^{\prime}} (k) = 0,~ 1 \leq \vert k \vert \leq N-1$};
		
		\node [] at (8.6+7.9+1+.7-2-1-1+11+1,7.45+1-3.5-.5) {\footnotesize $\sum_{m^{\prime}=1}^{2} r_{12,m^{\prime}} (k) = 0,~ -N+1 \leq  k  \leq N-1$};
		
		\node [] at (8.6+7.9+1+.7-2-1-1+3+1,7.45+1-3.5-.5+1) {$\Bigg \lbrace$};

		\draw[->,line width=1.6,green!100] (8.6-1-1,8.45-3-3.5)--+(3.5,0);
		\node [] at (8.6+7.9+1+.7-2-1-1-.5,8.45+1-4-3.5) {\footnotesize \hspace{.001pt}1st \hspace{-.8pt} time slot:$~\mathbf{x}_{2,1}$};
		\draw[->,line width=1.6,red!100] (8.6-1-1,7.45-3-3.5)--+(3.5,0);
		\node [] at (8.6+7.9+1+.7-2-1-1-.5,7.45+1-4-3.5) {\footnotesize 2nd time slot:$~\mathbf{x}_{2,2}$};

		\node [] at (8.6+7.9+1+.7-2-1-1+11,8.45+1-4-3.5+.5) {\footnotesize $\sum_{m^{\prime}=1}^{2} r_{22,m^{\prime}} (k) = 0,~ 1 \leq \vert k \vert \leq N-1$};
		
		\node [] at (8.6+7.9+1+.7-2-1-1+11+1,7.45+1-4-3.5-.5) {\footnotesize $\sum_{m^{\prime}=1}^{2} r_{21,m^{\prime}} (k) = 0,~ -N+1 \leq  k  \leq N-1$};

		\node [] at (8.6+7.9+1+.7-2-1-1+3+1,7.45+1-4-3.5-.5+1) {$\Bigg \lbrace$};	
		
		\node [] at (8.6+7.9+1+.7-2-1-1+3+1+6,7.45+1-4-3.5-.5+1-3) {\footnotesize Perfect MIMO Complementarity};	
		
	\end{tikzpicture}

	\caption{An illustration of MIMO radar with CSS and MIMO complementary conditions for the case of $N_T=M=2$.}
	\label{ht}
	\centering
\end{figure}

\begin{figure}[!t]
	\centering
	\begin{tikzpicture}[even odd rule,rounded corners=2pt,x=12pt,y=12pt,scale=.67,every node/.style={scale=.81}]
		
		\draw[thick] (-.5-7,-.5) rectangle ++(3.5,3) node[midway]{ $\mathbf{X}_{\mathcal{MC}}$ };
		\draw[thick] (5.5-7,-.5) rectangle ++(3.5,3) node[midway]{$\mathbf{X}^{T}_{\mathcal{MC}}$ };
		
		\draw[->,line width=1,black!100] (-5-7-1,1)--+(5.5,0);
		\draw[->,line width=1,black!100] (3-7,1)--+(2.5,0);
		\draw[->,line width=1,black!100] (9-7,1)--+(6.2,0);
		
		\node [] at (-2.9-7.5,1.7) {\footnotesize Input ($\mathbf{S}_{\mathcal{MC}}$)};
		\node [] at (11.6-6.4,1.8) {\footnotesize Output ($\mathbf{Y}_{\mathcal{MC}}$)};
		
		\node [] at (-8,-8.5) {\small a)~Network architecture};
		\node [] at (15.5,-8.5) {\small b)~Network weights};

		\draw[thick,rounded corners=0pt,fill=green!20] (18.5-3.5,-6) rectangle ++(5.8,3.5) node[midway]{ $\widetilde{\mathbf{X}}_{\mathcal{S}}(\mathbf{x}_{1,M})$ };
		\draw[thick,rounded corners=0pt] (18.5-3.5,-2.5) rectangle ++(5.8,3.5) node[midway]{ $\vdots$ };
		\draw[thick,rounded corners=0pt,fill=green!20] (18.5-3.5,1) rectangle ++(5.8,3.5) node[midway]{  $\widetilde{\mathbf{X}}_{\mathcal{S}}(\mathbf{x}_{1,2})$};
		\draw[thick,rounded corners=0pt,fill=green!20] (18.5-3.5,4.5) rectangle ++(5.8,3.5) node[midway]{ $\widetilde{\mathbf{X}}_{\mathcal{S}}(\mathbf{x}_{1,1})$ };

		\draw[thick,rounded corners=0pt,fill=green!20] (24.3-3.5,-6) rectangle ++(5.8,3.5) node[midway]{ $\widetilde{\mathbf{X}}_{\mathcal{S}}(\mathbf{x}_{2,M})$ };
		\draw[thick,rounded corners=0pt] (24.3-3.5,-2.5) rectangle ++(5.8,3.5) node[midway]{ $\vdots$ };
		\draw[thick,rounded corners=0pt,fill=green!20] (24.3-3.5,1) rectangle ++(5.8,3.5) node[midway]{  $\widetilde{\mathbf{X}}_{\mathcal{S}}(\mathbf{x}_{2,2})$};
		\draw[thick,rounded corners=0pt,fill=green!20] (24.3-3.5,4.5) rectangle ++(5.8,3.5) node[midway]{ $\widetilde{\mathbf{X}}_{\mathcal{S}}(\mathbf{x}_{2,1})$ };

		\draw[thick,rounded corners=0pt] (30.1-3.5,-6) rectangle ++(3.5,14) node[midway]{ $\hdots$ };

		\draw[thick,rounded corners=0pt,fill=green!20] (33.6-3.5,-6) rectangle ++(6,3.5) node[midway]{ $\widetilde{\mathbf{X}}_{\mathcal{S}}(\mathbf{x}_{N_{T},M})$ };
		\draw[thick,rounded corners=0pt] (33.6-3.5,-2.5) rectangle ++(6,3.5) node[midway]{ $\vdots$ };
		\draw[thick,rounded corners=0pt,fill=green!20] (33.6-3.5,1) rectangle ++(6,3.5) node[midway]{  $\widetilde{\mathbf{X}}_{\mathcal{S}}(\mathbf{x}_{N_{T},2})$};
		\draw[thick,rounded corners=0pt,fill=green!20] (33.6-3.5,4.5) rectangle ++(6,3.5) node[midway]{ $\widetilde{\mathbf{X}}_{\mathcal{S}}(\mathbf{x}_{N_{T},1})$ };

		\node [] at (12.5,1) { $\mathbf{X}_{\mathcal{MC}}$: };
		
	\end{tikzpicture}
	\caption{Two-layer BiSCorN to design aperiodoc/periodic complementary binary sequence sets within the MIMO radar context.}
	\label{h0t522}
	\centering
\end{figure}
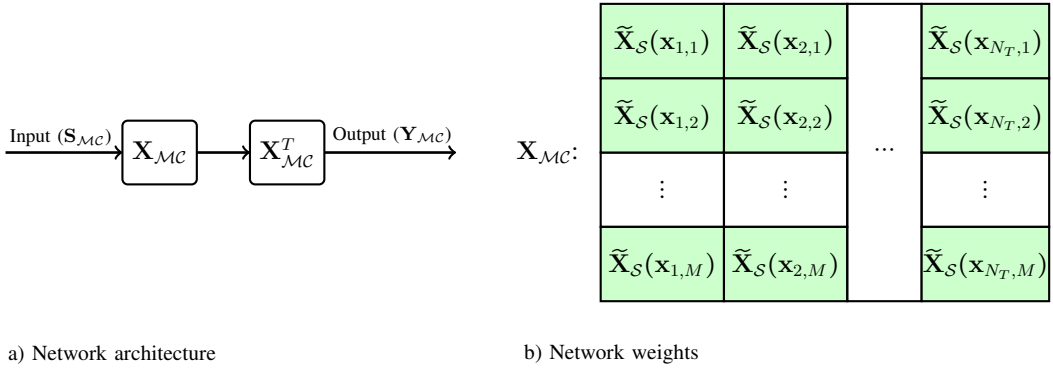

In this case, using $r_{ml,m^{\prime}} (k)= r_{lm,m^{\prime}} (-k)$ and relaxing the binary constraint, the relaxed WCISL minimization problem in \eqref{maxmin225631251} can be rewritten as
\begin{eqnarray}\label{maxmin22563125}
	\min_{\lbrace \widetilde{\mathbf{X}}_m \rbrace^{N_T}_{m=1} } &  \widetilde{\mathbf{w}}^T_{\mathcal{MC}} \hspace{1pt}\hspace{1pt}\widetilde{\mathbf{r}}_\mathcal{MC}   \\ \nonumber
	\mbox{s. t.}\;\;& \vert x_{m,m^{\prime}} (n) \vert \leq 1,~m=1, ..., N_T,~m^{\prime}=1,...,M,~n=1,...,N,
\end{eqnarray}
where $\widetilde{\mathbf{w}}_{\mathcal{MC}} =[{w}_{\mathcal{MC},b}, {w}_{\mathcal{MC},s} \times \mathbf{w}^T_{\mathcal{MC}}]^T \in \mathbb{R}^{N+1}_{+}$ with ${w}_{\mathcal{MC},b}+{w}_{\mathcal{MC},s}=1, \hspace{2pt} {w}_{\mathcal{MC},b} \geq 0, \hspace{2pt} {w}_{\mathcal{MC},s} \geq 0$ and
\begin{equation} \label{key24525}
	\widetilde{{r}}_{\mathcal{MC}} (k) =
	\begin{cases}
		\sum_{m=1}^{N_T} \left \vert  \sum_{m^{\prime}=1}^{M} r_{mm,m^{\prime}} (0) -MN \right \vert^2,~~k=1, \\
		\sum_{\substack{m,l=1\\m \neq l}}^{N_T} \left \vert  \sum_{m^{\prime}=1}^{M} r_{ml,m^{\prime}} (0) \right \vert^2,\hspace{2pt}~~~~~~~~~~k=2,
		\\
		2\sum_{m,l=1}^{N_T} \left \vert  \sum_{m^{\prime}=1}^{M} r_{ml,m^{\prime}} (k-1) \right \vert^2,~~~\hspace{1.5pt}~k=3,...,N+1.
	\end{cases}
\end{equation}
Again the problem in \eqref{maxmin22563125} can be handled by BiSCorN framework.
We first define
\begin{equation} \label{lj22}
	\mathbf{X}_{\mathcal{MC}}=
	\begin{cases}
		\mathbf{X}_{\mathcal{MC},AP}= \begin{bmatrix}
			\widetilde{\mathbf{X}}_{\mathcal{S},AP}(\mathbf{x}_{1,1}),&
			\hdots&
			\widetilde{\mathbf{X}}_{\mathcal{S},AP}(\mathbf{x}_{N_{T},1})\\
			\vdots& &\vdots \\
			\widetilde{\mathbf{X}}_{\mathcal{S},AP}(\mathbf{x}_{1,M}),&
			\hdots &
			\widetilde{\mathbf{X}}_{\mathcal{S},AP}(\mathbf{x}_{N_{T},M})
		\end{bmatrix}
		~ \in \mathbb{R}^{(2N-1)M \times NN_T }, \hspace{4pt} \textrm{Aperiodic},
		\\ \hspace{2pt}
		\mathbf{X}_{\mathcal{MC},P} \hspace{1pt}= \hspace{3pt}
		\begin{bmatrix}
			\widetilde{\mathbf{X}}_{\mathcal{S},P}(\mathbf{x}_{1,1}),&
			\hdots&
			\widetilde{\mathbf{X}}_{\mathcal{S},P}(\mathbf{x}_{N_{T},1})\\
			\vdots& &\vdots \\
			\widetilde{\mathbf{X}}_{\mathcal{S},P}(\mathbf{x}_{1,M}),&
			\hdots &
			\widetilde{\mathbf{X}}_{\mathcal{S},P}(\mathbf{x}_{N_{T},M})
		\end{bmatrix}
		~ \in \mathbb{R}^{NM \times NN_T }, \hspace{41pt} \textrm{Periodic},
	\end{cases}
\end{equation}
where $\widetilde{\mathbf{X}}_{\mathcal{S},AP}$ and $\widetilde{\mathbf{X}}_{\mathcal{S},P}$ are defined in Subsection~\ref{MIMOpro}. Then, letting $\mathbf{Y}_{\mathcal{MC}}=\mathbf{X}_{\mathcal{MC}}^T \mathbf{X}_{\mathcal{MC}} \hspace{1pt} \mathbf{S}_{\mathcal{MC}}$ and $MN\mathbf{S}_{\mathcal{MC}}$ the actual and the desired output from the network in Fig.~\ref{h0t522}, respectively, when the input is the matrix
\begin{equation} \label{nnm}
	\mathbf{S}_{\mathcal{MC}}={a_\mathcal{MC}}\left[
	\mathbf{e}_1,  \mathbf{e}_{N+1}, ..., \mathbf{e}_{(N_T -1)N+1} \right]
	\in \mathbb{R}^{NN_T \times N_T},
\end{equation}
the loss function is given by
\begin{align} \label{keyl}
	\mathcal{L}_{\mathcal{MC}} &=    \left \| \mathbf{Y}_{\mathcal{MC}} \left( \mathbf{S}_{\mathcal{MC}}, \lbrace \widetilde{\mathbf{X}}_m \rbrace^{N_T}_{m=1} \right) -MN\mathbf{S}_{\mathcal{MC}} \right \|_F^2= \mathbf{a}^T_{\mathcal{MC}} \hspace{1pt} \widetilde{\mathbf{r}}_{\mathcal{MC}},
\end{align}
where $\mathbf{a}_{\mathcal{MC}}=[{a}^2_{\mathcal{MC}},{a}^2_{\mathcal{MC}},...,{a}^2_{\mathcal{MC}}]^T \in \mathbb{R}^{N+1}_+$. By setting $\widetilde{\mathbf{w}}_{\mathcal{MC}}=\mathbf{a}_{\mathcal{MC}}$, the loss function in \eqref{keyl} corresponds to objective in \eqref{maxmin22563125}.  
\section{Dealing with LCZ}\label{lcz}
In this section, another application of the proposed learning framework is devised to deal with the design problem in \eqref{maxmin23} considering LCZ. Note that sequences with almost zero correlation lags within a certain interval are called LCZ; they have significant applications in cognitive radar systems \cite{stoicabook}. Handling LCZ is not possible for the aforementioned two-layer network in which we are forcing equal ISL weights, i.e., $\widetilde{\mathbf{w}}_{\mathcal{S}}=[a_{\mathcal{S}}^2, a_{\mathcal{S}}^2, ..., a_{\mathcal{S}}^2]^T$ for the network loss function in \eqref{key1} (see Subsection~\ref{SISOpro}). Therefore, herein, we resort to a single-layer network which we call it single-layer BiSCorN. 
Also, the single-layer BiSCorN provides the user with the degrees of freedom to freely adjust between binarization and sidelobe minimization; more precisely by arbitrary setting the binarization and sidelobe minimization coefficients as opposed to the two-layer case that these coefficients are fixed (see Subsection~\ref{SISOpro}). 
In what follows, single-layer BiSCorN is developed for various radar system configurations.

%
\subsection{SISO LCZ}\label{sisolcz}
\begin{figure*}
	\centering
	\begin{tikzpicture}[even odd rule,rounded corners=2pt,x=12pt,y=12pt,scale=.68,every node/.style={scale=.92}]

		\draw[->,line width=1,black!100] (-4.5,2)--+(4.5,0);
		\draw[->,line width=1,black!100] (28.2,2)--+(4.9,0);
		\node [] at (-2.4,2.7) {\scriptsize Input ($\mathbf{s}_{\mathcal{S}}$)};
		\node [] at (30.7,2.8) {\scriptsize Output ($\mathbf{y}_{\mathcal{S}}$)};

		\draw[thick,rounded corners=0pt,fill=blue!20,blue!20] (0,-5.5) rectangle ++(4.7,2.5); 
		\draw[thick,rounded corners=0pt,fill=blue!20,blue!20] (0+4.7,-5.5) rectangle ++(4.7,2.5) ; 	
		\draw[thick,rounded corners=0pt,fill=blue!20,blue!20] (0+2*4.7,-5.5) rectangle ++(4.7,2.5) ; 
		\draw[thick,rounded corners=0pt,fill=blue!20,blue!20] (0+3*4.7,-5.5) rectangle ++(4.7,2.5) ; 
		\draw[thick,rounded corners=0pt,fill=blue!20,blue!20] (0+4*4.7,-5.5) rectangle ++(4.7,2.5) ;

		\draw[thick,rounded corners=0pt,fill=blue!20,blue!20] (0,-5.5+2.5) rectangle ++(4.7,2.5); 
		\draw[thick,rounded corners=0pt,fill=blue!20,blue!20] (0+4.7,-5.5+2.5) rectangle ++(4.7,2.5) ; 	
		\draw[thick,rounded corners=0pt,fill=blue!20,blue!20] (0+2*4.7,-5.5+2.5) rectangle ++(4.7,2.5) ; 
		\draw[thick,rounded corners=0pt,fill=blue!20,blue!20] (0+3*4.7,-5.5+2.5) rectangle ++(4.7,2.5) ; 
		
		\draw[thick,rounded corners=0pt,fill=blue!20,blue!20] (0+5*4.7,-5.5+2.5) rectangle ++(4.7,2.5) ;

		\draw[thick,rounded corners=0pt,fill=blue!20,blue!20] (0,-5.5+2*2.5) rectangle ++(4.7,2.5) ; 
		\draw[thick,rounded corners=0pt,fill=blue!20,blue!20] (0+4.7,-5.5+2*2.5) rectangle ++(4.7,2.5) ; 	
		\draw[thick,rounded corners=0pt,fill=blue!20,blue!20] (0+2*4.7,-5.5+2*2.5) rectangle ++(4.7,2.5) ; 
		
		\draw[thick,rounded corners=0pt,fill=blue!20,blue!20] (0+4*4.7,-5.5+2*2.5) rectangle ++(4.7,2.5) ;
		\draw[thick,rounded corners=0pt,fill=blue!20,blue!20] (0+5*4.7,-5.5+2*2.5) rectangle ++(4.7,2.5) ;

		\draw[thick,rounded corners=0pt,fill=blue!20,blue!20] (0,-5.5+3*2.5) rectangle ++(4.7,2.5) ; 
		\draw[thick,rounded corners=0pt,fill=blue!20,blue!20] (0+4.7,-5.5+3*2.5) rectangle ++(4.7,2.5) ; 	
		
		\draw[thick,rounded corners=0pt,fill=blue!20,blue!20] (0+3*4.7,-5.5+3*2.5) rectangle ++(4.7,2.5) ; 
		\draw[thick,rounded corners=0pt,fill=blue!20,blue!20] (0+4*4.7,-5.5+3*2.5) rectangle ++(4.7,2.5) ;
		\draw[thick,rounded corners=0pt,fill=blue!20,blue!20] (0+5*4.7,-5.5+3*2.5) rectangle ++(4.7,2.5) ;

		\draw[thick,rounded corners=0pt,fill=blue!20,blue!20] (0,-5.5+4*2.5) rectangle ++(4.7,2.5) ; 
		
		\draw[thick,rounded corners=0pt,fill=blue!20,blue!20] (0+2*4.7,-5.5+4*2.5) rectangle ++(4.7,2.5) ; 
		\draw[thick,rounded corners=0pt,fill=blue!20,blue!20] (0+3*4.7,-5.5+4*2.5) rectangle ++(4.7,2.5) ; 
		\draw[thick,rounded corners=0pt,fill=blue!20,blue!20] (0+4*4.7,-5.5+4*2.5) rectangle ++(4.7,2.5) ;
		\draw[thick,rounded corners=0pt,fill=blue!20,blue!20] (0+5*4.7,-5.5+4*2.5) rectangle ++(4.7,2.5) ;

		\draw[thick,rounded corners=0pt,fill=blue!20,blue!20] (0+4.7,-5.5+5*2.5) rectangle ++(4.7,2.5) ; 	
		\draw[thick,rounded corners=0pt,fill=blue!20,blue!20] (0+2*4.7,-5.5+5*2.5) rectangle ++(4.7,2.5);
		\draw[thick,rounded corners=0pt,fill=blue!20,blue!20] (0+3*4.7,-5.5+5*2.5) rectangle ++(4.7,2.5) ; 
		\draw[thick,rounded corners=0pt,fill=blue!20,blue!20] (0+4*4.7,-5.5+5*2.5) rectangle ++(4.7,2.5) ;
		\draw[thick,rounded corners=0pt,fill=blue!20,blue!20] (0+5*4.7,-5.5+5*2.5) rectangle ++(4.7,2.5) ;

		\draw[thick,rounded corners=0pt,fill=green!20] (0,-5.5+5*2.5) rectangle ++(4.7,2.5) node[midway]{\tiny $ r(0) $}; 	
		\draw[thick,rounded corners=0pt,fill=green!20] (0+4.7,-5.5+4*2.5) rectangle ++(4.7,2.5) node[midway]{\tiny $ r(1) +N$}; 	
		\draw[thick,rounded corners=0pt,fill=green!20] (0+2*4.7,-5.5+3*2.5) rectangle ++(4.7,2.5) node[midway]{\tiny $ r(2) +N$};
		\draw[thick,rounded corners=0pt,fill=green!20,green!20] (0+3*4.7,-5.5+2*2.5) rectangle ++(4.7,2.5) ; 
		\draw[thick,rounded corners=0pt,fill=green!20,green!20] (0+4*4.7,-5.5+2.5) rectangle ++(4.7,2.5);	
		\draw[thick,rounded corners=0pt,fill=green!20] (0+5*4.7,-5.5) rectangle ++(4.7,2.5) node[midway]{\tiny $r(N-1)+N$};
		\draw[thick,rounded corners=0pt] (0,-5.5) rectangle ++(28.2,15);
		
		\node [] at (21.05,5.75) {\huge 0 };
		\node [] at (7.05,-1.75) { \huge 0 };
		\draw[red,thick] (10.93,3.27) circle (.37cm);
		\draw[->,line width=1,red!100] (10.9,4.1)--+(.4,3);
		\node [] at (13.4,8.1) {\scriptsize Aperiodic: $r_{{\mathcal{S}},AP} (2) = \sum_{n=1}^{N-2} x (n) x(n+2) $ };
		\draw[->,line width=1,red!100] (10,3.2)--+(-2.5,-1);
		\node [] at (9.6,1.1) { \scriptsize Periodic: $r_{{\mathcal{S}},P} (2) = \sum_{n=1}^{N}  x(n) 	x{({(n+2)}_{\textrm{\tiny{mod}}\hspace{1pt}N})} $ };
		
		\path (16.4,1.85) -- (21.4,-3.15) node [ font=\Large, midway, sloped] {$\dots$};
		
		\node [] at (14.1,-6.5) {\scriptsize Aperiodic/Periodic: $R_{\mathcal{S}} \in \mathbb{R}^{N \times N}$ };

	\end{tikzpicture}
	\caption{Single-layer BiSCorN to design aperiodoc/periodic binary sequences within the SISO radar context.}
	\label{ht1}
	\centering
\end{figure*}
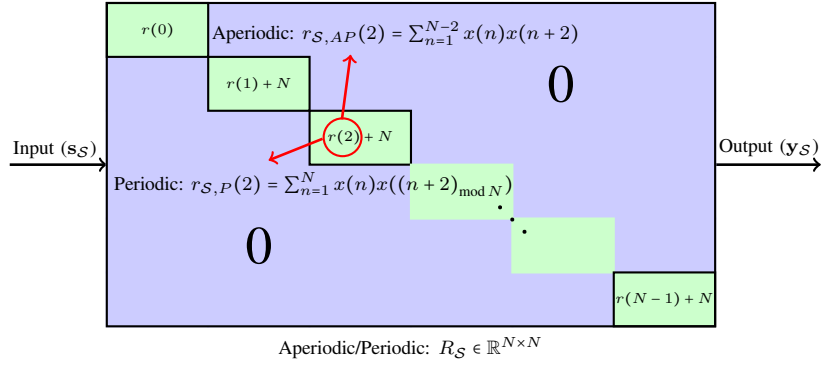

The SISO single-layer BiSCorN is presented to synthesize LCZ sequences as shown\footnote{Note that the architecture of the single-layer BiSCorN is similar for all system setups under study, and therefore only this architecture is shown for the SISO case for the sake of brevity.} in Fig.~\ref{ht1}. The output of this architecture can be written as
\begin{equation} \label{keyke}
\mathbf{y}_{\mathcal{S}} \left( \mathbf{s}_{\mathcal{S}}, \mathbf{x}  \right)=\mathbf{R}_{\mathcal{S}}  \hspace{2pt} \mathbf{s}_{\mathcal{S}},
\end{equation}
where $\mathbf{R}_{\mathcal{S}}= \textrm{Diag}  \left(\widehat{\mathbf{r}}_{\mathcal{S}} \right)$,  $\widehat{\mathbf{r}}_{\mathcal{S}}=[r(0),r(1)+N,r(2)+N, ... ,r(N-1)+N ]^T \in \mathbb{R}^N$, and $\mathbf{s}_{\mathcal{S}}\in \mathbb{R}^N$ is the network input, as illustrated in Fig.~\ref{ht1} for both aperiodic and periodic cases. Note that $\mathbf{R}_{\mathcal{S}}$ depends on $\mathbf{x}$ via $\widehat{\mathbf{r}}_{\mathcal{S}}$ ($\widehat{\mathbf{r}}_{\mathcal{S}}$ itself is the vector of correlation lags (see \eqref{first} and \eqref{second}). 
As opposed to the two-layer BiSCorN, the network input $\mathbf{s}_{\mathcal{S}}$ can be considered as a random vector in this subsection. Precisely, let $\mathbf{s}_{\mathcal{S}}=[{\widetilde{a}_{\mathcal{S},0}},{\widetilde{a}_{\mathcal{S},1}},...,{\widetilde{a}_{\mathcal{S},{N-1}}}]^T \sim \mathcal{N}({\boldsymbol{\mu}}_{\mathcal{S}},\textrm{Diag}({\mathbf{p}}_{\mathcal{S}}))$ with ${\boldsymbol{\mu}}_{\mathcal{S}}$ and ${\mathbf{p}}_{\mathcal{S}}$ being mean and variance of the network input (to be discussed shortly in Subsection~\ref{bibi}). Having sidelobe weights ${\mathbf{w}}_{\mathcal{S}} =[w_1,w_2, ..., w_{N-1}]^T \in \mathbb{R}^{N-1}_+$ (see Subsection~\ref{SISOpro}), the weight vector herein can be defined as $\widetilde{\mathbf{w}}_{\mathcal{S}}=[{{w}}_{\mathcal{S},b},{{w}}_{\mathcal{S},s}\times \mathbf{w}^T_{\mathcal{S}}]^T=\mathbb{E}_{\mathbf{s}_{\mathcal{S}}} \left[ \widetilde{\mathbf{a}}_\mathcal{S} \right]$ where $\widetilde{\mathbf{a}}_\mathcal{S}=[\widetilde{a}^2_{\mathcal{S},0},\widetilde{a}^2_{\mathcal{S},1},...,\widetilde{a}^2_{\mathcal{S},{N-1}}]^T \in \mathbb{R}^N_+$. Then, using \eqref{key} and \eqref{keyke}, the loss function is given by
\begin{align} \label{key3}
\mathcal{L}_{\mathcal{S}} & =\mathbb{E}_{\mathbf{s}_{\mathcal{S}}} \left[\left \| \mathbf{y}_{\mathcal{S}} \left( \mathbf{s}_{\mathcal{S}}, \mathbf{x}  \right) -N \mathbf{s}_{\mathcal{S}} \right \| _2^2 \right]  
=\widetilde{\mathbf{w}}_{\mathcal{S}} \hspace{2pt} \widetilde{\mathbf{r}}_{\mathcal{S}},
\end{align}
which is equal to the objective function of the design problem in \eqref{maxmin23}. 
\begin{rema}
Consider the single-layer architecture for the aperiodic case in which layer variables are the code entries (see Fig.~\ref{ht1}). They are related to modified correlation lags via the following expression:
\begin{equation}
\mathbf{R}_{\mathcal{S}}(k,k)=	\begin{cases}
		\sum_{n=1}^{N} {x(n)}^2 ,\hspace{65pt}~~~~~~~~k=1,
		\\
\	\sum_{n=1}^{N-k+1} x(n) \hspace{1pt} x(n+k-1) +N,~k=2,...,N.
	\end{cases}
\end{equation}
A similar expression can be written for the periodic network.
Indeed, exploiting the single-layer learning framework directly leads to the code $\mathbf{x}$.
\end{rema}
Note that despite having the same order of computational complexity (see Table~\ref{ty}), the single-layer BiSCorN is more time consuming in comparison with two-layer architecture. 
This can be explained
using the fact that in the two-layer architecture the network weights $\mathbf{X}_{\mathcal{S}}$ coincide with the sequence entries $x(n)$.
On the other hand, in the single-layer architecture the network weights $\mathbf{R}_{\mathcal{S}}$ are computed as the correlation lags of the sequence $\mathbf{x}$ which requires additional computational efforts and increases the run-time (see Fig.~\ref{h0t42} and Fig.~\ref{ht1}). However, the single-layer BiSCorN has some advantages that are summarized as the following:
\begin{itemize}
	\item In the single-layer architecture, both stochastic and deterministic inputs can be adopted. Since the ISL metric is multimodal, i.e., it has many local minima, 
	the stochastic input signal can improve the network performance.
	
	
	\item According to Lemma 1, for approximately large or increasing value of $\bar{w}_{\mathcal{S},b}$, we can obtain a substantially binary sequence with a low ISL.
	
	\item The LCZ binary sequences can be designed by single-layer BiSCorN.
	
\end{itemize}
\subsection{MIMO LCZ} \label{mimolcz}
Similar to Fig.~\ref{ht1} of Subsection~\ref{sisolcz}, the output of the MIMO single-layer BiSCorN (for both aperiodic and periodic cases) can be obtained as 
\begin{equation} \label{keyke1}
\mathbf{y}_{\mathcal{M}} \left( \mathbf{s}_{\mathcal{M}}, \widetilde{\mathbf{X}}_{\mathcal{M}}  \right)= \mathbf{R}_{\mathcal{M}} \hspace{2pt} \mathbf{s}_{\mathcal{M}},
\end{equation}
where $\mathbf{R}_{\mathcal{M}}= \textrm{Diag} \left(\widehat{\mathbf{r}}_{\mathcal{M}}\right)$ and $\widehat{\mathbf{r}}_{\mathcal{M}}= \left[\widehat{\mathbf{r}}_{\mathcal{M},1}^T,\widehat{\mathbf{r}}_{\mathcal{M},2}^T+N,2\hspace{2pt}\widehat{\mathbf{r}}_{\mathcal{M},3}^T+N\right]^T \in \mathbb{R}^{NN^2_T}$ with the following definitions
\begin{equation}\label{init1}
\widehat{\mathbf{r}}_{\mathcal{M},1}=[r_{11}(0),r_{22}(0),...,r_{N_T N_T}(0)]^T \in \mathbb{R}^{N_T},
\end{equation}
\begin{equation}
\widehat{\mathbf{r}}_{\mathcal{M},2}=[r_{12}(0),r_{13}(0),...,r_{1N_T}(0),...,r_{N_T 1}(0),r_{N_T\hspace{1pt} 2}(0),...,r_{N_T \hspace{2pt} N_T -1}(0)]^T \in \mathbb{R}^{N_T(N_T -1)},
\end{equation}
\begin{equation}
\widehat{\mathbf{r}}_{\mathcal{M},3}=\left[\widehat{\mathbf{r}}_{\mathcal{M},3}^T(1), \widehat{\mathbf{r}}_{\mathcal{M},3}^T(2),...,\widehat{\mathbf{r}}_{\mathcal{M},3}^T(N-1) \right]^T \in \mathbb{R}^{N^2_T(N -1)},
\end{equation}
\begin{equation}\label{last1}
	\widehat{\mathbf{r}}_{\mathcal{M},3}(k)=[r_{11}(k),r_{12}(k),...,r_{N_T \hspace{1pt} N_T}(k)]^T \in \mathbb{R}^{N^2_T},~~ k=1,...,N-1,
\end{equation}
and $\mathbf{s}_{\mathcal{M}} \in \mathbb{R}^{NN^2_T}$ is the random input vector. In the sequel, we let $\mathbf{s}_{\mathcal{M}}=\widehat{\mathbf{{a}}}_{\mathcal{M}}=[\widehat{\mathbf{{a}}}^T_{\mathcal{M},0},\widehat{\mathbf{{a}}}^T_{\mathcal{M},1},...,\widehat{\mathbf{{a}}}^T_{\mathcal{M},N}]^T$ $\sim \mathcal{N}({\boldsymbol{\mu}}_{\mathcal{M}},\textrm{Diag}({\mathbf{p}}_{\mathcal{M}}))$ with
\begin{equation} \label{init}
\widehat{\mathbf{{a}}}_{\mathcal{M},0}=[{\widehat{{a}}_{\mathcal{M},0}}, {\widehat{{a}}_{\mathcal{M},0}}, ..., {\widehat{{a}}_{\mathcal{M},0}} ]^T \in \mathbb{R}^{N_T}_+,
\end{equation}
\begin{equation}
\widehat{\mathbf{{a}}}_{\mathcal{M},1}=[{\widehat{{a}}_{\mathcal{M},1}}, {\widehat{{a}}_{\mathcal{M},1}}, ..., {\widehat{{a}}_{\mathcal{M},1}} ]^T \in \mathbb{R}^{N_T(N_T -1)}_+,
\end{equation}
\begin{equation}\label{last}
\widehat{\mathbf{{a}}}_{\mathcal{M},k}=[{\widehat{{a}}_{\mathcal{M},k}}, {\widehat{{a}}_{\mathcal{M},k}}, ..., {\widehat{{a}}_{\mathcal{M},k}} ]^T \in \mathbb{R}^{N^2_T}_+,~k=2,...,N.
\end{equation}
Therefore, using \eqref{key245}, \eqref{keyke1}, \eqref{init1}-\eqref{last1} to obtain the weight matrix ${\mathbf{R}_{\mathcal{M}}}$, and \eqref{init}-\eqref{last} to determine the input vector ${\mathbf{s}_{\mathcal{M}}}$, the loss function is given by
\begin{equation} \label{kkk2}
\mathcal{L}_{\mathcal{M}}=\mathbb{E}_{\mathbf{s}_{\mathcal{M}}} \left[ \left \| \mathbf{y}_{\mathcal{M}} \left( \mathbf{s}_{\mathcal{M}}, \widetilde{\mathbf{X}}_{\mathcal{M}}  \right) - N \mathbf{s}_{\mathcal{M}} \right \|^2_2 \right] =  \mathbb{E}_{\mathbf{s}_{\mathcal{M}}} \left[ \widetilde{\mathbf{a}}^{T}_{\mathcal{M}} \right] \hspace{2pt} \widetilde{\mathbf{r}}_{\mathcal{M}},
\end{equation}
where $\widetilde{\mathbf{a}}_{\mathcal{M}}=[\widehat{{a}}^2_{\mathcal{M},0},\widehat{{a}}^2_{\mathcal{M},1}, ..., \widehat{{a}}^2_{\mathcal{M},N}]^T \in \mathbb{R}^{N+1}_+$.
Therefore, selecting $\widetilde{\mathbf{w}}_{\mathcal{M}}=\mathbb{E}_{\mathbf{s}_{\mathcal{M}}} \left[\widetilde{\mathbf{a}}_{\mathcal{M}}\right]$, the objective in \eqref{maxmin22563} can be obtained. 
\subsection{Complementary LCZ} \label{csslcz}
The output of the single-layer complementary BiSCorN for both aperiodic and periodic cases can be obtained as
\begin{equation}\label{chrt}
\mathbf{y}_{\mathcal{C}} \left( \mathbf{s}_{\mathcal{C}}, \widetilde{\mathbf{X}}_{\mathcal{C}}  \right)=\mathbf{R}_{\mathcal{C}} \hspace{1pt}\mathbf{s}_{\mathcal{C}},
\end{equation}
where $ \mathbf{s}_{\mathcal{C}} \in \mathbb{R}^{N}$ is the input vector and $\mathbf{R}_{\mathcal{C}}=\textrm{Diag} \left(\widehat{\mathbf{r}}_{\mathcal{C}}\right)$ with 
\begin{equation}
\widehat{\mathbf{r}}_{\mathcal{C}}=\left[\sum_{m^{\prime}=1}^{M}   r_{m^{\prime}}(0), \sum_{m^{\prime}=1}^{M}   r_{m^{\prime}}(1) +MN, \sum_{m^{\prime}=1}^{M}   r_{m^{\prime}}(2)+MN, ...,\sum_{m^{\prime}=1}^{M}   r_{m^{\prime}}(N-1)+MN \right]^T \in \mathbb{R}^{N}.
\end{equation}
Defining the random input vector $\mathbf{s}_{\mathcal{C}}=[{\widetilde{{a}}_{\mathcal{C},0}},{\widetilde{{a}}_{\mathcal{C},1}},...,{\widetilde{{a}}_{\mathcal{C},N-1}}]^T \sim \mathcal{N}({\boldsymbol{\mu}}_{\mathcal{C}},\textrm{Diag}({\mathbf{p}}_{\mathcal{C}}))$ and considering \eqref{key2451} as well as \eqref{chrt}, lead to the following loss function
\begin{align} \label{cd1}
\mathcal{L}_{\mathcal{C}}=& \mathbb{E}_{\mathbf{s}_{\mathcal{C}}} \left[\left \| \mathbf{y}_{\mathcal{C}} \left( \mathbf{s}_{\mathcal{C}}, \widetilde{\mathbf{X}}_{\mathcal{C}}  \right) - MN\mathbf{s}_{\mathcal{C}} \right \|^2_2 \right] = \mathbb{E}_{\mathbf{s}_{\mathcal{C}}}  \left[ \widetilde{\mathbf{a}}_{\mathcal{C}}^{T} \right] \hspace{2pt} \widetilde{\mathbf{r}}_{\mathcal{C}},
\end{align}
where $\widetilde{\mathbf{a}}_{\mathcal{C}}=[\widetilde{{a}}^2_{\mathcal{C},0},\widetilde{{a}}^2_{\mathcal{C},1},...,\widetilde{{a}}^2_{\mathcal{C},N-1}]^T \in \mathbb{R}^{N}_+$. By selecting $\widetilde{\mathbf{w}}_{\mathcal{C}}=\mathbb{E}_{\mathbf{s}_{\mathcal{C}}}  \left[\widetilde{\mathbf{a}}_{\mathcal{C}} \right]$, the objective in \eqref{maxmin225631} can be acquired.

\subsection{MIMO Complementary LCZ}  \label{mimocss}
This single-layer architecture has input/output relationship as
\begin{equation}\label{kk1}
 \mathbf{y}_{\mathcal{MC}} \left( \mathbf{s}_{\mathcal{MC}}, \left\lbrace \widetilde{\mathbf{X}}_m \right\rbrace^{N_T}_{m=1}  \right)= \mathbf{R}_{\mathcal{MC}} \hspace{2pt} \mathbf{s}_{\mathcal{MC}},
\end{equation}
where $\mathbf{R}_{\mathcal{MC}}= \textrm{Diag} \left(\widehat{\mathbf{r}}_{\mathcal{MC}}\right)$ and $\widehat{\mathbf{r}}_{\mathcal{MC}}= \Big[\widehat{\mathbf{r}}_{\mathcal{MC},1}^T,\widehat{\mathbf{r}}_{\mathcal{MC},2}^T+MN,2\hspace{2pt}\widehat{\mathbf{r}}_{\mathcal{MC},3}^T+MN \Big]^T \in \mathbb{R}^{NN^2_T}$ with the following definitions
\begin{equation}\label{keyinit1}
\widehat{\mathbf{r}}_{\mathcal{MC},1}=\left[ \sum_{m^{\prime}=1}^{M} r_{11,m^{\prime}} (0), \sum_{m^{\prime}=1}^{M} r_{22,m^{\prime}} (0),...,\sum_{m^{\prime}=1}^{M} r_{N_T\hspace{1pt} N_T,m^{\prime}} (0) \right]^T \in \mathbb{R}^{N_T},
\end{equation}
\begin{align}
\widehat{\mathbf{r}}_{\mathcal{MC},2}=\Big[ &\sum_{m^{\prime}=1}^{M} r_{12,m^{\prime}} (0),\sum_{m^{\prime}=1}^{M} r_{13,m^{\prime}} (0),...,\sum_{m^{\prime}=1}^{M} r_{1N_T,m^{\prime}} (0),...,\\ \nonumber &\sum_{m^{\prime}=1}^{M} r_{N_T 1,m^{\prime}} (0),\sum_{m^{\prime}=1}^{M} r_{N_T 2,m^{\prime}} (0),..., \sum_{m^{\prime}=1}^{M} r_{N_T\hspace{1pt}N_T -1,m^{\prime}} (0) \Big]^T \in \mathbb{R}^{N_T(N_T -1)},
\end{align}
\begin{equation}
\widehat{\mathbf{r}}_{\mathcal{MC},3}=\left[ \widehat{\mathbf{r}}_{\mathcal{MC},3}^T(1),\widehat{\mathbf{r}}_{\mathcal{MC},3}^T(2),...,\widehat{\mathbf{r}}_{\mathcal{MC},3}^T(N-1) \right]^T \in \mathbb{R}^{N^2_T(N -1)},
\end{equation}
\begin{equation}\label{keylast1}
	\widehat{\mathbf{r}}_{\mathcal{MC},3}(k)=\left[ \sum_{m^{\prime}=1}^{M} r_{11,m^{\prime}} (k), \sum_{m^{\prime}=1}^{M} r_{12,m^{\prime}} (k),..., \sum_{m^{\prime}=1}^{M} r_{N_T \hspace{1pt}N_T,m^{\prime}} (k) \right]^T \in \mathbb{R}^{N^2_T},~ k=1,...,N-1,
\end{equation}
and $\mathbf{s}_{\mathcal{MC}} \in \mathbb{R}^{NN_T^2}$ is the random input vector.
Let us select the input vector as $\mathbf{s}_{\mathcal{MC}}=\widehat{\mathbf{{a}}}_{\mathcal{MC}} =[\widehat{\mathbf{{a}}}^T_{\mathcal{MC},0},\widehat{\mathbf{{a}}}^T_{\mathcal{MC},1},...,\widehat{\mathbf{{a}}}^T_{\mathcal{MC},N}]^T \sim \mathcal{N}({\boldsymbol{\mu}}_{\mathcal{MC}},\textrm{Diag}({\mathbf{p}}_{\mathcal{MC}}))$ with
\begin{equation}\label{keyinit}
\widehat{\mathbf{{a}}}_{\mathcal{MC},0}=[{\widehat{{a}}_{\mathcal{MC},0}}, {\widehat{{a}}_{\mathcal{MC},0}}, ..., {\widehat{{a}}_{\mathcal{MC},0}} ]^T \in \mathbb{R}^{N_T}_+,
\end{equation}
\begin{equation}
\widehat{\mathbf{{a}}}_{\mathcal{MC},1}=[{\widehat{{a}}_{\mathcal{MC},1}}, {\widehat{{a}}_{\mathcal{M},1}}, ..., {\widehat{{a}}_{\mathcal{MC},1}} ]^T \in \mathbb{R}^{N_T(N_T -1)}_+,
\end{equation}
\begin{equation}\label{keylast}
\widehat{\mathbf{{a}}}_{\mathcal{MC},k}=[{\widehat{{a}}_{\mathcal{MC},k}}, {\widehat{{a}}_{\mathcal{MC},k}}, ..., {\widehat{{a}}_{\mathcal{MC},k}} ]^T \in \mathbb{R}^{N^2_T}_+,~k=2,...,N.
\end{equation}
Then, by substituting the weight matrix ${\mathbf{{R}}}_{\mathcal{MC}}$ (see \eqref{keyinit1}-\eqref{keylast1}) and the input vector ${\mathbf{{s}}}_{\mathcal{MC}}$ (see \eqref{keyinit}-\eqref{keylast}) into \eqref{kk1} and using \eqref{key24525}, the loss function becomes
\begin{align}  \label{kkk}
	\mathcal{L}_{\mathcal{MC}}=& \mathbb{E}_{\mathbf{s}_{\mathcal{MC}}} \left[\left \| \mathbf{y}_{\mathcal{MC}} \left( \mathbf{s}_{\mathcal{MC}}, \left\lbrace \widetilde{\mathbf{X}}_m \right\rbrace^{N_T}_{m=1}  \right) - MN\mathbf{s}_{\mathcal{MC}} \right \|^2_2 \right] = \mathbb{E}_{\mathbf{s}_{\mathcal{MC}}}  \left[ \widetilde{\mathbf{a}}_{\mathcal{MC}}^{T} \right] \hspace{2pt} \widetilde{\mathbf{r}}_{\mathcal{MC}},
\end{align}
where $\widetilde{\mathbf{a}}_{\mathcal{MC}}=[\widehat{{a}}^2_{\mathcal{MC},0},\widehat{{a}}^2_{\mathcal{MC},1}, ..., \widehat{{a}}^2_{\mathcal{MC},N}]^T \in \mathbb{R}^{N+1}_+$.
Finally, the objective in \eqref{maxmin22563125} can be obtained by selecting $\widetilde{\mathbf{w}}_{\mathcal{MC}}=\mathbb{E}_{\mathbf{s}_{\mathcal{MC}}} \left[\widetilde{\mathbf{a}}_{\mathcal{MC}}\right]$.

\section{Implementation and Complexity Analysis}  \label{ext}
In this section, implementation aspects of the devised waveform learning framework and its computational complexity are addressed.  

\begin{figure}[!t]
	\centering
	\begin{tikzpicture}[even odd rule,rounded corners=2pt,x=12pt,y=12pt,scale=.6,every node/.style={scale=.95}]
		
		\draw[thick,fill=blue!20] (-30,0) rectangle ++(10,10) node[midway]{\footnotesize  BiSCorN};
		
		\node[draw=blue!20, rotate=270, minimum height=-1 cm, minimum width=1 cm] at (-29+.4, 5) {\footnotesize Initial $\mathbf{X}^{(1)}_{\mathcal{S}}$};
		
		\node[draw=blue!20, rotate=270, minimum height=-1 cm, minimum width=1 cm] at (-21-.43, 5) {\footnotesize Obtained $\mathbf{X}^{(1)}_{\mathcal{S}}$};
		
		\draw[->,line width=1,black!100] (-25,11.5)--+(0,-1.5);
		\node [] at (-24.7,12.1) {\footnotesize Input ($\widetilde{\mathbf{s}}_{\mathcal{S},1}$)};
		
		\draw[->,line width=1,black!100] (-25,0)--+(0,-1.5);
		\node [] at (-24.7,-2.1) {\footnotesize Output ($\widetilde{\mathbf{y}}_{\mathcal{S},1}$)};
		
		\draw[->,line width=1,black!100] (-39.2,5)--+(9.2,0);
		\node [] at (-34.7,5.9) {\footnotesize Initial Coefficient};
		\node [] at (-34.7,4) {\footnotesize $\mathbf{X}^{(0)}_{\mathcal{S}}$};
		
		\draw[->,line width=1,black!100] (-20,5)--+(2,0);

		\draw[thick,fill=blue!20] (-18,0) rectangle ++(10,10) node[midway]{\footnotesize BiSCorN};
		
		\node[draw=blue!20, rotate=270, minimum height=-1 cm, minimum width=1 cm] at (-17+.4, 5) {\footnotesize Initial $\mathbf{X}^{(2)}_{\mathcal{S}}$};
		
		\node[draw=blue!20, rotate=270, minimum height=-1 cm, minimum width=1 cm] at (-9-.43, 5) {\footnotesize Obtained $\mathbf{X}^{(2)}_{\mathcal{S}}$};

		\draw[->,line width=1,black!100] (-13,11.5)--+(0,-1.5);
		\node [] at (-12.7,12.1) {\footnotesize Input ($\widetilde{\mathbf{s}}_{\mathcal{S},2}$)};
		
		\draw[->,line width=1,black!100] (-13,0)--+(0,-1.5);
		\node [] at (-12.7,-2.1) {\footnotesize Output ($\widetilde{\mathbf{y}}_{\mathcal{S},2}$)};

		\draw[->,line width=1,black!100] (-8,5)--+(2,0);
		
		\node [] at (-5,4.8) {\large $\hdots$};
		
		\draw[->,line width=1,black!100] (-4,5)--+(2,0);

		\draw[thick,fill=blue!20] (-2,0) rectangle ++(10,10) node[midway]{\footnotesize BiSCorN};
		
		\node[draw=blue!20, rotate=270, minimum height=-1 cm, minimum width=1 cm] at (-1+.4, 5) {\footnotesize Initial $\mathbf{X}^{(d)}_{\mathcal{S}}$};
		
		\node[draw=blue!20, rotate=270, minimum height=-1 cm, minimum width=1 cm] at (7-.43, 5) {\footnotesize Obtained $\mathbf{X}^{(d)}_{\mathcal{S}}$};

		\draw[->,line width=1,black!100] (3,11.5)--+(0,-1.5);
		\node [] at (3.3,12.1) {\footnotesize Input ($\widetilde{\mathbf{s}}_{\mathcal{S},d}$)};
		
		\draw[->,line width=1,black!100] (3,0)--+(0,-1.5);
		\node [] at (3.3,-2.1) {\footnotesize Output ($\widetilde{\mathbf{y}}_{\mathcal{S},d}$)};
		
		\draw[->,line width=1,black!100] (8,5)--+(9.4,0);
		\node [] at (12.75,5.9) {\footnotesize Output Coefficient };
		\node [] at (12.6,4) {\footnotesize $\mathbf{X}_{\mathcal{S}}$};	
	\end{tikzpicture}
	\caption{Implementation of BiSCorN for SISO systems in a given epoch.}
	\label{aa1}
	\centering
\end{figure}
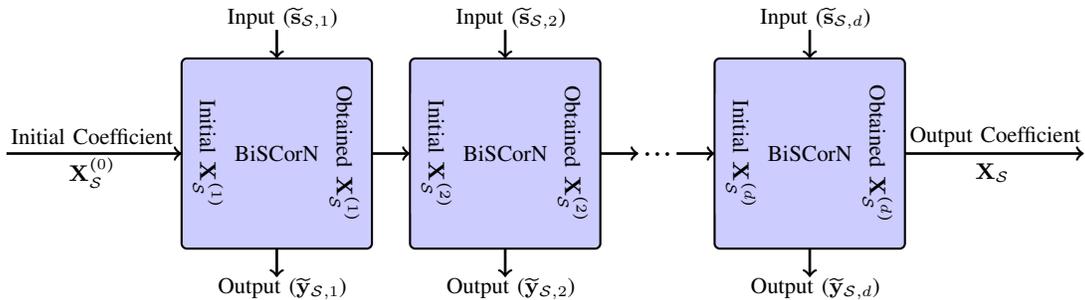

BiSCorN is implemented resorting to Tensor-flow \cite{tensor} with ADAM stochastic optimizer \cite{adam}. The learning process of the network involves both outer, namely epoch, and inner iterations. At each epoch, the network is fed with specific input signals; then, as the per associated loss function, network coefficients are updated iteratively to bring optimized waveform correlation features. 
For instance, with reference to the single-layer BiSCorN in SISO systems, at each epoch the network is fed sequentially exploiting the random input signal
\begin{equation} \label{lll}
	{\mathbf{s}}_{\mathcal{S}}=[{\widetilde{a}_{\mathcal{S},0}},{\widetilde{a}_{\mathcal{S},1}},...,{\widetilde{a}_{\mathcal{S},{N-1}}}]^T \in \mathbb{R}^{N}.
\end{equation}
Herein, the elements of the aforementioned vector are i.i.d with the distribution  $\mathcal{N}({\boldsymbol{\mu}}_{\mathcal{S}},\textrm{Diag}({\mathbf{p}}_{\mathcal{S}}))$.
The learning process performed at a given epoch is illustrated in Fig.~\ref{aa1}: first the network coefficient matrix $\mathbf{X}_{\mathcal{S}}$ is initialized by a starting matrix $\mathbf{X}^{(0)}_{\mathcal{S}}$. Then, as per input signal $\widetilde{\mathbf{s}}_{\mathcal{S},1}$, that is a batch of $b$ data/sequence with the structure given in \eqref{lll}, the loss function is minimized via an iterative algorithm (to be discussed shortly) to obtain the output signal $\widetilde{\mathbf{y}}_{\mathcal{S},1}$. Precisely, the loss function minimization is done via $\sum_{i=1}^{b} {\nabla}_{{\mathbf{x}}} \mathcal{L}^i_{\mathcal{S}}$ where $\mathcal{L}^i_{\mathcal{S}}$ is the loss function evaluated at $i^{th}$ input sequence. Next, the network coefficient matrix associated with $\widetilde{\mathbf{y}}_{\mathcal{S},1}$ is considered as the starting point for the minimization of the loss function corresponding to input signal $\widetilde{\mathbf{s}}_{\mathcal{S},2}$. After $d$ steps, the output coefficient matrix associated with the epoch is obtained. Finally, the coefficient matrix is employed to trigger the next epoch\footnote{For the first epoch, the network can be initialized with a random coefficient matrix.}. In practice, the process continues till a pre-defined stop criterion is satisfied.

In the case of two-layer BiSCorN for SISO systems, the input signal of each batch is composed of $b$ deterministic vector\footnote{It is obvious that selecting $b>1$ is reasonable only for random input signals, and in the case of deterministic input, the available choice is $b=1$.} each of the form $\mathbf{s}_{\mathcal{S}}=[{a_{\mathcal{S}}},0,0,...,0]^T$.

Note that the process associated with other system configurations is similar and not included here.

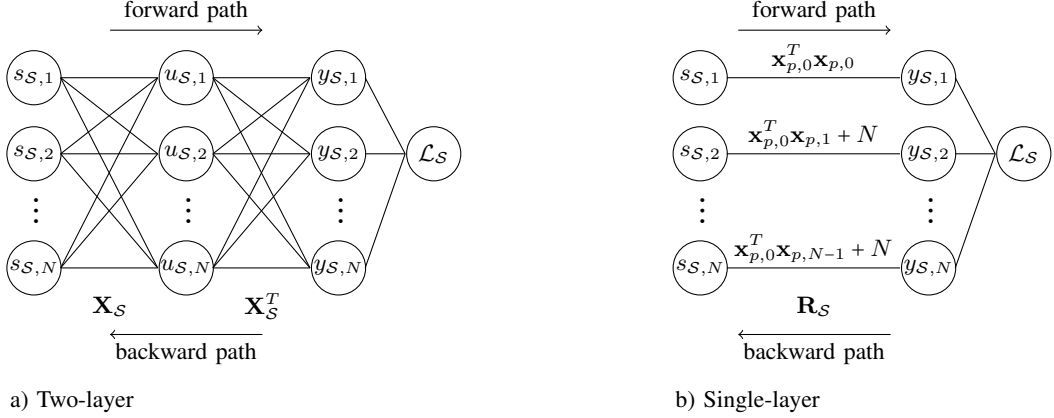
\begin{figure}[!t]
	\centering
	\begin{tikzpicture}[even odd rule,rounded corners=2pt,x=12pt,y=12pt,scale=.6,every node/.style={scale=1}]

		\draw (0,0) circle (.6cm) node {\footnotesize $s_{\mathcal{S},1}$};
		
		\draw (0,-4) circle (.6cm) node {\footnotesize $s_{\mathcal{S},2}$};
		
		\node [] at (.12,-7) {\large $\vdots$};
		
		\draw (0,-10) circle (.6cm) node {\footnotesize $s_{\mathcal{S},N}$};

		\draw[black!100] (1.4,0)--+(5.1,0);
		\draw[black!100] (1.4,0)--+(5.1,-4);
		\draw[black!100] (1.4,0)--+(5.1,-10);
		
		\draw[black!100] (1.4,-4)--+(5.1,0);
		\draw[black!100] (1.4,-4)--+(5.1,4);
		\draw[black!100] (1.4,-4)--+(5.1,-6);
		
		\draw[black!100] (1.4,-10)--+(5.1,0);
		\draw[black!100] (1.4,-10)--+(5.1,6);
		\draw[black!100] (1.4,-10)--+(5.1,10);

		\draw (8,0) circle (.6cm) node {\footnotesize $u_{\mathcal{S},1}$};
		
		\draw (8,-4) circle (.6cm) node {\footnotesize $u_{\mathcal{S},2}$};
		
		\node [] at (8.12,-7) {\large $\vdots$};
		
		\draw (8,-10) circle (.6cm) node {\footnotesize $u_{\mathcal{S},N}$};

		\draw[black!100] (1.4+8,0)--+(5.1,0);
		\draw[black!100] (1.4+8,0)--+(5.1,-4);
		\draw[black!100] (1.4+8,0)--+(5.1,-10);
		
		\draw[black!100] (1.4+8,-4)--+(5.1,0);
		\draw[black!100] (1.4+8,-4)--+(5.1,4);
		\draw[black!100] (1.4+8,-4)--+(5.1,-6);
		
		\draw[black!100] (1.4+8,-10)--+(5.1,0);
		\draw[black!100] (1.4+8,-10)--+(5.1,6);
		\draw[black!100] (1.4+8,-10)--+(5.1,10);

		\draw (16,0) circle (.6cm) node {\footnotesize $y_{\mathcal{S},1}$};
		
		\draw (16,-4) circle (.6cm) node {\footnotesize $y_{\mathcal{S},2}$};
		
		\node [] at (16.12,-7) {\large $\vdots$};
		
		\draw (16,-10) circle (.6cm) node {\footnotesize $y_{\mathcal{S},N}$};

		\draw[black!100] (1.4+16,0)--+(2.1,-4);
		\draw[black!100] (1.4+16,-4)--+(2.1,0);
		\draw[black!100] (1.4+16,-10)--+(2.1,6);

		\draw (21,-4) circle (.6cm) node {\footnotesize $\mathcal{L}_{\mathcal{S}}$};

		\node [] at (4,-12) {\footnotesize $\mathbf{X}_{\mathcal{S}}$};
		\node [] at (4+8,-12) {\footnotesize $\mathbf{X}^T_{\mathcal{S}}$};

		\draw[->,black!100] (4,2.5)--+(8,0);
		\node [] at (8,3.5) {\footnotesize forward path};
		\draw[<-,black!100] (4,2.5-16)--+(8,0);
		\node [] at (8,2.5-17) {\footnotesize backward path};

		\draw (0+35,0) circle (.6cm) node {\footnotesize $s_{\mathcal{S},1}$};
		
		\draw (0+35,-4) circle (.6cm) node {\footnotesize $s_{\mathcal{S},2}$};
		
		\node [] at (.12+35,-7) {\large $\vdots$};
		
		\draw (0+35,-10) circle (.6cm) node {\footnotesize $s_{\mathcal{S},N}$};

		\draw[black!100] (1.4+35,0)--+(9.1,0);
		\draw[black!100] (1.4+35,-4)--+(9.1,0);
		\draw[black!100] (1.4+35,-10)--+(9.1,0);
		
		\node [] at (1.4+36+3.55,1) {\footnotesize $\mathbf{x}^T_{p,0} \mathbf{x}_{p,0}$};
		
		\node [] at (1.4+36+3.55,1-4) {\footnotesize $\mathbf{x}^T_{p,0} \mathbf{x}_{p,1} +N$};
		
		\node [] at (1.4+36+3.55,1-10) {\footnotesize $\mathbf{x}^T_{p,0} \mathbf{x}_{p,N-1} +N$};

		\draw (12+35,0) circle (.6cm) node {\footnotesize $y_{\mathcal{S},1}$};
		
		\draw (12+35,-4) circle (.6cm) node {\footnotesize $y_{\mathcal{S},2}$};
		
		\node [] at (12.12+35,-7) {\large $\vdots$};
		
		\draw (12+35,-10) circle (.6cm) node {\footnotesize $y_{\mathcal{S},N}$};

		\draw[black!100] (1.4+12+35,0)--+(2.1,-4);
		\draw[black!100] (1.4+12+35,-4)--+(2.1,0);
		\draw[black!100] (1.4+12+35,-10)--+(2.1,6);

		\draw (17+35,-4) circle (.6cm) node {\footnotesize $\mathcal{L}_{\mathcal{S}}$};

		\node [] at (4+37,-12) {\footnotesize $\mathbf{R}_{\mathcal{S}}$};

		\draw[->,black!100] (37,2.5)--+(8,0);
		\node [] at (4+37,3.5) {\footnotesize forward path};
		\draw[<-,black!100] (37,2.5-16)--+(8,0);
		\node [] at (4+37,2.5-17) {\footnotesize backward path};

		\node [] at (2,2.5-17-2.5) {\footnotesize a)~Two-layer};
		\node [] at (2+35.5,2.5-17-2.5) {\footnotesize b)~Single-layer};
	\end{tikzpicture}
	\caption{Computational graph of BiSCorN for SISO systems in periodic case.}
	\label{aa2}
	\centering
\end{figure}

According to ADAM algorithm, the network coefficients can be updated using the loss function gradient as well as element-wise square power of the loss function gradient at each batch (see \cite{adam} for more details). Therefore, the dominant computational burden is associated with the gradient computation of $b$ loss functions corresponding to $b$ input signals (see Fig.~\ref{aa1}). The gradient calculation in Tensor-flow is done by means of back-propagation algorithm, which computes the gradient using the chain rule \cite{goodfellow2016deep}. Note that in Tensor-flow, first, the information propagates forward through the network. Then, Tensor-flow allows the information backtracks from the loss function to the desired layers to compute the gradients. For example, in the periodic SISO two-layer BiSCorN, the forward propagation is accomplished by $\mathbf{y}_{\mathcal{S}}=\mathbf{X}^T_{\mathcal{S}} \mathbf{u}_{\mathcal{S}}=\mathbf{X}^T_{\mathcal{S}} \mathbf{X}_{\mathcal{S}} \mathbf{s}_{\mathcal{S}}$ as shown in Fig.~\ref{aa2}.a. Then, in the backward path, by using the chain rule, the gradient of the loss function w.r.t. $\mathbf{x}$ can be obtained as follows\footnote{The loss function depends on the network output $\mathbf{y}_{\mathcal{S}}$, and $\mathbf{y}_{\mathcal{S}}$ itself is a function of $\mathbf{x}$ (see \eqref{key12}).} \cite{goodfellow2016deep}
\begin{table}
	\caption{Comparison Between the Per-Iteration Computational Complexity of BiSCorN with Other Methods for Periodic Binary Sequence Design.}
	\centering
		\begin{tikzpicture}[even odd rule,rounded corners=0pt,x=12pt,y=12pt,scale=.5]
			

			\draw[rounded corners=0pt]
			(13.5,-2.5) rectangle ++(12,2.5) node[midway]{\scriptsize BiSCorN (2-layer)};
			\draw[rounded corners=0pt]
			(25.5,-2.5) rectangle ++(12,2.5) node[midway]{\scriptsize BiSCorN (1-layer)};
			\draw[rounded corners=0pt]
			(37.5,-2.5) rectangle ++(12,2.5) node[midway]{\scriptsize CAN};
			\draw[rounded corners=0pt]
			(49.5,-2.5) rectangle ++(12,2.5) node[midway]{\scriptsize WeCAN};
			\draw[rounded corners=0pt]
			(61.5,-2.5) rectangle ++(12,2.5) node[midway]{\scriptsize CD};

			\draw[rounded corners=0pt]
			(13.5,-5) rectangle ++(12,2.5) node[midway]{\scriptsize $\mathcal{O}(N^2)$};
			\draw[rounded corners=0pt]
			(25.5,-5) rectangle ++(12,2.5) node[midway]{\scriptsize $\mathcal{O}(N^2)$};
			\draw[rounded corners=0pt]
			(37.5,-5) rectangle ++(12,2.5) node[midway]{\scriptsize $\mathcal{O}(N \textrm{log}(N))$};
			\draw[rounded corners=0pt]
			(49.5,-5) rectangle ++(12,2.5) node[midway]{\scriptsize $\mathcal{O}((N^2  \textrm{log}(N))$};
			\draw[rounded corners=0pt]
			(61.5,-5) rectangle ++(12,2.5) node[midway]{\scriptsize $\mathcal{O}(N^3)$};

			\draw[rounded corners=0pt]
			(5.5,-5) rectangle ++(8,5) node[midway]{\footnotesize SISO};	

			\draw[rounded corners=0pt]
	    	(13.5,-2.5-5) rectangle ++(12,2.5) node[midway]{\scriptsize BiSCorN (2-layer)};
	    	\draw[rounded corners=0pt]
	    	(25.5,-2.5-5) rectangle ++(12,2.5) node[midway]{\scriptsize BiSCorN (1-layer)};
	    	\draw[rounded corners=0pt]
	    	(37.5,-2.5-5) rectangle ++(12,2.5) node[midway]{\scriptsize Multi-CAN};
	    	\draw[rounded corners=0pt]
	    	(49.5,-2.5-5) rectangle ++(12,2.5) node[midway]{\scriptsize WeMulti-CAN};
	    	\draw[rounded corners=0pt]
	    	(61.5,-2.5-5) rectangle ++(12,2.5) node[midway]{\scriptsize BiST};

			\draw[rounded corners=0pt]
		(13.5,-5-5) rectangle ++(12,2.5) node[midway]{\scriptsize $\mathcal{O}(N^2 N^2_T)$};
		\draw[rounded corners=0pt]
		(25.5,-5-5) rectangle ++(12,2.5) node[midway]{\scriptsize $\mathcal{O}(N^2 N^2_T)$};
		\draw[rounded corners=0pt]
		(37.5,-5-5) rectangle ++(12,2.5) node[midway]{\scriptsize $\mathcal{O}(NN_T \textrm{log}(N))$};
		\draw[rounded corners=0pt]
		(49.5,-5-5) rectangle ++(12,2.5) node[midway]{\scriptsize $\mathcal{O}(N^2 N_T \textrm{log}(N))$};
		\draw[rounded corners=0pt]
		(61.5,-5-5) rectangle ++(12,2.5) node[midway]{\scriptsize $\mathcal{O}(N^3 N_T)$};

	    	\draw[rounded corners=0pt]
	    	(5.5,-7.5-2.5) rectangle ++(8,5) node[midway]{\footnotesize MIMO};
			

\draw[rounded corners=0pt]
(13.5,-2.5-7.5-2.5) rectangle ++(20,2.5) node[midway]{\scriptsize BiSCorN (2-layer)};
\draw[rounded corners=0pt]
(25.5+8,-2.5-7.5-2.5) rectangle ++(20,2.5) node[midway]{\scriptsize BiSCorN (1-layer)};
\draw[rounded corners=0pt]
(37.5+16,-2.5-7.5-2.5) rectangle ++(20,2.5) node[midway]{\scriptsize CANARY};

\draw[rounded corners=0pt]
(13.5,-5-7.5-2.5) rectangle ++(20,2.5) node[midway]{\scriptsize $\mathcal{O}(N^2 M)$};
\draw[rounded corners=0pt]
(25.5+8,-5-7.5-2.5) rectangle ++(20,2.5) node[midway]{\scriptsize $\mathcal{O}(N^2 M)$};
\draw[rounded corners=0pt]
(37.5+16,-5-7.5-2.5) rectangle ++(20,2.5) node[midway]{\scriptsize $\mathcal{O}(NM \textrm{log}(N))$};

\draw[rounded corners=0pt]
(5.5,-7.5-7.5) rectangle ++(8,5) node[midway]{\footnotesize CSS};
			

\draw[rounded corners=0pt]
(13.5,-2.5-7.5-7.5) rectangle ++(30,2.5) node[midway]{\scriptsize BiSCorN (2-layer)};
\draw[rounded corners=0pt]
(25.5+18,-2.5-7.5-7.5) rectangle ++(30,2.5) node[midway]{\scriptsize BiSCorN (1-layer)};

\draw[rounded corners=0pt]
(13.5,-5-7.5-7.5) rectangle ++(30,2.5) node[midway]{\scriptsize $\mathcal{O}(N^2 N^2_T M)$};
\draw[rounded corners=0pt]
(25.5+18,-5-7.5-7.5) rectangle ++(30,2.5) node[midway]{\scriptsize $\mathcal{O}(N^2 N^2_T M)$};

\draw[rounded corners=0pt]
(5.5,-7.5-7.5-5) rectangle ++(8,5) node[midway]{\footnotesize MIMO-CSS};			
		\end{tikzpicture}  \label{ty}
\end{table}
\begin{equation} \label{mmnn}
	{\nabla}_{{\mathbf{x}}} \mathcal{L}_{\mathcal{S}}= \mathbf{J}^T_{\mathbf{u}_\mathcal{S},\mathbf{x}}\hspace{1pt} \mathbf{J}^T_{\mathbf{y}_\mathcal{S},\mathbf{u}_\mathcal{S}} {\nabla}_{{\mathbf{y}_\mathcal{S}}} \mathcal{L}_{\mathcal{S}},
\end{equation}	
where 
\begin{equation}
	\mathbf{J}_{\mathbf{y}_{\mathcal{S}},\mathbf{u}_{\mathcal{S}}}= 
	\begin{bmatrix}
		\frac{\partial y_{\mathcal{S},1}}{\partial u_{\mathcal{S},1}},&
		\hdots&
		\frac{\partial y_{\mathcal{S},1}}{\partial u_{\mathcal{S},N}}\\
		\vdots& &\vdots \\
		\frac{\partial y_{\mathcal{S},N}}{\partial u_{\mathcal{S},1}},&
		\hdots &
		\frac{\partial y_{\mathcal{S},N}}{\partial u_{\mathcal{S},N}}
	\end{bmatrix} \in \mathbb{R}^{N \times N}, \hspace{20pt}
	\mathbf{J}_{\mathbf{u}_{\mathcal{S}},\mathbf{x}}= 
	\begin{bmatrix}
		\frac{\partial u_{\mathcal{S},1}}{\partial x(1)},&
		\hdots&
		\frac{\partial u_{\mathcal{S},1}}{\partial x(N)}\\
		\vdots& &\vdots \\
		\frac{\partial u_{\mathcal{S},N}}{\partial x(1)},&
		\hdots &
		\frac{\partial u_{\mathcal{S},N}}{\partial x(N)}
	\end{bmatrix} \in \mathbb{R}^{N \times N},
\end{equation}
are the Jacobian matrices of $\mathbf{y}_{\mathcal{S}}$ w.r.t. $\mathbf{u}_{\mathcal{S}}$ and $\mathbf{u}_{\mathcal{S}}$ w.r.t. $\mathbf{x}$, respectively\footnote{Note that for evaluation of these parameters, Tensor-flow uses the values which are stored efficiently in the forward path.}. It can be observed that both forward and backward propagation have the computational complexity of $\mathcal{O}({N^2})$. 
Next, consider the single-layer case in Fig.~\ref{aa2}.b: first, the $i^{th}$ entry of the input signal $s_{\mathbf{S},i}$ is multiplied by 
\begin{equation}
\begin{cases}
		\mathbf{x}^T_{p,0} \hspace{2pt} \mathbf{x}_{p,i-1},~~~~~~~~i=1,
		\\
		\mathbf{x}^T_{p,0}\hspace{2pt}  \mathbf{x}_{p,i-1} +N,~~~i=2,...,N,
	\end{cases}
\end{equation}
in the forward path. Then, the backward path consists of a Jacobian-gradient product (similar to \eqref{mmnn}) for gradient computation.
Therefore, using the back-propagation scheme, regardless of the specific BiSCorN procedure, the per-iteration computational complexity is $\mathcal{O}(N^2)$. It is worth pointing out that the computational complexity of other types of BiSCorN can be obtained similarly. For comparison, in Table~\ref{ty} the per-iteration computational complexity of the BiSCorN, CAN \cite{CAN}, WeCAN \cite{CAN}, CD \cite{CD}, Multi-CAN \cite{he2009designing}, WeMulti-CAN \cite{he2009designing}, and CANARY \cite{soltanalian2013fast} are reported for the case of periodic sequence design.
\section{Performance Analysis} \label{per}
\begin{figure}[!t]
	\includegraphics[scale=.55]{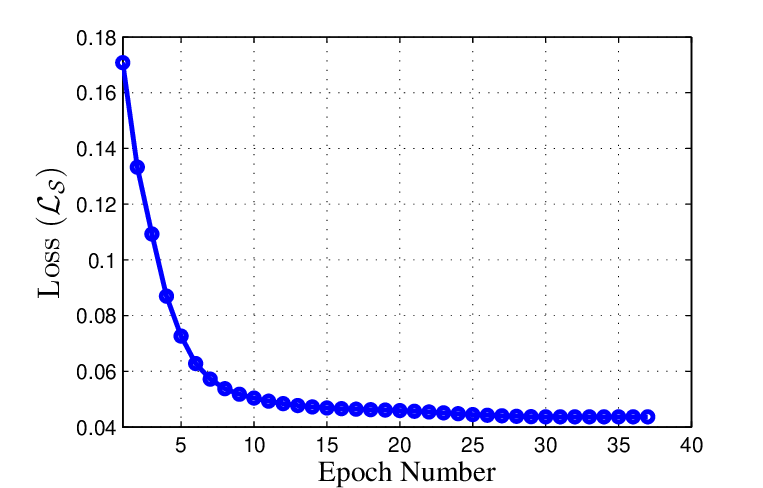}
	\centering
	\caption{Convergence behavior of the proposed aperiodic SISO two-layer BiSCorN for the case of $N=400$.}
	\label{PA45l}
\end{figure}

In this section, the performance of the BiSCorN is evaluated to design binary sequences with good correlation properties. Various system architectures (e.g., SISO, MIMO, and complementary sets) for single and two layer BiSCorN are considered in both periodic and aperiodic cases. 
The input signals fed the network with $d=400$ batches each of size $b=1$ and $b=512$ for deterministic and stochastic data, respectively (see Section~\ref{ext}).
The ISL, WISL, PSL, and CISL metrics are reported in logarithmic scale, e.g., $\textrm{ISL (dB)}= 10 \textrm{log}_{10} \hspace{1pt}\textrm{ISL}$.
The proposed network as well as benchmark methods are initialized using random sequences in all numerical cases. These initializations are realized using a uniform distribution over all feasible sets of sequence. 
Also, all methods are terminated using a stopping criteria with threshold $\epsilon=10^{-5}$ unless otherwise explicitly stated.
Note that the quality of the solutions depends on the starting point; hence, we report the average performance metric (ISL/PSL) over 10 independent trials (except Fig.~\ref{PA45l}, Fig.~\ref{h5t210}, Fig.~\ref{ht11}, Fig.~\ref{PA5l}, some curves of Fig.~\ref{PA4011}, and Fig.~\ref{PA4k012}).
\subsection{SISO}	
The CAN algorithm and its periodic and weighted versions with quantization into two levels, i.e., CAN~(B), PeCAN~(B), and WeCAN~(B)\footnote{Notice that in CAN~(B), PeCAN~(B), and WeCAN~(B) the quantization is performed on the (continuous phase) output sequence of the CAN, PeCAN, and WeCAN algorithms, respectively.}, as well as the CD method for binary sequence design, i.e., CD~(B), are adopted as benchmarks in this subsection\footnote{Note that \cite{CD} and \cite{lin2019efficient} are very close in terms of ISL and so we report the results of \cite{CD}.}. Note that CAN almost usually meets the Welch lower bound on ISL for sequence design with constant modulus but arbitrary phases \cite{welch}. 
\subsubsection{Convergence}	
The convergence of the proposed aperiodic SISO two-layer BiSCorN is shown in Fig.~\ref{PA45l} via considering the values of the loss function in \eqref{key1} for sequence length $N=400$ versus the number of epochs.
It is observed that the loss function
values have a monotonic descent property, as expected.  
\subsubsection{Binarization Versus Sidelobe Minimization} \label{bibi}
%
\begin{table}
	\caption{The ISL Values and the Number of Non-Binary Elements in the Output Aperiodic Sequence of the SISO Single-Layer BiSCorN with Deterministic and Stochastic Inputs for Gradually Increasing the Weight Ratio $\textrm{WR}$ and Different Values of $N$.}
	\centering
	\begin{tikzpicture}[even odd rule,rounded corners=0pt,x=12pt,y=12pt,scale=.5]
		

		\draw[rounded corners=0pt]
		(13.5,-2.5) rectangle ++(33,2.5) node[midway]{\scriptsize $N=50$};
		\draw[rounded corners=0pt]
		(46.5,-2.5) rectangle ++(33,2.5) node[midway]{\scriptsize $N=100$};

		\draw[rounded corners=0pt]
		(13.5,-5) rectangle ++(9,2.5) node[midway]{\scriptsize Avg. ISL (dB)};
		\draw[rounded corners=0pt]
		(22.5,-5) rectangle ++(9,2.5) node[midway]{\scriptsize Best ISL (dB)};
		\draw[rounded corners=0pt]
		(31.5,-5) rectangle ++(15,2.5) node[midway]{\scriptsize Binary Elements (Percent)};
		\draw[rounded corners=0pt]
		(46.5,-5) rectangle ++(9,2.5) node[midway]{\scriptsize Avg. ISL (dB)};
		\draw[rounded corners=0pt]
		(55.5,-5) rectangle ++(9,2.5) node[midway]{\scriptsize Best ISL (dB)};
		\draw[rounded corners=0pt]
		(64.5,-5) rectangle ++(15,2.5) node[midway]{\scriptsize Binary Elements (Percent)};
		
		\draw[rounded corners=0pt]
		(2,-5) rectangle ++(11.5,5) node[midway]{\scriptsize Deterministic Input};		
		
		\draw[rounded corners=0pt]
		(2,-7.5) rectangle ++(11.5,2.5) node[midway]{\scriptsize ${\textrm{WR}}^{(1)}=1$};	
		
		\draw[rounded corners=0pt]
		(13.5,-7.5) rectangle ++(9,2.5) node[midway]{\scriptsize 27.98};	
		
		\draw[rounded corners=0pt]
		(22.5,-7.5) rectangle ++(9,2.5) node[midway]{\scriptsize 27.09};
		
		\draw[rounded corners=0pt]
		(31.5,-7.5) rectangle ++(15,2.5) node[midway]{\scriptsize 76.20};

		\draw[rounded corners=0pt]
		(46.5,-7.5) rectangle ++(9,2.5) node[midway]{\scriptsize 34.28};
		
		\draw[rounded corners=0pt]
		(55.5,-7.5) rectangle ++(9,2.5) node[midway]{\scriptsize 34.01};
		
		\draw[rounded corners=0pt]
		(64.5,-7.5) rectangle ++(15,2.5) node[midway]{\scriptsize 75.37};

		\draw[rounded corners=0pt]
		(2,-10) rectangle ++(11.5,2.5) node[midway]{\scriptsize ${\textrm{WR}}^{(2)}=2$};	
		
		\draw[rounded corners=0pt]
		(13.5,-10) rectangle ++(9,2.5) node[midway]{\scriptsize 27.93};	
		
		\draw[rounded corners=0pt]
		(22.5,-10) rectangle ++(9,2.5) node[midway]{\scriptsize 27.09};
		
		\draw[rounded corners=0pt]
		(31.5,-10) rectangle ++(15,2.5) node[midway]{\scriptsize 90.00};

		\draw[rounded corners=0pt]
		(46.5,-10) rectangle ++(9,2.5) node[midway]{\scriptsize 33.80};
		
		\draw[rounded corners=0pt]
		(55.5,-10) rectangle ++(9,2.5) node[midway]{\scriptsize 33.32};
		
		\draw[rounded corners=0pt]
		(64.5,-10) rectangle ++(15,2.5) node[midway]{\scriptsize 89.38};

		\draw[rounded corners=0pt]
		(2,-12.5) rectangle ++(11.5,2.5) node[midway]{\scriptsize ${\textrm{WR}}^{(3)}=4$};	
		
		\draw[rounded corners=0pt]
		(13.5,-12.5) rectangle ++(9,2.5) node[midway]{\scriptsize 27.84};	
		
		\draw[rounded corners=0pt]
		(22.5,-12.5) rectangle ++(9,2.5) node[midway]{\scriptsize 27.09};
		
		\draw[rounded corners=0pt]
		(31.5,-12.5) rectangle ++(15,2.5) node[midway]{\scriptsize 95.40};

		\draw[rounded corners=0pt]
		(46.5,-12.5) rectangle ++(9,2.5) node[midway]{\scriptsize 33.62};
		
		\draw[rounded corners=0pt]
		(55.5,-12.5) rectangle ++(9,2.5) node[midway]{\scriptsize 33.32};
		
		\draw[rounded corners=0pt]
		(64.5,-12.5) rectangle ++(15,2.5) node[midway]{\scriptsize 95.50};

		\draw[rounded corners=0pt]
		(2,-15) rectangle ++(11.5,2.5) node[midway]{\scriptsize ${\textrm{WR}}^{(4)}=8$};	
		
		\draw[rounded corners=0pt]
		(13.5,-15) rectangle ++(9,2.5) node[midway]{\scriptsize 27.84};	
		
		\draw[rounded corners=0pt]
		(22.5,-15) rectangle ++(9,2.5) node[midway]{\scriptsize 27.09};
		
		\draw[rounded corners=0pt]
		(31.5,-15) rectangle ++(15,2.5) node[midway]{\scriptsize 97.80};

		\draw[rounded corners=0pt]
		(46.5,-15) rectangle ++(9,2.5) node[midway]{\scriptsize 33.59};
		
		\draw[rounded corners=0pt]
		(55.5,-15) rectangle ++(9,2.5) node[midway]{\scriptsize 33.32};
		
		\draw[rounded corners=0pt]
		(64.5,-15) rectangle ++(15,2.5) node[midway]{\scriptsize 98.63};

		\draw[rounded corners=0pt]
		(2,-17.5) rectangle ++(11.5,2.5) node[midway]{\scriptsize ${\textrm{WR}}^{(5)}=16$};	
		
		\draw[rounded corners=0pt]
		(13.5,-17.5) rectangle ++(9,2.5) node[midway]{\scriptsize 27.84};	
		
		\draw[rounded corners=0pt]
		(22.5,-17.5) rectangle ++(9,2.5) node[midway]{\scriptsize 27.09};
		
		\draw[rounded corners=0pt]
		(31.5,-17.5) rectangle ++(15,2.5) node[midway]{\scriptsize 98.00};

		\draw[rounded corners=0pt]
		(46.5,-17.5) rectangle ++(9,2.5) node[midway]{\scriptsize 33.59};
		
		\draw[rounded corners=0pt]
		(55.5,-17.5) rectangle ++(9,2.5) node[midway]{\scriptsize 33.32};
		
		\draw[rounded corners=0pt]
		(64.5,-17.5) rectangle ++(15,2.5) node[midway]{\scriptsize 98.88};

\draw[rounded corners=0pt]
(2,-20) rectangle ++(11.5,2.5) node[midway]{\scriptsize ${\textrm{WR}}^{(6)}=32$};	

\draw[rounded corners=0pt]
(13.5,-20) rectangle ++(9,2.5) node[midway]{\scriptsize 27.84};	

\draw[rounded corners=0pt]
(22.5,-20) rectangle ++(9,2.5) node[midway]{\scriptsize 27.09};

\draw[rounded corners=0pt]
(31.5,-20) rectangle ++(15,2.5) node[midway]{\scriptsize 98.00};

\draw[rounded corners=0pt]
(46.5,-20) rectangle ++(9,2.5) node[midway]{\scriptsize 33.59};

\draw[rounded corners=0pt]
(55.5,-20) rectangle ++(9,2.5) node[midway]{\scriptsize 33.32};

\draw[rounded corners=0pt]
(64.5,-20) rectangle ++(15,2.5) node[midway]{\scriptsize 98.88};
		
		
		\draw[rounded corners=0pt]
		(13.5,-2.5-18-2.5) rectangle ++(33,2.5) node[midway]{\scriptsize $N=50$};
		\draw[rounded corners=0pt]
		(46.5,-2.5-18-2.5) rectangle ++(33,2.5) node[midway]{\scriptsize $N=100$};

		\draw[rounded corners=0pt]
		(13.5,-5-18-2.5) rectangle ++(9,2.5) node[midway]{\scriptsize Avg. ISL (dB)};
		\draw[rounded corners=0pt]
		(22.5,-5-18-2.5) rectangle ++(9,2.5) node[midway]{\scriptsize Best ISL (dB)};
		\draw[rounded corners=0pt]
		(31.5,-5-18-2.5) rectangle ++(15,2.5) node[midway]{\scriptsize Binary Elements (Percent)};
		\draw[rounded corners=0pt]
		(46.5,-5-18-2.5) rectangle ++(9,2.5) node[midway]{\scriptsize Avg. ISL (dB)};
		\draw[rounded corners=0pt]
		(55.5,-5-18-2.5) rectangle ++(9,2.5) node[midway]{\scriptsize Best ISL (dB)};
		\draw[rounded corners=0pt]
		(64.5,-5-18-2.5) rectangle ++(15,2.5) node[midway]{\scriptsize Binary Elements (Percent)};
		
		\draw[rounded corners=0pt]
		(2,-5-18-2.5) rectangle ++(11.5,5) node[midway]{\scriptsize Stochastic Input};		
		
		\draw[rounded corners=0pt]
		(2,-7.5-18-2.5) rectangle ++(11.5,2.5) node[midway]{\scriptsize ${\textrm{WR}}^{(1)}=1$};	
		
		\draw[rounded corners=0pt]
		(13.5,-7.5-18-2.5) rectangle ++(9,2.5) node[midway]{\scriptsize 27.81};	
		
		\draw[rounded corners=0pt]
		(22.5,-7.5-18-2.5) rectangle ++(9,2.5) node[midway]{\scriptsize 26.96};
		
		\draw[rounded corners=0pt]
		(31.5,-7.5-18-2.5) rectangle ++(15,2.5) node[midway]{\scriptsize 74.80};

		\draw[rounded corners=0pt]
		(46.5,-7.5-18-2.5) rectangle ++(9,2.5) node[midway]{\scriptsize 33.89};
		
		\draw[rounded corners=0pt]
		(55.5,-7.5-18-2.5) rectangle ++(9,2.5) node[midway]{\scriptsize 33.06};
		
		\draw[rounded corners=0pt]
		(64.5,-7.5-18-2.5) rectangle ++(15,2.5) node[midway]{\scriptsize 76.88};

		\draw[rounded corners=0pt]
		(2,-10-18-2.5) rectangle ++(11.5,2.5) node[midway]{\scriptsize ${\textrm{WR}}^{(2)}=2$};	
		
		\draw[rounded corners=0pt]
		(13.5,-10-18-2.5) rectangle ++(9,2.5) node[midway]{\scriptsize 27.52};	
		
		\draw[rounded corners=0pt]
		(22.5,-10-18-2.5) rectangle ++(9,2.5) node[midway]{\scriptsize 26.96};
		
		\draw[rounded corners=0pt]
		(31.5,-10-18-2.5) rectangle ++(15,2.5) node[midway]{\scriptsize 88.80};

		\draw[rounded corners=0pt]
		(46.5,-10-18-2.5) rectangle ++(9,2.5) node[midway]{\scriptsize 33.58};
		
		\draw[rounded corners=0pt]
		(55.5,-10-18-2.5) rectangle ++(9,2.5) node[midway]{\scriptsize 32.90};
		
		\draw[rounded corners=0pt]
		(64.5,-10-18-2.5) rectangle ++(15,2.5) node[midway]{\scriptsize 89.75};

		\draw[rounded corners=0pt]
		(2,-12.5-18-2.5) rectangle ++(11.5,2.5) node[midway]{\scriptsize ${\textrm{WR}}^{(3)}=4$};	
		
		\draw[rounded corners=0pt]
		(13.5,-12.5-18-2.5) rectangle ++(9,2.5) node[midway]{\scriptsize 27.45};	
		
		\draw[rounded corners=0pt]
		(22.5,-12.5-18-2.5) rectangle ++(9,2.5) node[midway]{\scriptsize 26.96};
		
		\draw[rounded corners=0pt]
		(31.5,-12.5-18-2.5) rectangle ++(15,2.5) node[midway]{\scriptsize 94.80};

		\draw[rounded corners=0pt]
		(46.5,-12.5-18-2.5) rectangle ++(9,2.5) node[midway]{\scriptsize 33.47};
		
		\draw[rounded corners=0pt]
		(55.5,-12.5-18-2.5) rectangle ++(9,2.5) node[midway]{\scriptsize 32.90};
		
		\draw[rounded corners=0pt]
		(64.5,-12.5-18-2.5) rectangle ++(15,2.5) node[midway]{\scriptsize 96.62};

		\draw[rounded corners=0pt]
		(2,-15-18-2.5) rectangle ++(11.5,2.5) node[midway]{\scriptsize ${\textrm{WR}}^{(4)}=8$};	
		
		\draw[rounded corners=0pt]
		(13.5,-15-18-2.5) rectangle ++(9,2.5) node[midway]{\scriptsize 27.45};	
		
		\draw[rounded corners=0pt]
		(22.5,-15-18-2.5) rectangle ++(9,2.5) node[midway]{\scriptsize 26.96};
		
		\draw[rounded corners=0pt]
		(31.5,-15-18-2.5) rectangle ++(15,2.5) node[midway]{\scriptsize 97.40};

		\draw[rounded corners=0pt]
		(46.5,-15-18-2.5) rectangle ++(9,2.5) node[midway]{\scriptsize 33.46};
		
		\draw[rounded corners=0pt]
		(55.5,-15-18-2.5) rectangle ++(9,2.5) node[midway]{\scriptsize 32.90};
		
		\draw[rounded corners=0pt]
		(64.5,-15-18-2.5) rectangle ++(15,2.5) node[midway]{\scriptsize 98.38};

		\draw[rounded corners=0pt]
		(2,-17.5-18-2.5) rectangle ++(11.5,2.5) node[midway]{\scriptsize ${\textrm{WR}}^{(5)}=16$};	
		
		\draw[rounded corners=0pt]
		(13.5,-17.5-18-2.5) rectangle ++(9,2.5) node[midway]{\scriptsize 27.45};	
		
		\draw[rounded corners=0pt]
		(22.5,-17.5-18-2.5) rectangle ++(9,2.5) node[midway]{\scriptsize 26.96};
		
		\draw[rounded corners=0pt]
		(31.5,-17.5-18-2.5) rectangle ++(15,2.5) node[midway]{\scriptsize 97.80};

		\draw[rounded corners=0pt]
		(46.5,-17.5-18-2.5) rectangle ++(9,2.5) node[midway]{\scriptsize 33.46};
		
		\draw[rounded corners=0pt]
		(55.5,-17.5-18-2.5) rectangle ++(9,2.5) node[midway]{\scriptsize 32.90};
		
		\draw[rounded corners=0pt]
		(64.5,-17.5-18-2.5) rectangle ++(15,2.5) node[midway]{\scriptsize 98.88};

		\draw[rounded corners=0pt]
		(2,-20-18-2.5) rectangle ++(11.5,2.5) node[midway]{\scriptsize ${\textrm{WR}}^{(6)}=32$};	
		
		\draw[rounded corners=0pt]
		(13.5,-20-18-2.5) rectangle ++(9,2.5) node[midway]{\scriptsize 27.45};	
		
		\draw[rounded corners=0pt]
		(22.5,-20-18-2.5) rectangle ++(9,2.5) node[midway]{\scriptsize 26.96};
		
		\draw[rounded corners=0pt]
		(31.5,-20-18-2.5) rectangle ++(15,2.5) node[midway]{\scriptsize 97.80};

		\draw[rounded corners=0pt]
		(46.5,-20-18-2.5) rectangle ++(9,2.5) node[midway]{\scriptsize 33.46};
		
		\draw[rounded corners=0pt]
		(55.5,-20-18-2.5) rectangle ++(9,2.5) node[midway]{\scriptsize 22.90};
		
		\draw[rounded corners=0pt]
		(64.5,-20-18-2.5) rectangle ++(15,2.5) node[midway]{\scriptsize 98.88};
	\end{tikzpicture}  \label{ty11}
\end{table}	
According to Lemma~1, the design problem in \eqref{maxmin21} can be dealt with via a stop-start procedure applied to problem in \eqref{maxmin23} with fixed $\bar{w}_{\mathcal{S},s}$ and increasing $\bar{w}_{\mathcal{S},b}$. To illustrate this, we consider both deterministic and random cases for single-layer BiSCorN. We define the weight ratio (WR) as ${\textrm{WR}}^{(i)}=\frac{\bar{w}^{(i)}_{\mathcal{S},b}}{\bar{w}_{\mathcal{S},s}}$. The results for deterministic case ${\textrm{WR}}^{(i)}=2^{(i-1)}$, $i \geq 1$, and different values of $N$ are reported in Table~\ref{ty11}. That is, the table includes the ISL values of the synthesized binary sequences (see Remark~1) as well as the number of non-binary elements\footnote{An entry $x(n)$ of a sequence is considered as binary if $\left \vert \hspace{1pt}\vert x(n) \vert -1 \hspace{1pt}\right \vert \leq 10^{-8}$.}. As expected, in terms of average performance, increasing WR leads to less non-binary elements; however, for WR greater than 4, the resulting ISL values are kept fixed. Considering the best performance over 10 trials, we observed no change in ISL values for WR grater than 2. As an example, the resulting sequence before synthesis stage is shown in Fig.~\ref{h5t210} for $N=50$ and different values of WR.  It can be seen that by increasing WR,
the number of non-binary elements decreases; also, the absolute value of non-binary elements approaches $\lbrace -1, +1\rbrace$ as can be seen by considering $\textrm{WR}=8,16,32$. As to random case, from Lemma~1 and Subsection~\ref{sisolcz} one can write
\begin{figure}
	\centering
	\begin{tikzpicture}[even odd rule,rounded corners=0pt,x=12pt,y=12pt,scale=.65,every node/.style={scale=1.1}]

		\draw[fill=blue!30] (0,16) rectangle ++(2,2) node[midway]{\tiny 1};
		\draw[fill=blue!30] (2,16) rectangle ++(2,2)  node[midway]{\tiny -1};
		\draw[fill=blue!30] (4,16) rectangle ++(2,2) node[midway]{\tiny 1};
		\draw[fill=blue!30] (6,16) rectangle ++(2,2) node[midway]{\tiny 1};
		\draw[fill=blue!30] (8,16) rectangle ++(2,2)  node[midway]{\tiny -1};
		\draw[fill=blue!30] (10,16) rectangle ++(2,2)  node[midway]{\tiny -1};
		\draw[fill=blue!30] (12,16) rectangle ++(2,2)  node[midway]{\tiny 1};
		\draw[fill=green!30] (14,16) rectangle ++(2,2)  
		node[midway]{\tiny \textbf{0.33}};
		\draw[fill=blue!30] (16,16) rectangle ++(2,2)  node[midway]{\tiny -1};
		\draw[fill=blue!30] (18,16) rectangle ++(2,2)  node[midway]{\tiny 1};
		\draw[fill=blue!30] (20,16) rectangle ++(2,2)  node[midway]{\tiny 1};
		\draw[fill=blue!30] (22,16) rectangle ++(2,2)  node[midway]{\tiny -1};
		\draw[fill=blue!30] (24,16) rectangle ++(2,2)  node[midway]{\tiny 1};
		\draw[fill=green!30] (26,16) rectangle ++(2,2)  
		node[midway]{\tiny \textbf{-0.94}};
		\draw[fill=green!30] (28,16) rectangle ++(2,2)  node[midway]{\tiny \textbf{0.86}};
		\draw[fill=green!30] (30,16) rectangle ++(2,2)  node[midway]{\tiny \textbf{0.76}};
		\draw[fill=blue!30] (32,16) rectangle ++(2,2)  node[midway]{\tiny -1};
		\draw[fill=blue!30] (34,16) rectangle ++(2,2)  node[midway]{\tiny 1};
		\draw[fill=blue!30] (36,16) rectangle ++(2,2)  node[midway]{\tiny 1};
		\draw[fill=blue!30] (38,16) rectangle ++(2,2)  node[midway]{\tiny 1};
		\draw[fill=blue!30] (40,16) rectangle ++(2,2)  node[midway]{\tiny 1};
		\draw[fill=blue!30] (42,16) rectangle ++(2,2)  node[midway]{\tiny -1};
		\draw[fill=blue!30] (44,16) rectangle ++(2,2)  node[midway]{\tiny -1};
		\draw[fill=blue!30] (46,16) rectangle ++(2,2)  node[midway]{\tiny -1};
		\draw[fill=blue!30] (48,16) rectangle ++(2,2)  node[midway]{\tiny -1};
		
		\draw[fill=blue!30] (0,14) rectangle ++(2,2) node[midway]{\tiny 1};
		\draw[fill=blue!30] (2,14) rectangle ++(2,2)  node[midway]{\tiny 1};
		\draw[fill=blue!30] (4,14) rectangle ++(2,2) node[midway]{\tiny 1};
		\draw[fill=blue!30] (6,14) rectangle ++(2,2) node[midway]{\tiny 1};
		\draw[fill=green!30] (8,14) rectangle ++(2,2)  node[midway]{\tiny \textbf{0.18}};
		\draw[fill=green!30] (10,14) rectangle ++(2,2)  node[midway]{\tiny \textbf{0.75}};
		\draw[fill=blue!30] (12,14) rectangle ++(2,2)  node[midway]{\tiny 1};
		\draw[fill=blue!30] (14,14) rectangle ++(2,2)  node[midway]{\tiny 1};
		\draw[fill=blue!30] (16,14) rectangle ++(2,2)  node[midway]{\tiny 1};
		\draw[fill=blue!30] (18,14) rectangle ++(2,2)  node[midway]{\tiny -1};
		\draw[fill=green!30] (20,14) rectangle ++(2,2)  node[midway]{\tiny \textbf{0.10}};
		\draw[fill=green!30] (22,14) rectangle ++(2,2)  node[midway]{\tiny \textbf{-0.82}};
		\draw[fill=blue!30] (24,14) rectangle ++(2,2)  node[midway]{\tiny 1};
		\draw[fill=blue!30] (26,14) rectangle ++(2,2)  node[midway]{\tiny -1};
		\draw[fill=green!30] (28,14) rectangle ++(2,2)  node[midway]{\tiny \textbf{0.18}};
		\draw[fill=blue!30] (30,14) rectangle ++(2,2)  node[midway]{\tiny -1};
		\draw[fill=blue!30] (32,14) rectangle ++(2,2)  node[midway]{\tiny 1};
		\draw[fill=green!30] (34,14) rectangle ++(2,2)  node[midway]{\tiny \textbf{-0.08}};
		\draw[fill=blue!30] (36,14) rectangle ++(2,2)  node[midway]{\tiny -1};
		\draw[fill=blue!30] (38,14) rectangle ++(2,2)  node[midway]{\tiny -1};
		\draw[fill=blue!30] (40,14) rectangle ++(2,2)  node[midway]{\tiny -1};
		\draw[fill=blue!30] (42,14) rectangle ++(2,2)  node[midway]{\tiny 1};
		\draw[fill=green!30] (44,14) rectangle ++(2,2)  node[midway]{\tiny \textbf{-0.39}};
		\draw[fill=green!30] (46,14) rectangle ++(2,2)  node[midway]{\tiny \textbf{-0.48}};
		\draw[fill=blue!30] (48,14) rectangle ++(2,2)  node[midway]{\tiny -1};

		\draw[fill=blue!30] (0,12) rectangle ++(2,2) node[midway]{\tiny 1};
		\draw[fill=blue!30] (2,12) rectangle ++(2,2)  node[midway]{\tiny -1};
		\draw[fill=blue!30] (4,12) rectangle ++(2,2) node[midway]{\tiny 1};
		\draw[fill=blue!30] (6,12) rectangle ++(2,2) node[midway]{\tiny 1};
		\draw[fill=blue!30] (8,12) rectangle ++(2,2)  node[midway]{\tiny -1};
		\draw[fill=blue!30] (10,12) rectangle ++(2,2)  node[midway]{\tiny -1};
		\draw[fill=blue!30] (12,12) rectangle ++(2,2)  node[midway]{\tiny 1};
		\draw[fill=blue!30] (14,12) rectangle ++(2,2)  node[midway]{\tiny 1};
		\draw[fill=blue!30] (16,12) rectangle ++(2,2)  node[midway]{\tiny -1};
		\draw[fill=blue!30] (18,12) rectangle ++(2,2)  node[midway]{\tiny {1}};
		\draw[fill=blue!30] (20,12) rectangle ++(2,2)  node[midway]{\tiny 1};
		\draw[fill=blue!30] (22,12) rectangle ++(2,2)  node[midway]{\tiny {-1}};
		\draw[fill=green!30] (24,12) rectangle ++(2,2)  node[midway]{\tiny \textbf{-0.55}};
		\draw[fill=blue!30] (26,12) rectangle ++(2,2)  node[midway]{\tiny -1};
		\draw[fill=blue!30] (28,12) rectangle ++(2,2)  node[midway]{\tiny 1};
		\draw[fill=blue!30] (30,12) rectangle ++(2,2)  node[midway]{\tiny 1};
		\draw[fill=green!30] (32,12) rectangle ++(2,2)  node[midway]{\tiny \textbf{-0.77}};
		\draw[fill=blue!30] (34,12) rectangle ++(2,2)  node[midway]{\tiny 1};
		\draw[fill=blue!30] (36,12) rectangle ++(2,2)  node[midway]{\tiny 1};
		\draw[fill=blue!30] (38,12) rectangle ++(2,2)  node[midway]{\tiny 1};
		\draw[fill=blue!30] (40,12) rectangle ++(2,2)  node[midway]{\tiny 1};
		\draw[fill=blue!30] (42,12) rectangle ++(2,2)  node[midway]{\tiny -1};
		\draw[fill=blue!30] (44,12) rectangle ++(2,2)  node[midway]{\tiny -1};
		\draw[fill=blue!30] (46,12) rectangle ++(2,2)  node[midway]{\tiny -1};
		\draw[fill=blue!30] (48,12) rectangle ++(2,2)  node[midway]{\tiny -1};

		\draw[fill=blue!30] (0,10) rectangle ++(2,2) node[midway]{\tiny {1}};
		\draw[fill=blue!30] (2,10) rectangle ++(2,2)  node[midway]{\tiny 1};
		\draw[fill=blue!30] (4,10) rectangle ++(2,2) node[midway]{\tiny 1};
		\draw[fill=blue!30] (6,10) rectangle ++(2,2) node[midway]{\tiny {1}};
		\draw[fill=blue!30] (8,10) rectangle ++(2,2)  node[midway]{\tiny -1};
		\draw[fill=blue!30] (10,10) rectangle ++(2,2)  node[midway]{\tiny 1};
		\draw[fill=blue!30] (12,10) rectangle ++(2,2)  node[midway]{\tiny 1};
		\draw[fill=blue!30] (14,10) rectangle ++(2,2)  node[midway]{\tiny 1};
		\draw[fill=blue!30] (16,10) rectangle ++(2,2)  node[midway]{\tiny 1};
		\draw[fill=blue!30] (18,10) rectangle ++(2,2)  node[midway]{\tiny {-1}};
		\draw[fill=blue!30] (20,10) rectangle ++(2,2)  node[midway]{\tiny 1};
		\draw[fill=blue!30] (22,10) rectangle ++(2,2)  node[midway]{\tiny -1};
		\draw[fill=blue!30] (24,10) rectangle ++(2,2)  node[midway]{\tiny 1};
		\draw[fill=blue!30] (26,10) rectangle ++(2,2)  node[midway]{\tiny {-1}};
		\draw[fill=green!30] (28,10) rectangle ++(2,2)  node[midway]{\tiny \textbf{0.14}};
		\draw[fill=blue!30] (30,10) rectangle ++(2,2)  node[midway]{\tiny -1};
		\draw[fill=blue!30] (32,10) rectangle ++(2,2)  node[midway]{\tiny 1};
		\draw[fill=blue!30] (34,10) rectangle ++(2,2)  node[midway]{\tiny -1};
		\draw[fill=blue!30] (36,10) rectangle ++(2,2)  node[midway]{\tiny -1};
		\draw[fill=blue!30] (38,10) rectangle ++(2,2)  node[midway]{\tiny -1};
		\draw[fill=blue!30] (40,10) rectangle ++(2,2)  node[midway]{\tiny -1};
		\draw[fill=green!30] (42,10) rectangle ++(2,2)  node[midway]{\tiny \textbf{0.09}};
		\draw[fill=green!30] (44,10) rectangle ++(2,2)  node[midway]{\tiny \textbf{0.77}};
		\draw[fill=blue!30] (46,10) rectangle ++(2,2)  node[midway]{\tiny -1};
		\draw[fill=blue!30] (48,10) rectangle ++(2,2)  node[midway]{\tiny -1};

		\draw[fill=blue!30] (0,8) rectangle ++(2,2) node[midway]{\tiny 1};
		\draw[fill=blue!30] (2,8) rectangle ++(2,2)  node[midway]{\tiny -1};
		\draw[fill=blue!30] (4,8) rectangle ++(2,2) node[midway]{\tiny 1};
		\draw[fill=blue!30] (6,8) rectangle ++(2,2) node[midway]{\tiny 1};
		\draw[fill=blue!30] (8,8) rectangle ++(2,2)  node[midway]{\tiny {-1}};
		\draw[fill=blue!30] (10,8) rectangle ++(2,2)  node[midway]{\tiny -1};
		\draw[fill=blue!30] (12,8) rectangle ++(2,2)  node[midway]{\tiny 1};
		\draw[fill=blue!30] (14,8) rectangle ++(2,2)  node[midway]{\tiny {1}};
		\draw[fill=blue!30] (16,8) rectangle ++(2,2)  node[midway]{\tiny -1};
		\draw[fill=blue!30] (18,8) rectangle ++(2,2)  node[midway]{\tiny 1};
		\draw[fill=blue!30] (20,8) rectangle ++(2,2)  node[midway]{\tiny 1};
		\draw[fill=blue!30] (22,8) rectangle ++(2,2)  node[midway]{\tiny -1};
		\draw[fill=blue!30] (24,8) rectangle ++(2,2)  node[midway]{\tiny -1};
		\draw[fill=blue!30] (26,8) rectangle ++(2,2)  node[midway]{\tiny -1};
		\draw[fill=blue!30] (28,8) rectangle ++(2,2)  node[midway]{\tiny 1};
		\draw[fill=blue!30] (30,8) rectangle ++(2,2)  node[midway]{\tiny 1};
		\draw[fill=blue!30] (32,8) rectangle ++(2,2)  node[midway]{\tiny -1};
		\draw[fill=blue!30] (34,8) rectangle ++(2,2)  node[midway]{\tiny 1};
		\draw[fill=blue!30] (36,8) rectangle ++(2,2)  node[midway]{\tiny 1};
		\draw[fill=blue!30] (38,8) rectangle ++(2,2)  node[midway]{\tiny 1};
		\draw[fill=blue!30] (40,8) rectangle ++(2,2)  node[midway]{\tiny 1};
		\draw[fill=blue!30] (42,8) rectangle ++(2,2)  node[midway]{\tiny -1};
		\draw[fill=blue!30] (44,8) rectangle ++(2,2)  node[midway]{\tiny -1};
		\draw[fill=blue!30] (46,8) rectangle ++(2,2)  node[midway]{\tiny -1};
		\draw[fill=blue!30] (48,8) rectangle ++(2,2)  node[midway]{\tiny -1};

		\draw[fill=blue!30] (0,6) rectangle ++(2,2) node[midway]{\tiny 1};
		\draw[fill=blue!30] (2,6) rectangle ++(2,2)  node[midway]{\tiny 1};
		\draw[fill=blue!30] (4,6) rectangle ++(2,2) node[midway]{\tiny 1};
		\draw[fill=blue!30] (6,6) rectangle ++(2,2) node[midway]{\tiny 1};
		\draw[fill=blue!30] (8,6) rectangle ++(2,2)  node[midway]{\tiny -1};
		\draw[fill=blue!30] (10,6) rectangle ++(2,2)  node[midway]{\tiny {1}};
		\draw[fill=blue!30] (12,6) rectangle ++(2,2)  node[midway]{\tiny 1};
		\draw[fill=blue!30] (14,6) rectangle ++(2,2)  node[midway]{\tiny 1};
		\draw[fill=blue!30] (16,6) rectangle ++(2,2)  node[midway]{\tiny 1};
		\draw[fill=blue!30] (18,6) rectangle ++(2,2)  node[midway]{\tiny {-1}};
		\draw[fill=blue!30] (20,6) rectangle ++(2,2)  node[midway]{\tiny 1};
		\draw[fill=blue!30] (22,6) rectangle ++(2,2)  node[midway]{\tiny -1};
		\draw[fill=blue!30] (24,6) rectangle ++(2,2)  node[midway]{\tiny 1};
		\draw[fill=blue!30] (26,6) rectangle ++(2,2)  node[midway]{\tiny -1};
		\draw[fill=green!30] (28,6) rectangle ++(2,2)  node[midway]{\tiny \textbf{0.19}};
		\draw[fill=blue!30] (30,6) rectangle ++(2,2)  node[midway]{\tiny {-1}};
		\draw[fill=blue!30] (32,6) rectangle ++(2,2)  node[midway]{\tiny 1};
		\draw[fill=blue!30] (34,6) rectangle ++(2,2)  node[midway]{\tiny -1};
		\draw[fill=blue!30] (36,6) rectangle ++(2,2)  node[midway]{\tiny -1};
		\draw[fill=blue!30] (38,6) rectangle ++(2,2)  node[midway]{\tiny -1};
		\draw[fill=blue!30] (40,6) rectangle ++(2,2)  node[midway]{\tiny -1};
		\draw[fill=green!30] (42,6) rectangle ++(2,2)  node[midway]{\tiny \textbf{-0.74}};
		\draw[fill=blue!30] (44,6) rectangle ++(2,2)  node[midway]{\tiny 1};
		\draw[fill=blue!30] (46,6) rectangle ++(2,2)  node[midway]{\tiny -1};
		\draw[fill=blue!30] (48,6) rectangle ++(2,2)  node[midway]{\tiny -1};

		\draw[fill=blue!30] (0,4) rectangle ++(2,2) node[midway]{\tiny 1};
		\draw[fill=blue!30] (2,4) rectangle ++(2,2)  node[midway]{\tiny -1};
		\draw[fill=blue!30] (4,4) rectangle ++(2,2) node[midway]{\tiny 1};
		\draw[fill=blue!30] (6,4) rectangle ++(2,2) node[midway]{\tiny 1};
		\draw[fill=blue!30] (8,4) rectangle ++(2,2)  node[midway]{\tiny {-1}};
		\draw[fill=blue!30] (10,4) rectangle ++(2,2)  node[midway]{\tiny -1};
		\draw[fill=blue!30] (12,4) rectangle ++(2,2)  node[midway]{\tiny 1};
		\draw[fill=blue!30] (14,4) rectangle ++(2,2)  node[midway]{\tiny {1}};
		\draw[fill=blue!30] (16,4) rectangle ++(2,2)  node[midway]{\tiny -1};
		\draw[fill=blue!30] (18,4) rectangle ++(2,2)  node[midway]{\tiny 1};
		\draw[fill=blue!30] (20,4) rectangle ++(2,2)  node[midway]{\tiny 1};
		\draw[fill=blue!30] (22,4) rectangle ++(2,2)  node[midway]{\tiny -1};
		\draw[fill=blue!30] (24,4) rectangle ++(2,2)  node[midway]{\tiny -1};
		\draw[fill=blue!30] (26,4) rectangle ++(2,2)  node[midway]{\tiny -1};
		\draw[fill=blue!30] (28,4) rectangle ++(2,2)  node[midway]{\tiny 1};
		\draw[fill=blue!30] (30,4) rectangle ++(2,2)  node[midway]{\tiny 1};
		\draw[fill=blue!30] (32,4) rectangle ++(2,2)  node[midway]{\tiny -1};
		\draw[fill=blue!30] (34,4) rectangle ++(2,2)  node[midway]{\tiny 1};
		\draw[fill=blue!30] (36,4) rectangle ++(2,2)  node[midway]{\tiny 1};
		\draw[fill=blue!30] (38,4) rectangle ++(2,2)  node[midway]{\tiny 1};
		\draw[fill=blue!30] (40,4) rectangle ++(2,2)  node[midway]{\tiny 1};
		\draw[fill=blue!30] (42,4) rectangle ++(2,2)  node[midway]{\tiny -1};
		\draw[fill=blue!30] (44,4) rectangle ++(2,2)  node[midway]{\tiny -1};
		\draw[fill=blue!30] (46,4) rectangle ++(2,2)  node[midway]{\tiny -1};
		\draw[fill=blue!30] (48,4) rectangle ++(2,2)  node[midway]{\tiny -1};

		\draw[fill=blue!30] (0,2) rectangle ++(2,2) node[midway]{\tiny 1};
		\draw[fill=blue!30] (2,2) rectangle ++(2,2)  node[midway]{\tiny 1};
		\draw[fill=blue!30] (4,2) rectangle ++(2,2) node[midway]{\tiny 1};
		\draw[fill=blue!30] (6,2) rectangle ++(2,2) node[midway]{\tiny 1};
		\draw[fill=blue!30] (8,2) rectangle ++(2,2)  node[midway]{\tiny -1};
		\draw[fill=blue!30] (10,2) rectangle ++(2,2)  node[midway]{\tiny {1}};
		\draw[fill=blue!30] (12,2) rectangle ++(2,2)  node[midway]{\tiny 1};
		\draw[fill=blue!30] (14,2) rectangle ++(2,2)  node[midway]{\tiny 1};
		\draw[fill=blue!30] (16,2) rectangle ++(2,2)  node[midway]{\tiny 1};
		\draw[fill=blue!30] (18,2) rectangle ++(2,2)  node[midway]{\tiny {-1}};
		\draw[fill=blue!30] (20,2) rectangle ++(2,2)  node[midway]{\tiny 1};
		\draw[fill=blue!30] (22,2) rectangle ++(2,2)  node[midway]{\tiny -1};
		\draw[fill=blue!30] (24,2) rectangle ++(2,2)  node[midway]{\tiny 1};
		\draw[fill=blue!30] (26,2) rectangle ++(2,2)  node[midway]{\tiny -1};
		\draw[fill=green!30] (28,2) rectangle ++(2,2)  node[midway]{\tiny \textbf{0.78}};
		\draw[fill=blue!30] (30,2) rectangle ++(2,2)  node[midway]{\tiny {-1}};
		\draw[fill=blue!30] (32,2) rectangle ++(2,2)  node[midway]{\tiny 1};
		\draw[fill=blue!30] (34,2) rectangle ++(2,2)  node[midway]{\tiny -1};
		\draw[fill=blue!30] (36,2) rectangle ++(2,2)  node[midway]{\tiny -1};
		\draw[fill=blue!30] (38,2) rectangle ++(2,2)  node[midway]{\tiny -1};
		\draw[fill=blue!30] (40,2) rectangle ++(2,2)  node[midway]{\tiny -1};
		\draw[fill=blue!30] (42,2) rectangle ++(2,2)  node[midway]{\tiny {-1}};
		\draw[fill=blue!30] (44,2) rectangle ++(2,2)  node[midway]{\tiny 1};
		\draw[fill=blue!30] (46,2) rectangle ++(2,2)  node[midway]{\tiny -1};
		\draw[fill=blue!30] (48,2) rectangle ++(2,2)  node[midway]{\tiny -1};

		\draw[fill=blue!30] (0,0) rectangle ++(2,2) node[midway]{\tiny 1};
		\draw[fill=blue!30] (2,0) rectangle ++(2,2)  node[midway]{\tiny -1};
		\draw[fill=blue!30] (4,0) rectangle ++(2,2) node[midway]{\tiny 1};
		\draw[fill=blue!30] (6,0) rectangle ++(2,2) node[midway]{\tiny 1};
		\draw[fill=blue!30] (8,0) rectangle ++(2,2)  node[midway]{\tiny {-1}};
		\draw[fill=blue!30] (10,0) rectangle ++(2,2)  node[midway]{\tiny -1};
		\draw[fill=blue!30] (12,0) rectangle ++(2,2)  node[midway]{\tiny 1};
		\draw[fill=blue!30] (14,0) rectangle ++(2,2)  node[midway]{\tiny {1}};
		\draw[fill=blue!30] (16,0) rectangle ++(2,2)  node[midway]{\tiny -1};
		\draw[fill=blue!30] (18,0) rectangle ++(2,2)  node[midway]{\tiny 1};
		\draw[fill=blue!30] (20,0) rectangle ++(2,2)  node[midway]{\tiny 1};
		\draw[fill=blue!30] (22,0) rectangle ++(2,2)  node[midway]{\tiny -1};
		\draw[fill=blue!30] (24,0) rectangle ++(2,2)  node[midway]{\tiny -1};
		\draw[fill=blue!30] (26,0) rectangle ++(2,2)  node[midway]{\tiny -1};
		\draw[fill=blue!30] (28,0) rectangle ++(2,2)  node[midway]{\tiny 1};
		\draw[fill=blue!30] (30,0) rectangle ++(2,2)  node[midway]{\tiny 1};
		\draw[fill=blue!30] (32,0) rectangle ++(2,2)  node[midway]{\tiny -1};
		\draw[fill=blue!30] (34,0) rectangle ++(2,2)  node[midway]{\tiny 1};
		\draw[fill=blue!30] (36,0) rectangle ++(2,2)  node[midway]{\tiny 1};
		\draw[fill=blue!30] (38,0) rectangle ++(2,2)  node[midway]{\tiny 1};
		\draw[fill=blue!30] (40,0) rectangle ++(2,2)  node[midway]{\tiny 1};
		\draw[fill=blue!30] (42,0) rectangle ++(2,2)  node[midway]{\tiny -1};
		\draw[fill=blue!30] (44,0) rectangle ++(2,2)  node[midway]{\tiny -1};
		\draw[fill=blue!30] (46,0) rectangle ++(2,2)  node[midway]{\tiny -1};
		\draw[fill=blue!30] (48,0) rectangle ++(2,2)  node[midway]{\tiny -1};

		\draw[fill=blue!30] (0,-2) rectangle ++(2,2) node[midway]{\tiny 1};
		\draw[fill=blue!30] (2,-2) rectangle ++(2,2)  node[midway]{\tiny 1};
		\draw[fill=blue!30] (4,-2) rectangle ++(2,2) node[midway]{\tiny 1};
		\draw[fill=blue!30] (6,-2) rectangle ++(2,2) node[midway]{\tiny 1};
		\draw[fill=blue!30] (8,-2) rectangle ++(2,2)  node[midway]{\tiny -1};
		\draw[fill=blue!30] (10,-2) rectangle ++(2,2)  node[midway]{\tiny {1}};
		\draw[fill=blue!30] (12,-2) rectangle ++(2,2)  node[midway]{\tiny 1};
		\draw[fill=blue!30] (14,-2) rectangle ++(2,2)  node[midway]{\tiny 1};
		\draw[fill=blue!30] (16,-2) rectangle ++(2,2)  node[midway]{\tiny 1};
		\draw[fill=blue!30] (18,-2) rectangle ++(2,2)  node[midway]{\tiny {-1}};
		\draw[fill=blue!30] (20,-2) rectangle ++(2,2)  node[midway]{\tiny 1};
		\draw[fill=blue!30] (22,-2) rectangle ++(2,2)  node[midway]{\tiny -1};
		\draw[fill=blue!30] (24,-2) rectangle ++(2,2)  node[midway]{\tiny 1};
		\draw[fill=blue!30] (26,-2) rectangle ++(2,2)  node[midway]{\tiny -1};
		\draw[fill=green!30] (28,-2) rectangle ++(2,2)  node[midway]{\tiny \textbf{0.94}};
		\draw[fill=blue!30] (30,-2) rectangle ++(2,2)  node[midway]{\tiny {-1}};
		\draw[fill=blue!30] (32,-2) rectangle ++(2,2)  node[midway]{\tiny 1};
		\draw[fill=blue!30] (34,-2) rectangle ++(2,2)  node[midway]{\tiny -1};
		\draw[fill=blue!30] (36,-2) rectangle ++(2,2)  node[midway]{\tiny -1};
		\draw[fill=blue!30] (38,-2) rectangle ++(2,2)  node[midway]{\tiny -1};
		\draw[fill=blue!30] (40,-2) rectangle ++(2,2)  node[midway]{\tiny -1};
		\draw[fill=blue!30] (42,-2) rectangle ++(2,2)  node[midway]{\tiny {-1}};
		\draw[fill=blue!30] (44,-2) rectangle ++(2,2)  node[midway]{\tiny 1};
		\draw[fill=blue!30] (46,-2) rectangle ++(2,2)  node[midway]{\tiny -1};
		\draw[fill=blue!30] (48,-2) rectangle ++(2,2)  node[midway]{\tiny -1};

		\draw[fill=blue!30] (0,-4) rectangle ++(2,2) node[midway]{\tiny 1};
		\draw[fill=blue!30] (2,-4) rectangle ++(2,2)  node[midway]{\tiny -1};
		\draw[fill=blue!30] (4,-4) rectangle ++(2,2) node[midway]{\tiny 1};
		\draw[fill=blue!30] (6,-4) rectangle ++(2,2) node[midway]{\tiny 1};
		\draw[fill=blue!30] (8,-4) rectangle ++(2,2)  node[midway]{\tiny {-1}};
		\draw[fill=blue!30] (10,-4) rectangle ++(2,2)  node[midway]{\tiny -1};
		\draw[fill=blue!30] (12,-4) rectangle ++(2,2)  node[midway]{\tiny 1};
		\draw[fill=blue!30] (14,-4) rectangle ++(2,2)  node[midway]{\tiny {1}};
		\draw[fill=blue!30] (16,-4) rectangle ++(2,2)  node[midway]{\tiny -1};
		\draw[fill=blue!30] (18,-4) rectangle ++(2,2)  node[midway]{\tiny 1};
		\draw[fill=blue!30] (20,-4) rectangle ++(2,2)  node[midway]{\tiny 1};
		\draw[fill=blue!30] (22,-4) rectangle ++(2,2)  node[midway]{\tiny -1};
		\draw[fill=blue!30] (24,-4) rectangle ++(2,2)  node[midway]{\tiny -1};
		\draw[fill=blue!30] (26,-4) rectangle ++(2,2)  node[midway]{\tiny -1};
		\draw[fill=blue!30] (28,-4) rectangle ++(2,2)  node[midway]{\tiny 1};
		\draw[fill=blue!30] (30,-4) rectangle ++(2,2)  node[midway]{\tiny 1};
		\draw[fill=blue!30] (32,-4) rectangle ++(2,2)  node[midway]{\tiny -1};
		\draw[fill=blue!30] (34,-4) rectangle ++(2,2)  node[midway]{\tiny 1};
		\draw[fill=blue!30] (36,-4) rectangle ++(2,2)  node[midway]{\tiny 1};
		\draw[fill=blue!30] (38,-4) rectangle ++(2,2)  node[midway]{\tiny 1};
		\draw[fill=blue!30] (40,-4) rectangle ++(2,2)  node[midway]{\tiny 1};
		\draw[fill=blue!30] (42,-4) rectangle ++(2,2)  node[midway]{\tiny -1};
		\draw[fill=blue!30] (44,-4) rectangle ++(2,2)  node[midway]{\tiny -1};
		\draw[fill=blue!30] (46,-4) rectangle ++(2,2)  node[midway]{\tiny -1};
		\draw[fill=blue!30] (48,-4) rectangle ++(2,2)  node[midway]{\tiny -1};

		\draw[fill=blue!30] (0,-6) rectangle ++(2,2) node[midway]{\tiny 1};
		\draw[fill=blue!30] (2,-6) rectangle ++(2,2)  node[midway]{\tiny 1};
		\draw[fill=blue!30] (4,-6) rectangle ++(2,2) node[midway]{\tiny 1};
		\draw[fill=blue!30] (6,-6) rectangle ++(2,2) node[midway]{\tiny 1};
		\draw[fill=blue!30] (8,-6) rectangle ++(2,2)  node[midway]{\tiny -1};
		\draw[fill=blue!30] (10,-6) rectangle ++(2,2)  node[midway]{\tiny {1}};
		\draw[fill=blue!30] (12,-6) rectangle ++(2,2)  node[midway]{\tiny 1};
		\draw[fill=blue!30] (14,-6) rectangle ++(2,2)  node[midway]{\tiny 1};
		\draw[fill=blue!30] (16,-6) rectangle ++(2,2)  node[midway]{\tiny 1};
		\draw[fill=blue!30] (18,-6) rectangle ++(2,2)  node[midway]{\tiny {-1}};
		\draw[fill=blue!30] (20,-6) rectangle ++(2,2)  node[midway]{\tiny 1};
		\draw[fill=blue!30] (22,-6) rectangle ++(2,2)  node[midway]{\tiny -1};
		\draw[fill=blue!30] (24,-6) rectangle ++(2,2)  node[midway]{\tiny 1};
		\draw[fill=blue!30] (26,-6) rectangle ++(2,2)  node[midway]{\tiny -1};
		\draw[fill=green!30] (28,-6) rectangle ++(2,2)  node[midway]{\tiny \textbf{0.98}};
		\draw[fill=blue!30] (30,-6) rectangle ++(2,2)  node[midway]{\tiny {-1}};
		\draw[fill=blue!30] (32,-6) rectangle ++(2,2)  node[midway]{\tiny 1};
		\draw[fill=blue!30] (34,-6) rectangle ++(2,2)  node[midway]{\tiny -1};
		\draw[fill=blue!30] (36,-6) rectangle ++(2,2)  node[midway]{\tiny -1};
		\draw[fill=blue!30] (38,-6) rectangle ++(2,2)  node[midway]{\tiny -1};
		\draw[fill=blue!30] (40,-6) rectangle ++(2,2)  node[midway]{\tiny -1};
		\draw[fill=blue!30] (42,-6) rectangle ++(2,2)  node[midway]{\tiny {-1}};
		\draw[fill=blue!30] (44,-6) rectangle ++(2,2)  node[midway]{\tiny 1};
		\draw[fill=blue!30] (46,-6) rectangle ++(2,2)  node[midway]{\tiny -1};
		\draw[fill=blue!30] (48,-6) rectangle ++(2,2)  node[midway]{\tiny -1};

		\draw[thick,line width=1.2] (0,-6) rectangle ++(50,4) ; 	
		\draw[thick,line width=1.2] (0,-2) rectangle ++(50,4) ; 	
		\draw[thick,line width=1.2] (0,2) rectangle ++(50,4) ; 		
		\draw[thick,line width=1.2] (0,6) rectangle ++(50,4) ;
		\draw[thick,line width=1.2] (0,10) rectangle ++(50,4) ;
		\draw[thick,line width=1.2] (0,14) rectangle ++(50,4) ;
		\draw[thick,line width=1.2] (-7,-6) rectangle ++(7,4) node[midway]{\footnotesize $\textrm{WR}^{(6)}=32$};
		\draw[thick,line width=1.2] (-7,-2) rectangle ++(7,4) node[midway]{\footnotesize $\textrm{WR}^{(5)}=16$};
		\draw[thick,line width=1.2] (-7,2) rectangle ++(7,4) node[midway]{\footnotesize $\textrm{WR}^{(4)}=8$};
		\draw[thick,line width=1.2] (-7,6) rectangle ++(7,4) node[midway]{\footnotesize $\textrm{WR}^{(3)}=4$};
		\draw[thick,line width=1.2] (-7,10) rectangle ++(7,4) node[midway]{\footnotesize $\textrm{WR}^{(2)}=2$};
		\draw[thick,line width=1.2] (-7,14) rectangle ++(7,4) node[midway]{\footnotesize $\textrm{WR}^{(1)}=1$};	
		
	\end{tikzpicture}
	\caption{The output aperiodic sequence of SISO single-layer BiSCorN with deterministic input without quantization for gradually increasing the weight ratio $\textrm{WR}$ and $N=50$.}
	\label{h5t210} 
	\centering
\end{figure}
\begin{equation} \;\label{jjk}
\widetilde{\mathbf{w}}=	\begin{cases}
		\bar{{{w}}}_{\mathcal{S},b}={{\mu}}^2_{\mathcal{S},k}+ p_{\mathcal{S},k},~~~~~~~~~k=1,
		\\
		w_{k-1} \bar{{w}}_{\mathcal{S},s}={{\mu}}^2_{\mathcal{S},k}+ p_{\mathcal{S},k},~~ 2 \leq k \leq N.
	\end{cases}
\end{equation}
Now, given $\widetilde{\mathbf{w}}$, \eqref{jjk} is a system of linear equations with multiple choices for $\boldsymbol{{\mu}}_{\mathcal{S}}$ and $\mathbf{{p}}_{\mathcal{S}}$. We set ${{\mu}}_{\mathcal{S},k}$ equal to $0.5$,  ${{p}}_{\mathcal{S},k}$ equal to $\frac{{{\mu}}_{\mathcal{S},k}}{5}$, for $2 \leq k \leq N$, and then gradually increase ${{\mu}}_{\mathcal{S},1}$ with ${{p}}_{\mathcal{S},1}=\frac{{{\mu}}_{\mathcal{S},1}}{5}$ to obtain WRs in Table~\ref{ty11}. Note that network inputs for various steps are statistically independent.
The results for the random case are also given in Table~\ref{ty11}. Similar considerations as those for the deterministic case hold true. We herein remark the fact that employing BiSCorN in a stop-start procedure is associated with higher computational burden while the improvement in ISL values is not significant. Therefore, in the sequel, we present results employing BiSCorN without stop-start procedure. More precisely, for SISO and CSS we set $\textrm{WR}=1$, i.e., $w_{\mathcal{S},b}=w_{\mathcal{S},s}$ and $w_{\mathcal{C},b}=w_{\mathcal{C},s}$; then, for MIMO and MIMO-CSS cases we set $\textrm{WR}=N_T$, i.e., $w_{\mathcal{M},b}=N_T\hspace{1pt} w_{\mathcal{M},s}$ and $w_{\mathcal{MC},b}=N_T\hspace{1pt} w_{\mathcal{MC},s}$ based on our numerical observation (see Subsection~\ref{keynew1}).
\subsubsection{SISO ISL Minimization}
\begin{figure}[!t]
	\includegraphics[scale=.45]{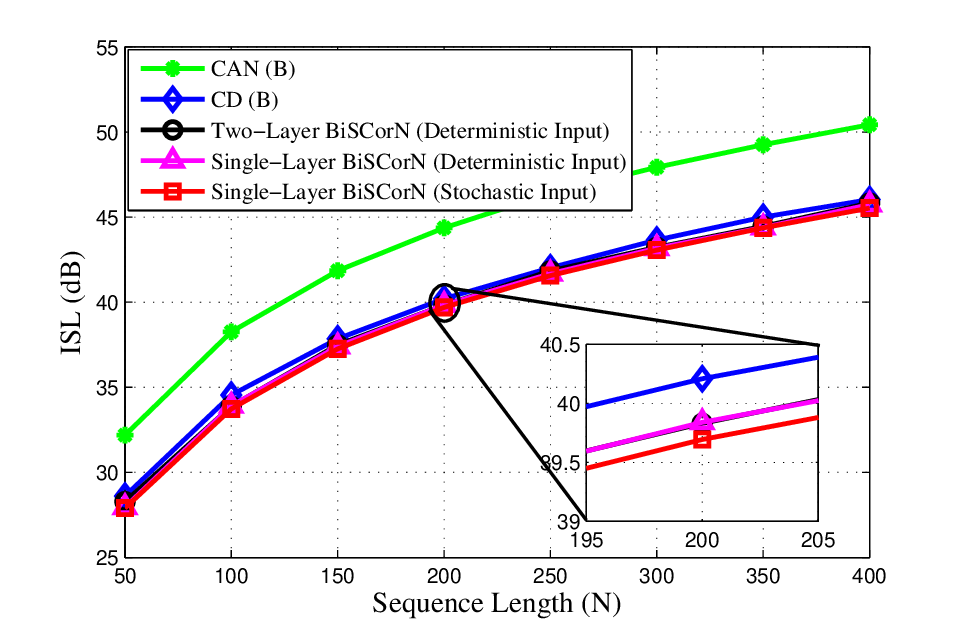}
	\centering
	\caption{ISL of binary sequences obtained by BiSCorN, CAN~(B), and CD~(B) versus sequence length $N$ for the aperiodic ISL minimization problem.}
	\label{PA4}
\end{figure}
\begin{figure}
	\centering
	\begin{tikzpicture}[even odd rule,rounded corners=0pt,x=12pt,y=12pt,scale=.69,every node/.style={scale=.8}]

		\draw[-,line width=1,black!100] (10,20)--+(0,-10);

	\draw[fill=White!30] (0,30) rectangle ++(10,2) node[midway]{\footnotesize Sequence};
	\draw[fill=White!30] (10,30) rectangle ++(4,2) node[midway]{\footnotesize Barker };

		\draw[fill=green!30] (0,10) rectangle ++(2,2) node[midway]{\footnotesize \textbf{-0.81}}; 
		\draw[fill=blue!30] (2,10) rectangle ++(2,2) node[midway]{\footnotesize -1};
		\draw[fill=blue!30] (4,10) rectangle ++(2,2) node[midway]{\footnotesize -1}; 
		\draw[fill=blue!30] (6,10) rectangle ++(2,2) node[midway]{\footnotesize 1};
		\draw[fill=green!30] (8,10) rectangle ++(2,2) node[midway]{\footnotesize \textbf{-0.81}};
		\draw[fill=white!30] (10,10) rectangle ++(4,2) node[midway]{\ding{51}};

		\draw[fill=green!30] (0,12) rectangle ++(2,2) node[midway]{\footnotesize \textbf{0.81}}; 
		\draw[fill=blue!30] (2,12) rectangle ++(2,2) node[midway]{\footnotesize -1};
		\draw[fill=blue!30] (4,12) rectangle ++(2,2) node[midway]{\footnotesize 1}; 
		\draw[fill=blue!30] (6,12) rectangle ++(2,2) node[midway]{\footnotesize 1};
		\draw[fill=green!30] (8,12) rectangle ++(2,2) node[midway]{\footnotesize \textbf{0.81}};
		\draw[fill=white!30] (10,12) rectangle ++(4,2) node[midway]{\ding{51}};

		\draw[fill=green!30] (0,14) rectangle ++(2,2) node[midway]{\footnotesize \textbf{0.81}}; 
		\draw[fill=blue!30] (2,14) rectangle ++(2,2) node[midway]{\footnotesize -1};
		\draw[fill=blue!30] (4,14) rectangle ++(2,2) node[midway]{\footnotesize 1}; 
		\draw[fill=blue!30] (6,14) rectangle ++(2,2) node[midway]{\footnotesize 1};
		\draw[fill=green!30] (8,14) rectangle ++(2,2) node[midway]{\footnotesize \textbf{0.81}};
		\draw[fill=white!30] (10,14) rectangle ++(4,2) node[midway]{\ding{51}};

		\draw[fill=green!30] (0,16) rectangle ++(2,2) node[midway]{\footnotesize \textbf{-0.81}}; 
		\draw[fill=blue!30] (2,16) rectangle ++(2,2) node[midway]{\footnotesize -1};
		\draw[fill=blue!30] (4,16) rectangle ++(2,2) node[midway]{\footnotesize -1}; 
		\draw[fill=blue!30] (6,16) rectangle ++(2,2) node[midway]{\footnotesize 1};
		\draw[fill=green!30] (8,16) rectangle ++(2,2) node[midway]{\footnotesize \textbf{-0.81}};
		\draw[fill=white!30] (10,16) rectangle ++(4,2) node[midway]{\ding{51}};

		\draw[fill=green!30] (0,18) rectangle ++(2,2) node[midway]{\footnotesize \textbf{0.81}}; 
		\draw[fill=blue!30] (2,18) rectangle ++(2,2) node[midway]{\footnotesize -1};
		\draw[fill=blue!30] (4,18) rectangle ++(2,2) node[midway]{\footnotesize 1}; 
		\draw[fill=blue!30] (6,18) rectangle ++(2,2) node[midway]{\footnotesize 1};
		\draw[fill=green!30] (8,18) rectangle ++(2,2) node[midway]{\footnotesize \textbf{0.81}};
		\draw[fill=white!30] (10,18) rectangle ++(4,2) node[midway]{\ding{51}};

		\draw[fill=green!30] (0,20) rectangle ++(2,2) node[midway]{\footnotesize \textbf{0.81}}; 
		\draw[fill=blue!30] (2,20) rectangle ++(2,2) node[midway]{\footnotesize 1};
		\draw[fill=blue!30] (4,20) rectangle ++(2,2) node[midway]{\footnotesize 1}; 
		\draw[fill=blue!30] (6,20) rectangle ++(2,2) node[midway]{\footnotesize -1};
		\draw[fill=green!30] (8,20) rectangle ++(2,2) node[midway]{\footnotesize \textbf{0.81}};
		\draw[fill=white!30] (10,20) rectangle ++(4,2) node[midway]{\ding{51}};

		\draw[fill=green!30] (0,22) rectangle ++(2,2) node[midway]{\footnotesize \textbf{0.94}}; 
		\draw[fill=green!30] (2,22) rectangle ++(2,2) node[midway]{\footnotesize \textbf{0.27}};
		\draw[fill=blue!30] (4,22) rectangle ++(2,2) node[midway]{\footnotesize 1}; 
		\draw[fill=blue!30] (6,22) rectangle ++(2,2) node[midway]{\footnotesize 1};
		\draw[fill=blue!30] (8,22) rectangle ++(2,2) node[midway]{\footnotesize -1};
		\draw[fill=red!30] (10,22) rectangle ++(4,2) node[midway]{\ding{55}};

		\draw[fill=green!30] (0,24) rectangle ++(2,2) node[midway]{\footnotesize \textbf{0.81}}; 
		\draw[fill=blue!30] (2,24) rectangle ++(2,2) node[midway]{\footnotesize 1};
		\draw[fill=blue!30] (4,24) rectangle ++(2,2) node[midway]{\footnotesize 1}; 
		\draw[fill=blue!30] (6,24) rectangle ++(2,2) node[midway]{\footnotesize -1};
		\draw[fill=green!30] (8,24) rectangle ++(2,2) node[midway]{\footnotesize \textbf{0.81}};
		\draw[fill=white!30] (10,24) rectangle ++(4,2) node[midway]{\ding{51}};

		\draw[fill=green!30] (0,26) rectangle ++(2,2) node[midway]{\footnotesize \textbf{-0.81}}; 
		\draw[fill=blue!30] (2,26) rectangle ++(2,2) node[midway]{\footnotesize -1};
		\draw[fill=blue!30] (4,26) rectangle ++(2,2) node[midway]{\footnotesize -1}; 
		\draw[fill=blue!30] (6,26) rectangle ++(2,2) node[midway]{\footnotesize 1};
		\draw[fill=green!30] (8,26) rectangle ++(2,2) node[midway]{\footnotesize \textbf{-0.81}};
		\draw[fill=white!30] (10,26) rectangle ++(4,2) node[midway]{\ding{51}};

		\draw[fill=blue!30] (0,28) rectangle ++(2,2) node[midway]{\footnotesize 1}; 
		\draw[fill=blue!30] (2,28) rectangle ++(2,2) node[midway]{\footnotesize -1};
		\draw[fill=blue!30] (4,28) rectangle ++(2,2) node[midway]{\footnotesize -1}; 
		\draw[fill=green!30] (6,28) rectangle ++(2,2) node[midway]{\footnotesize \textbf{-0.27}};
		\draw[fill=green!30] (8,28) rectangle ++(2,2) node[midway]{\footnotesize \textbf{-0.94}};
		\draw[fill=red!30] (10,28) rectangle ++(4,2) node[midway]{\ding{55}};
	\end{tikzpicture}
	\caption{The output sequence of the two-layer BiSCorN (without quantization) for $N=5$. The various rows in the figure are associated with different random initializations.}
	\label{ht11}
	\centering
\end{figure}
\begin{figure}[!t]
	\includegraphics[scale=.52]{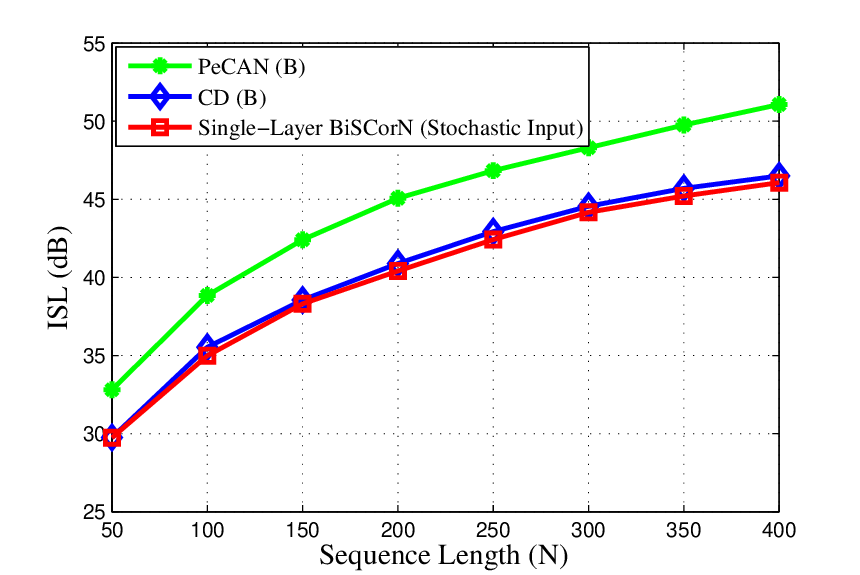}
	\centering
	\caption{ISL of binary sequences obtained by BiSCorN, PeCAN~(B), and CD~(B) versus sequence length $N$ for the periodic ISL minimization problem.}
	\label{PA5}
\end{figure}
The ISL versus the sequence length $N$ is shown in Fig. \ref{PA4} for the aperiodic ISL minimization problem. We observe that the proposed BiSCorN has better performance than the benchmark methods, i.e., CAN~(B) and CD~(B), for the considered values of sequence length $N$. Specifically, the maximum ISL improvements of the single-layer BiSCorN for the case of stochastic input w.r.t. the CAN~(B) and CD~(B) are 4.90 dB and 0.82 dB, respectively.

Fig. \ref{PA4} also shows the performance gain granted by the stochastic version of the input signal $\mathbf{s}_{\mathcal{S}}$ compared to the deterministic one; this can be explained using the fact that the ISL minimization is a multimodal problem and therefore the stochastic version of $\mathbf{s}_{\mathcal{S}}$ can help BiSCorN to achieve a better performance.

It is also numerically observed that the BiSCorN can produce Barker codes for the sequence lengths $N=2,3,4,5,7,11,13$ which confirm the effectiveness of the method. For instance, in Fig.~\ref{ht11}, we employ the two-layer BiSCorN with $N =5$ and consider $10$ random initializations. It can be seen that in most of  the instances, the resulting binary sequences are coincide with Barker codes.
 
Next, the ISL of the single-layer BiSCorN with stochastic input as well as that of benchmark methods is illustrated versus the sequence length $N$ for periodic ISL minimization in Fig. \ref{PA5}. In this case, the maximum performance gains of the proposed network w.r.t. PeCAN~(B) and also CD~(B) algorithms are 4.99 dB and 0.55 dB, respectively.
\subsubsection{SISO ISL Minimization with LCZ}	
\begin{figure}
	\hfill
	\subfigure[$\widetilde{a}_{\mathcal{S},k}=1,~ 0 \leq k \leq \ceil{0.1 \times N},~~\widetilde{a}_{\mathcal{S},k}=0,~\textrm{otherwise}$.]{\includegraphics[width=6.9cm,height=4.45cm]{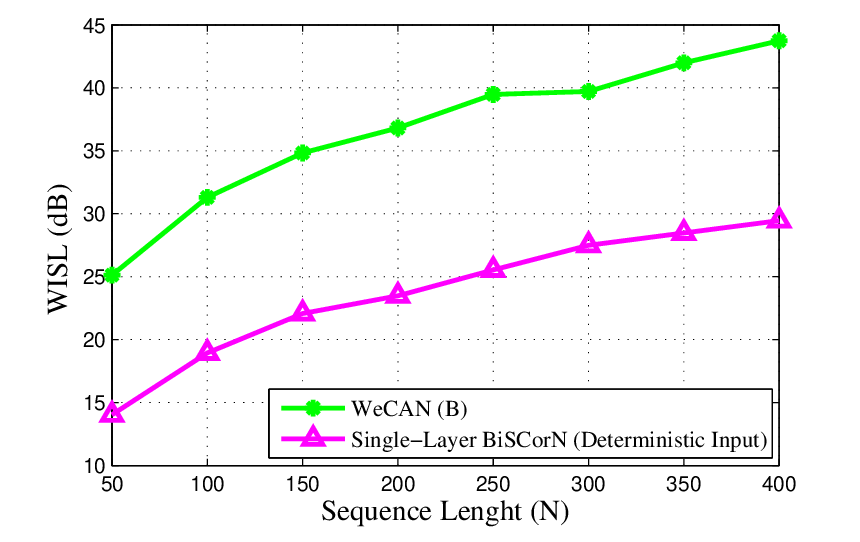}}
	\hfill
	\subfigure[$\widetilde{a}_{\mathcal{S},k}=1,~ 0 \leq k \leq \ceil{0.2 \times N},~~\widetilde{a}_{\mathcal{S},k}=0,~\textrm{otherwise}$.]{\includegraphics[width=6.9cm,height=4.5cm]{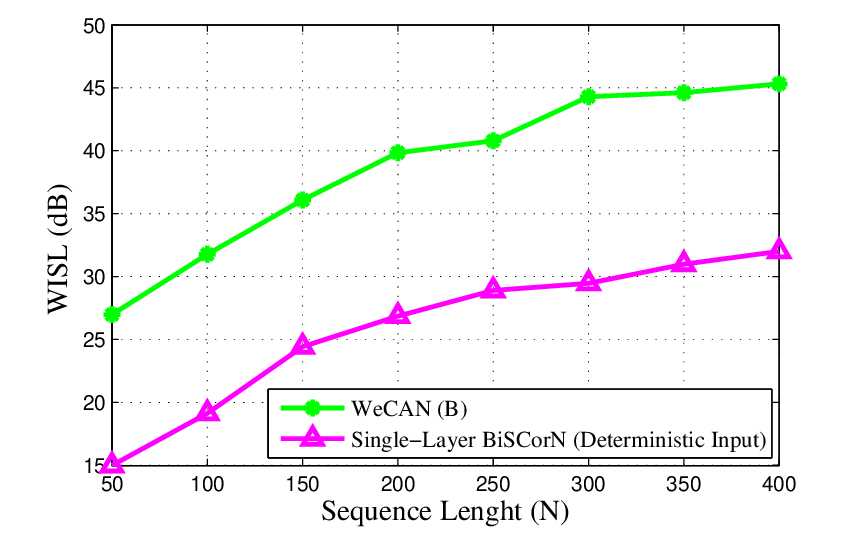}}
	\hfill
	\caption{WISL of binary sequences obtained by BiSCorN and WeCAN~(B) versus sequence length $N$ for the aperiodic WISL minimization problem: (a)~$\textrm{ROS}=10\%$,~(b)~$\textrm{ROS}=20\%$.}
	
	\label{httttttt0}
\end{figure}
\begin{figure}[!t]
	\includegraphics[scale=.48]{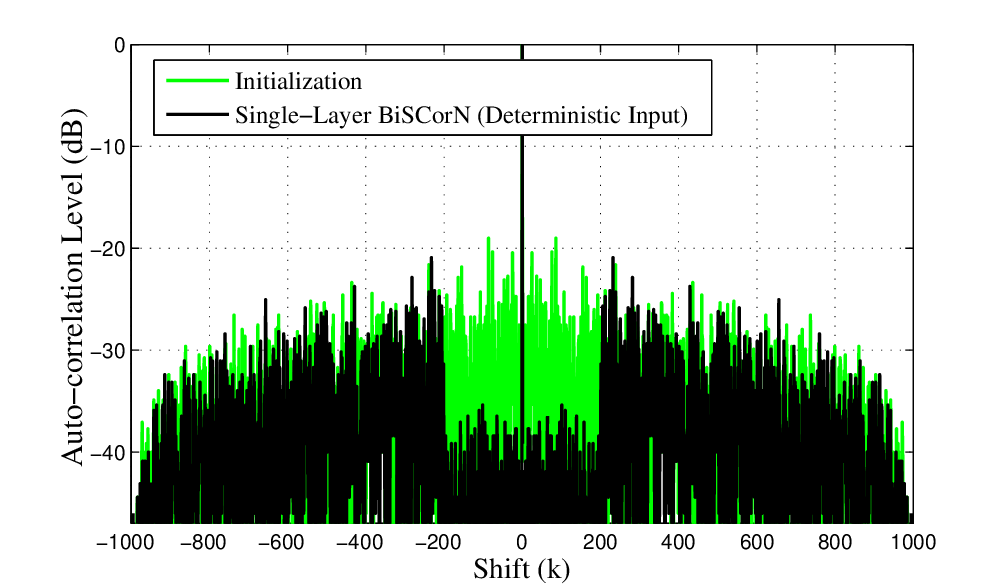}
	\centering
	\caption{Auto-correlation level of binary sequence obtained by BiSCorN for the aperiodic WISL minimization problem with $N=1000$.}
	\label{PA5l}
\end{figure}

The performance of the proposed SISO BiSCorN is analyzed to design LCZ binary sequences. Let Region of Suppress (ROS) represents the set of lags associated with LCZ.
Thus, the input signal $\mathbf{s}_{\mathcal{S}}$ must be compliant with the ROS and at the same time to balance between binarization and sidelobe minimization. Fig.~\ref{httttttt0} shows WISL of the single-layer BiSCorN with deterministic input compared to the WeCAN~(B) for WISL minimization problem. Precisely, in Fig.~\ref{httttttt0}.a and Fig.~\ref{httttttt0}.b, respectively, the aim is to suppress the sidelobes $\left \lbrace r(k)\vert_{k=1}^{\ceil{0.1\times N}} \right\rbrace$ and $\left\lbrace r(k)\vert_{k=1}^{\ceil{0.2 \times N}}\right \rbrace$ (corresponding to the ROS with $10\%$ and $20\%$ of all sidelobes).
It can be seen from Fig.~\ref{httttttt0} that BiSCorN achieves significant smaller WISL than the benchmark method, i.e., WeCAN~(B).
Also, as an example the aperiodic auto-correlation function of the code obtained by single-layer BiSCorN along with the employed random initial sequence are illustrated	in Fig.~\ref{PA5l} for sequence length $N=1000$ and the ROS with $20\%$ of all sidelobes. 
This figure clearly unveils the effectiveness of the proposed BiSCorN.
\subsubsection{PSL Evaluation} 
\begin{figure}[!t]
	\includegraphics[scale=.45]{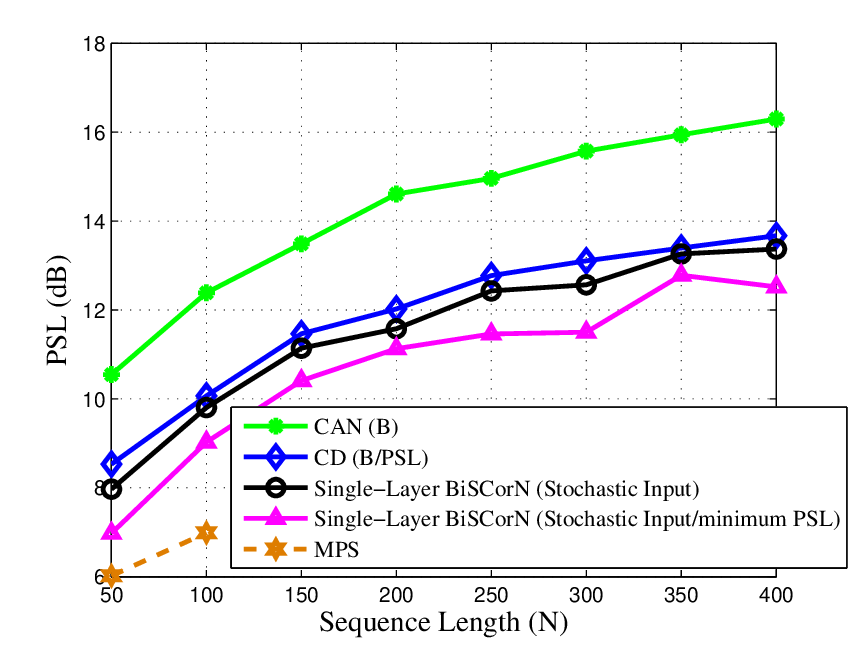}
	\centering
	\caption{PSL of aperiodic binary sequences obtained by BiSCorN, CAN~(B), and CD~(B/PSL) versus the sequence length $N$.}
	\label{PA4011}
\end{figure}

The PSL is now considered as a comparison metric that can be defined as
\begin{equation*}
\textrm{PSL}= \max_{k} ~\vert r(k) \vert, ~~1 \leq k \leq N-1.
\end{equation*}
The PSL values of the proposed ISL minimization algorithm BiSCorN,
CAN~(B), and CD~(B) with pure PSL objective function, namely CD~(B/PSL), are reported in Fig.~\ref{PA4011} for the aperiodic case. Interestingly, it can be observed form Fig.~\ref{PA4011} that the single-layer BiSCorN with stochastic input has smaller PSL than the benchmark methods for all values of the sequence length. The maximum PSL gains of the BiSCorN w.r.t. the CAN~(B) and CD~(B/PSL) are 3.03 dB and 0.56 dB, respectively. Fig.~\ref{PA4011} also includes the PSL values for Minimum Peak Sidelobe (MPS) sequences \cite{nunn2008best} (known up to length 105) as well as the best PSL values associated with BiSCorN considering 10 independent trials.
\subsection{MIMO}
The quantized Multi-CAN \cite{he2009designing} and its weighted version, i.e., Multi-CAN (B) and WeMulti-CAN (B), as well as BiST method \cite{alaee2019designing} are considered as benchmarks in this subsection. Note that the lower bound for ISL in the MIMO case is $N^{2} N_T (N_T -1)$ \cite{song2016sequence}. 
\subsubsection{MIMO ISL Minimization}
\begin{figure}
	\hfill
	\subfigure[]{\includegraphics[scale=.5]{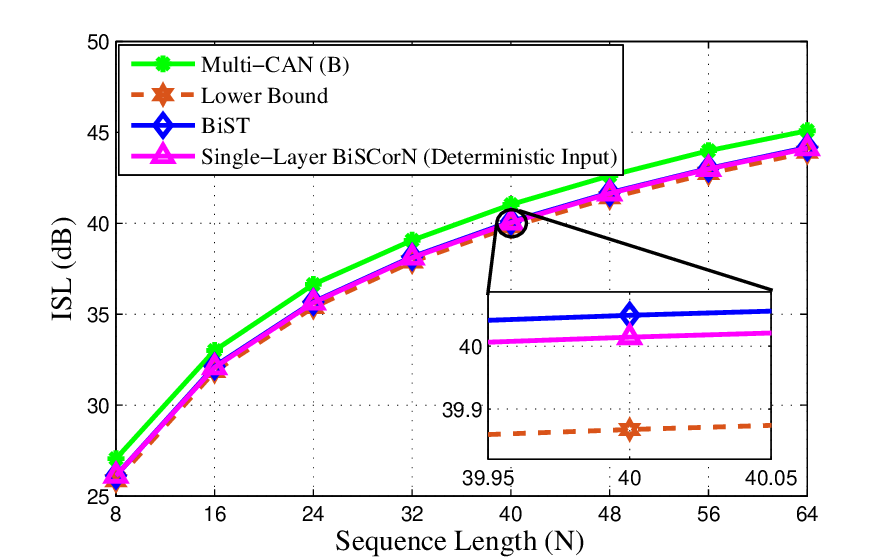}}
	\hfill
	\subfigure[]{\includegraphics[scale=.5]{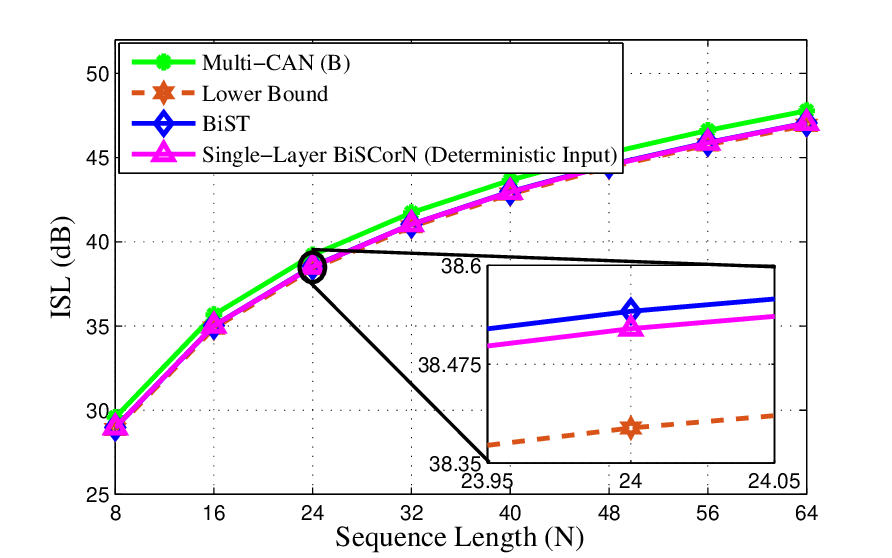}}
	\hfill
	\caption{ISL of binary sequences obtained by BiSCorN, Multi-CAN (B), and BiST versus the sequence length $N$ for the aperiodic ISL minimization problem: (a)~$N_T=3$,~(b)~$N_T=4$.}
	\label{PA4012}
\end{figure}

In Fig.~\ref{PA4012}, the aperiodic ISL versus the sequence length $N$ is depicted for different number of transmit antennas $N_T \in\lbrace3,4\rbrace$. The figure shows that the BiSCorN (single-layer with deterministic input) exhibits better performance than BiST and Multi-CAN (B) for both the values of $N_T$ and also for the values of the sequence length $N$. The maximum ISL gains of the BiSCorN w.r.t. Multi-CAN (B) and BiST are 1.01 dB and 0.06 dB for $N_T=3$ and 0.75 dB and 0.03 dB for $N_T=4$. Interestingly, the ISL values of BiSCorN sequence sets are neighboring to around 0.2 dB and 0.1 dB of the lower bound for $N_T=3$ and $N_T=4$ respectively.
\subsubsection{MIMO ISL Minimization with LCZ}
\begin{figure}[!t]
	\includegraphics[scale=.51]{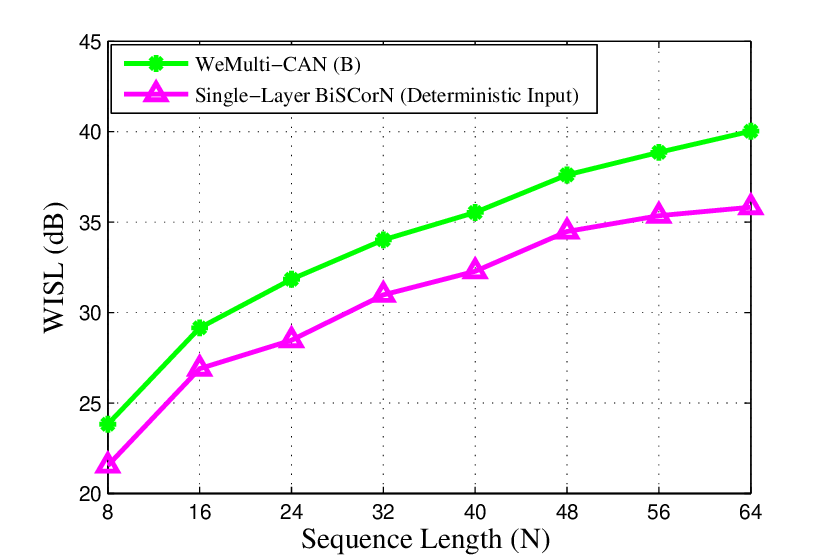}
	\centering
	\caption{WISL of binary sequences obtained by BiSCorN and WeMulti-CAN~(B) versus the sequence length $N$ for the aperiodic WISL minimization problem with $\textrm{ROS}=20\%$ and $N_T=3$.}
	\label{PA4k0}
\end{figure}

The performance of the MIMO BiSCorN to design LCZ sequences is now investigated. Fig.~\ref{PA4k0} illustrates the aperiodic WISL of the BiSCorN (single-layer with deterministic input) in comparison with WeMulti-CAN (B) for WISL minimization problem. In this case, the number of transmit antennas is set to $N_T=3$ and the ROS is equal to $20\%$ of all sidelobes i.e., $\left\lbrace r(k)\vert_{k=1}^{\ceil{0.2 \times N}}\right \rbrace$. The superior performance of the BiSCorN compared to WeMulti-CAN (B) can be observed from Fig.~\ref{PA4k0}.
\subsubsection{MIMO Binarization and Sidelobe Minimization Weights}\label{keynew1}
\begin{figure}
	\hfill
	\subfigure[]{\includegraphics[width=7.4cm,height=4.5cm]{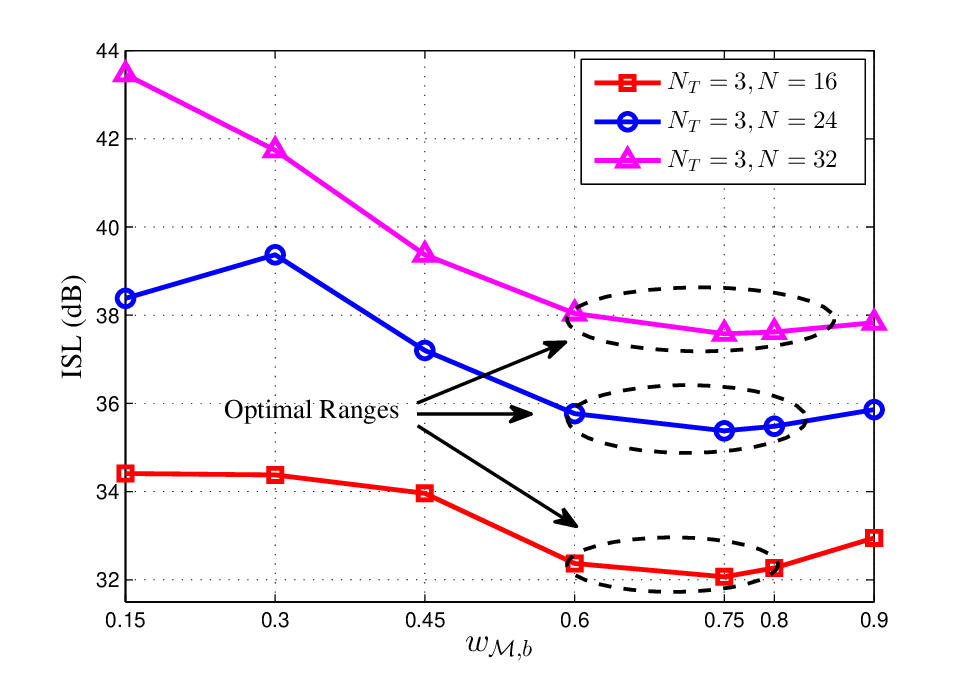}}
	\hfill
	\subfigure[]{\includegraphics[width=7.4cm,height=4.45cm]{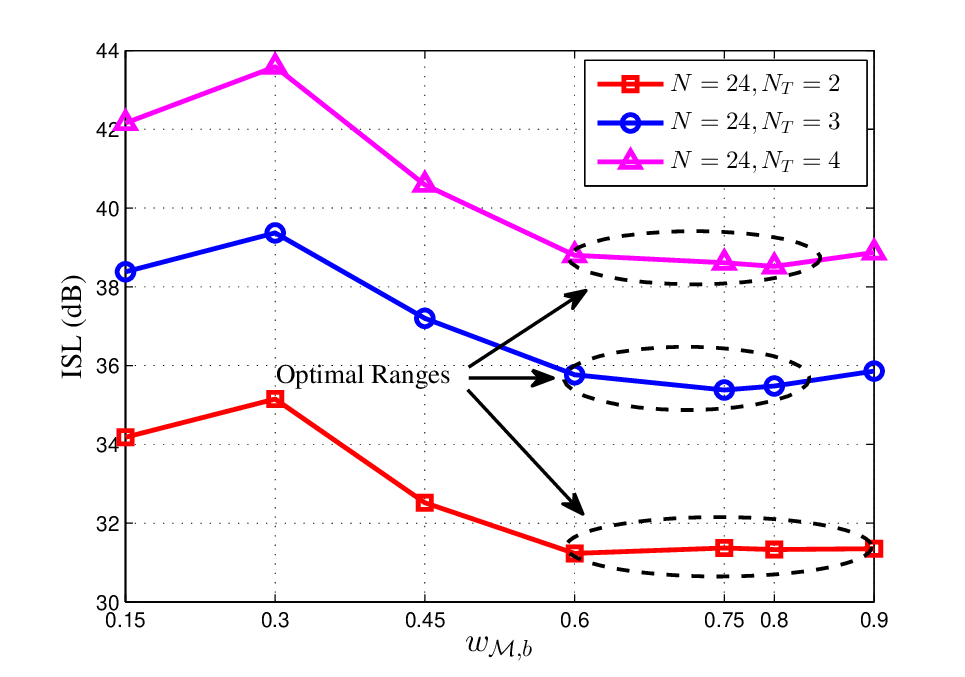}}
	\hfill
	\caption{ISL of binary sequences obtained by BiSCorN versus binarization coefficient $w_{\mathcal{M},b}$ for the aperiodic ISL minimization problem: (a)~$N_T=3$ and $N\in \lbrace 8,16,32 \rbrace$,~(b)~$N=24$ and $N_T \in \lbrace 2,3,4 \rbrace$.}
	
	\label{kjk}
\end{figure}

As mentioned, we consider BiSCorN without stop-start procedure, viz. a fixed $\textrm{WR}=\frac{w_{\mathcal{M},b}}{w_{\mathcal{M},s}}$; then, herein, we study the effect of WR on the performance. For this purpose, the ISL values obtained by single-layer
BiSCorN for the case of aperiodic MIMO versus $w_{\mathcal{M},b}$ are plotted in Fig.~\ref{kjk}. This figure includes the behavior for different values of $N$ and $N_T$. It can be observed that for the current setup, a value of $w_{\mathcal{M},b}$ in the
range of $[0.6 , 0.8]$ is a good choice. Considering $\textrm{WR}=\frac{w_{\mathcal{M},b}}{w_{\mathcal{M},s}}$ and $w_{\mathcal{M},b} + w_{\mathcal{M},s} =1$, the aforementioned interval of $w_{\mathcal{M},b}$ translates to $\textrm{WR} = N_T$.
\subsection{CSS}
The binary version of the CANARY method \cite{soltanalian2013fast}, i.e., CANARY (B)\footnote{Note that the parameters $\lambda=0.5$ and $\epsilon=10^{-15}$ (for stopping criteria) are set for the CANARY (B) algorithm in this subsection as suggested in \cite{soltanalian2013fast}.} is adopted as a benchmark in this subsection. 
\subsubsection{CISL Minimization}
\begin{figure}[!t]
	\includegraphics[scale=.43]{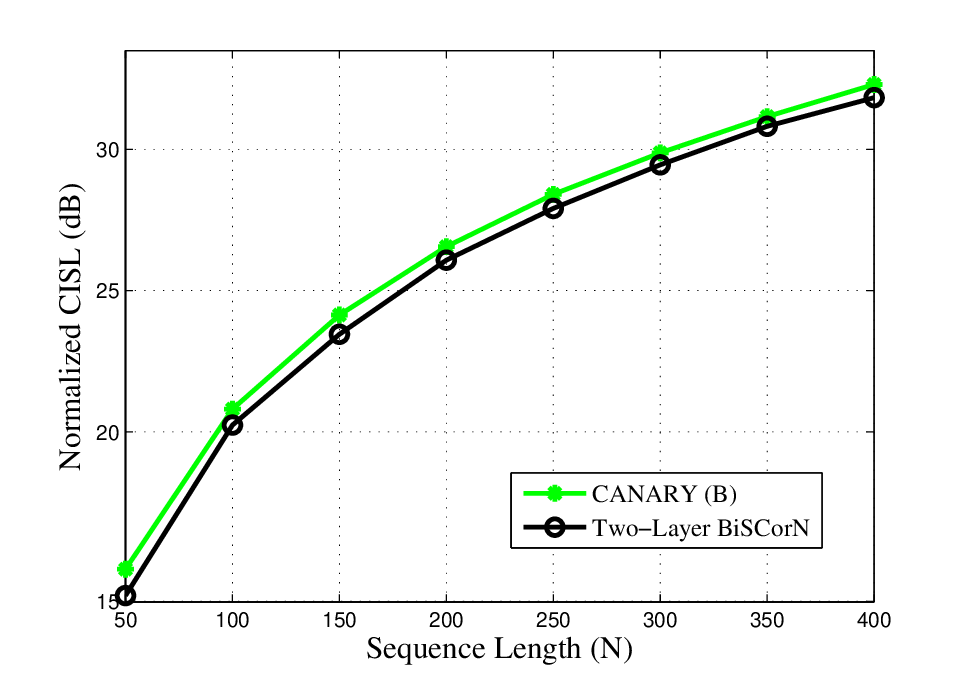}
	\centering
	\caption{Normalized CISL of aperiodic binary complementary sequences obtained by BiSCorN and CANARY~(B) versus sequence length $N$ for the case of $M=3$.}
	\label{PA4k01}
\end{figure}

In Fig.~\ref{PA4k01}, the performance of the two-layer BiSCorN and CANARY (B) is considered for producing aperiodic binary complementary sequences where the normalized CISL (dB), i.e.,
\begin{equation}
\textrm{Normalized CISL (dB)}=20\hspace{2pt} \textrm{log}_{10} \frac{\textrm{CISL}}{M},~~ 1 \leq k \leq N-1,
\end{equation}
is used as comparison metric. In this case, the number of complementary sequences is set to $M=3$. The results
show that the BiSCorN outperforms CANARY (B) providing
maximum gain of 0.94 dB.
\subsubsection{CISL Minimization with LCZ}
\begin{figure}[!t]
	\includegraphics[width=8.2cm,height=5cm]{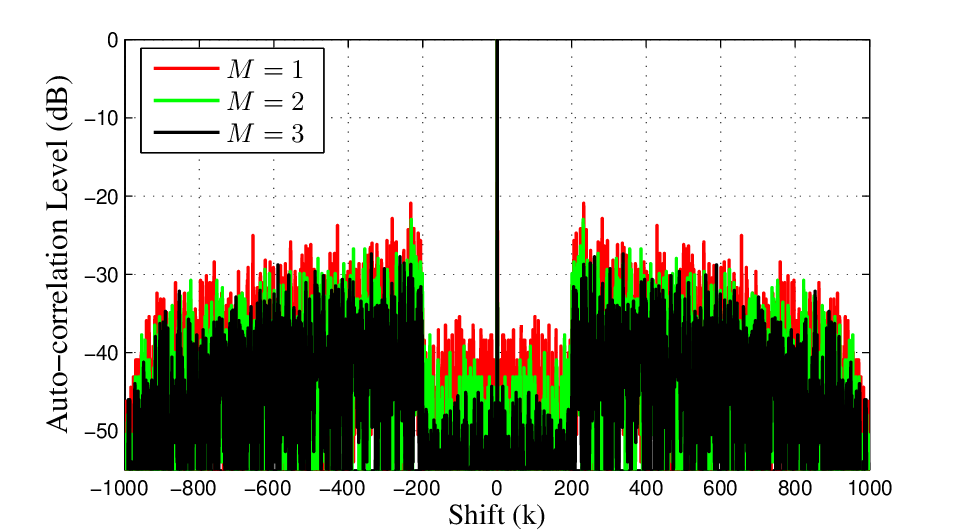}
	\centering
	\caption{Auto-correlation level of aperiodic binary complementary sequences obtained by two-layer BiSCorN for a different number of complementary sequences $M=1,2,3$ with $N=1000$ and ROS=$20\%$.}
	\label{PA4k012}
\end{figure}

In Fig.~\ref{PA4k012} the aperiodic auto-correlation function of BiSCorN (single-layer with deterministic input) using different numbers of complementary sequences $M=1,2,3$ is illustrated for a sequence length $N=1000$ and a ROS=$20\%$. In this case, the auto-correlation sums are normalized and expressed in dB, i.e.,
\begin{equation}
\textrm{Auto-correlation Level (dB)}=20\hspace{2pt} \textrm{log}_{10} \frac{w_{\mathcal{C},k}\left \vert \sum_{m^{\prime}=1}^{M} r_{m^{\prime}}(k)  \right \vert}{\sum_{m^{\prime}=1}^{M} r_{m^{\prime}}(0)},~~ 1 \leq k \leq N-1.
\end{equation} 
As expected, it can be observed that the auto-correlation level decreases as $M$
increases. 
\subsection{MIMO-CSS}
\begin{figure}[!t]
	\includegraphics[scale=.48]{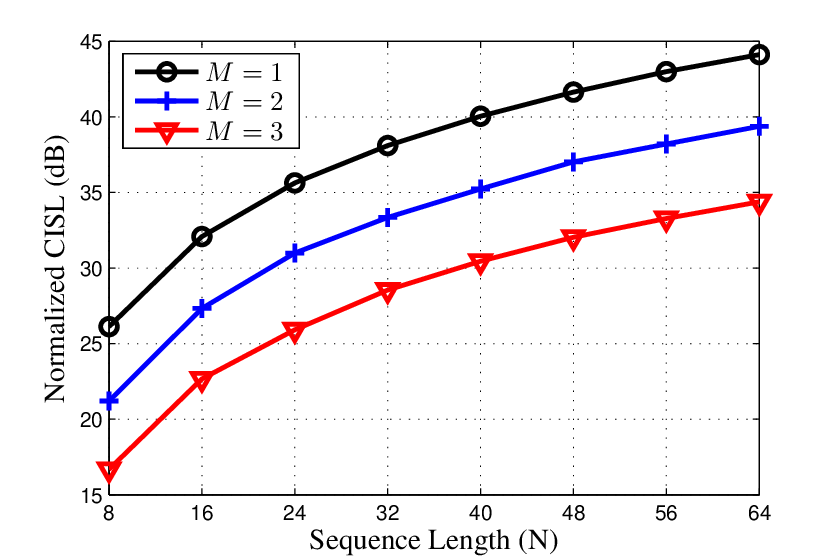}
	\centering
	\caption{Normalized CISL of aperiodic binary complementary sequences obtained by MIMO two-layer BiSCorN versus the sequence length $N$ for different number of complementary sequences $M=1,2,3$.}
	\label{PA4k0112}
\end{figure}

The performance of MIMO-CSS systems (see Subsection~\ref{mimocss}) is evaluated in this subsection. Fig.~\ref{PA4k0112} shows normalized aperiodic CISL values of the MIMO two-layer BiSCorN versus the sequence length $N$ for different numbers of complementary sequences $M=1,2,3$.
It is observed from this experiment that the MIMO complementary BiSCorN, i.e., the cases with $M=2,3$, achieves better performance than MIMO BiSCorN with $M=1$. Specifically, the maximum CISL gains of MIMO complementary BiSCorN for $M=3$ and $M=2$ w.r.t. the MIMO BiSCorN with $M=1$ are 9.75 dB and 4.91 dB, respectively.
\section{Conclusion} \label{con}
The design of binary sequences with good aperiodic/periodic correlation
properties has been considered, adopting the ISL as performance metric. In this respect, leveraging the asymptotic equivalence between the original design and its relaxed version, a novel learning-based framework denoted as BiSCorN, has been proposed for synthesizing binary (possibly complementary)
sequences in SISO and MIMO systems. Then, network architectures have been devised to deal with a modified version of the ISL metric, i.e.,
weighted ISL. ADAM optimizer has been employed for the implementation of the BiSCorN learning process. Finally, a synthesis stage has been exploited to obtain feasible/binary solutions to the original problem, i.e., binary codes.

Numerical results have been provided to illustrate the effectiveness of the proposed learning-based approaches. In particular, it has been observed that the conceived architectures may outperform state-of-the-art algorithms for both SISO and MIMO systems in terms of ISL with a reasonable computational complexity. Future works may deal with designing binary sequences with good ambiguity function features, so as to deal with the presence of possible Doppler shifts.

%
%





%
\appendices 
\section{Proof of Lemma 1} \label{app4}
Let us proceed by contradiction to prove the first item of the lemma. To this end, let us assume that the absolute value of at least one entry of $\mathbf{x}^{(i)}$ does not converge to one. This is tantamount to supposing that there exists a subsequence $\mathbf{x}^{(k_i)}$ of $\mathbf{x}^{(i)}$ such that
\begin{eqnarray}
	\lim_{i \,\to\, + \infty } &  \displaystyle
	\min_{n=1,...,N} & \left \vert x^{(k_i)} (n) \right \vert < 1,
\end{eqnarray}
which implies the existence of $0<\delta <1$ such that
\begin{eqnarray}\label{limit}
	\min_{n=1,...,N} & \left \vert x^{(k_i)} (n) \right \vert < (1-\delta),\,\,\mbox{provided that}\,\, i\,\,\mbox{is large enough}.
\end{eqnarray}
Before proceeding further, let us observe that 
any optimal solution to the original Problem \eqref{maxmin21} is a feasible point to  Problem $\mathcal{P}^{(k_i)}$, achieving ${w}_{\mathcal{S},s}^{(k_i)} v^{\star}$ as objective value, with $v^{\star}$ the optimal value of \eqref{maxmin21}. Moreover, based on (\ref{limit}), it follows 
\begin{equation}
	r^{(k_i)}(0)=\sum_{n=1}^{N} \vert x^{(k_i)}(n) \vert^2 \leq N-1+(1-\delta)^2 ~ \,\Leftrightarrow\,~ N-r^{(k_i)}(0) \geq 1-(1-\delta)^2.
\end{equation}
As a consequence, denoting by $v \left( \mathcal{P}^{(k_i)}  \right)$  the optimal value to Problem $\mathcal{P}^{(k_i)}$, the following inequality holds
\begin{equation}
	{w}_{\mathcal{S},s}^{(k_i)} v^{\star} \geq v \left(\mathcal{P}^{(k_i)}  \right) \geq  w_{\mathcal{S},b}^{(k_i)} \left \vert 1-(1-\delta)^2 \right \vert^2,
\end{equation}
being $w_{\mathcal{S},s}^{(k_i)}\sum_{k=1}^{N-1} w_{\mathcal{S},k} \vert r (k) \vert^2\geq 0$. Hence,
\begin{equation}
	v^{\star} \geq  \frac{\bar{w}^{(k_i)}_{\mathcal{S},b}}{\bar{w}_{\mathcal{S},s}} \left \vert 1-(1-\delta)^2 \right \vert^2,
\end{equation}
which is evidently contradiction as $i \,\to\, + \infty$, since the right-hand side of the inequality  goes to infinity while the left-hand side is a finite value. Therefore, $\left \vert x^{(i)} (n) \right \vert \,\to\, 1$ for any $n$. 

Let us now focus on the second item of the lemma. In this respect, let $\mathbf{x}^{(k_i)}$ be a convergent subsequence of $\mathbf{x}^{(i)}$ with a limit point $\bar{\mathbf{x}}$ (its existence is guaranteed by the boundness of $\mathbf{x}^{(i)}$). Hence, denoting by $\mathbf{r}^{(k_i)}$ and $\bar{\mathbf{r}}$ the auto-correlation vectors associated with $\mathbf{x}^{(k_i)}$ and $\bar{\mathbf{x}}$, respectively, it follows that\footnote{Note that $\mathbf{r}^{(k_i)}$ converges to $\bar{\mathbf{r}}$ since each lag of the auto-correlation function is a continuous function of the sequence $\mathbf{x}$.}
\begin{equation} \label{key145}
\sum_{m=1}^{N-1} w_{\mathcal{S},m} \left \vert \bar{{r}}(m) \right \vert^2 \geq  v^{\star}.
\end{equation}
Indeed, according to the first item of this lemma, at the limit point each entry of the sequence is unit modular  and, therefore, the limit solution is a feasible point for the original problem in \eqref{maxmin21}. Now, one can write 
	\begin{equation}  \label{key124}
		w_{\mathcal{S},s}^{(k_i)} \sum_{m=1}^{N-1} w_{\mathcal{S},m}  \left \vert {r}^{(k_i)}(m) \right \vert ^2 \leq w_{\mathcal{S},b}^{(k_i)} \left \vert {r}^{(k_i)}(0) -N \right \vert^2 + w_{\mathcal{S},s}^{(k_i)} \sum_{m=1}^{N-1} w_{\mathcal{S},m}  \left \vert {r}^{(k_i)}(m) \right \vert^2,~\forall i,
	\end{equation}
since $w_{\mathcal{S},b}^{(k_i)} \left \vert {r}^{(k_i)}(0) -N \right \vert^2 \geq 0$. Then, note that any optimal solution to Problem \eqref{maxmin21} is a suboptimal solution to $\mathcal{P}^{(k_i)}$, hence, 
	\begin{equation}  \label{key1241}
		 w_{\mathcal{S},b}^{(k_i)} \left \vert {r}^{(k_i)}(0) -N \right \vert^2 + w_{\mathcal{S},s}^{(k_i)} \sum_{m=1}^{N-1} w_{\mathcal{S},m}  \left \vert {r}^{(k_i)}(m) \right \vert^2 \leq w_{\mathcal{S},s}^{(k_i)} v^{\star},~\forall i.
	\end{equation}
Finally, \eqref{key124} along with \eqref{key1241}	implies that
	\begin{equation}  \label{key123}
		\sum_{m=1}^{N-1} w_{\mathcal{S},m}  \left \vert \bar{{r}}(m) \right \vert^2 \leq v^{\star}.
	\end{equation}
	By considering \eqref{key145} and \eqref{key123}, it follows that $\bar{\mathbf{x}}$ is a binary sequence achieving the optimal auto-correlation sidelobes in terms of the WISL metric.
\section{The Derivation of \eqref{al} and \eqref{cd}} \label{app2}
First, considering \eqref{lj2}, the matrix $\mathbf{X}_{\mathcal{C}}^T \mathbf{X}_{\mathcal{C}}$ can be expressed as the following
\begin{equation} \label{12k}
	{\begin{bmatrix} 
			\sum_{m^{\prime}=1}^{M}r_{m^{\prime}}(0),& \sum_{m^{\prime}=1}^{M}r_{m^{\prime}}(1),& \sum_{m^{\prime}=1}^{M}r_{m^{\prime}}(2),& ..., & \sum_{m^{\prime}=1}^{M}r_{m^{\prime}}(N-1)\\
			\sum_{m^{\prime}=1}^{M}r_{m^{\prime}}(1),&\sum_{m^{\prime}=1}^{M}r_{m^{\prime}}(0),&\sum_{m^{\prime}=1}^{M}r_{m^{\prime}}(1),& ..., & \sum_{m^{\prime}=1}^{M}r_{m^{\prime}}(N-2)\\
			\vdots& \vdots&\vdots&  & \vdots\\
			\sum_{m^{\prime}=1}^{M}r_{m^{\prime}}(N-1),& \sum_{m^{\prime}=1}^{M}r_{m^{\prime}}(N-2),& \sum_{m^{\prime}=1}^{M}r_{m^{\prime}}(N-3),& ..., & \sum_{m^{\prime}=1}^{M}r_{m^{\prime}}(0)
	\end{bmatrix}} \in \mathbb{R}^{N \times N}.
\end{equation}
Note that $\mathbf{X}_{\mathcal{S}}^T \mathbf{X}_{\mathcal{S}}$ can be obtained from the matrix in \eqref{12k} setting $M=1$.  
Then, using \eqref{12k}, the output signal for SISO system is given by
\begin{equation*} 
	\mathbf{y}_{\mathcal{S}}=	\mathbf{X}^{T}_{\mathcal{S}}	\mathbf{X}_{\mathcal{S}}	\mathbf{s}_{\mathcal{S}}=
	{\begin{bmatrix} 
			s_{\mathcal{S},1} r(0)+ s_{\mathcal{S},2} r(1)+...+s_{\mathcal{S},N} r(N-1)\\
			s_{\mathcal{S},1} r(1)+s_{\mathcal{S},2} r(0)+...s_{\mathcal{S},N} r(N-2)\\
			\vdots\\
			s_{\mathcal{S},1} r(N-1)+s_{\mathcal{S},2} r(N-2)+...+s_{\mathcal{S},N} r(0)
	\end{bmatrix}} \in \mathbb{R}^{N}.
\end{equation*}
Now, it is verified that the $k^{th}$ entry of $\mathbf{y}_{\mathcal{S}}$ can be expressed as in \eqref{al}. 

Next, considering the input signal $\mathbf{s}_{\mathcal{C}}=[{a}_{\mathcal{C}},0,0,...,0]\in \mathbb{R}^{N}$ from Subsection~\ref{CSSpro} and using the matrix in \eqref{12k}, the output vector for the CSS case can be obtained as 
\begin{equation} \label{1200010}
	\mathbf{y}_{\mathcal{C}}=\mathbf{X}_{\mathcal{C}}^T \mathbf{X}_{\mathcal{C}} \hspace{1pt} \mathbf{s}_{\mathcal{C}}= {a_{\mathcal{C}}} \hspace{2pt}
	{\begin{bmatrix} 
			\sum_{m^{\prime}=1}^{M}r_{m^{\prime}}(0)\\
			\sum_{m^{\prime}=1}^{M}r_{m^{\prime}}(1)\\
			\vdots\\
			\sum_{m^{\prime}=1}^{M}r_{m^{\prime}}(N-1)
	\end{bmatrix}}\in \mathbb{R}^{N}.
\end{equation}
Finally, using \eqref{1200010} and \eqref{key2451}, the loss function in \eqref{cd} can be achieved.
\section{The Derivation of \eqref{13} and \eqref{keyl}} \label{app3}
Considering \eqref{lj22} we can write
\begin{equation} \label{1200}
	\mathbf{X}_{\mathcal{MC}}^T \mathbf{X}_{\mathcal{MC}}=
	{\begin{bmatrix} 
			\mathbf{R}_{11},& \mathbf{R}_{12},& ..., & \mathbf{R}_{1N_T}\\
			\mathbf{R}_{21},& \mathbf{R}_{22},& ..., & \mathbf{R}_{2N_T}\\
			\vdots& \vdots&  & \vdots\\
			\mathbf{R}_{N_{T}1},& \mathbf{R}_{N_{T}2},& ..., & \mathbf{R}_{N_{T}N_{T}}
	\end{bmatrix}} \in \mathbb{R}^{NN_{T} \times NN_{T}},
\end{equation}
where $\mathbf{R}_{ij}= \sum_{m^{\prime}=1}^{M} \mathbf{R}_{ij,m^{\prime}},~i,j=1,...,N_T,$ and
\begin{equation} \label{12000}
	\mathbf{R}_{ij,m^{\prime}} =
	{\begin{bmatrix} 
			r_{ij,m^{\prime}}(0),& \sqrt{2}\hspace{2pt} r_{ij,m^{\prime}}(1),& \sqrt{2}\hspace{2pt}r_{ij,m^{\prime}}(2),& ..., & \sqrt{2}\hspace{2pt}r_{ij,m^{\prime}}(N-1)\\
			\sqrt{2}\hspace{2pt}r_{ij,m^{\prime}}(1),& 2\hspace{2pt}r_{ij,m^{\prime}}(0),& 2\hspace{2pt}r_{ij,m^{\prime}}(1),& ..., & 2\hspace{2pt}r_{ij,m^{\prime}}(N-2)\\
			\vdots& \vdots&\vdots&  & \vdots\\
			\sqrt{2}\hspace{2pt}r_{ij,m^{\prime}}(N-1),& 2\hspace{2pt}r_{ij,m^{\prime}}(N-2),&2\hspace{2pt} r_{ij,m^{\prime}}(N-3),& ..., & 2\hspace{2pt}r_{ij,m^{\prime}}(0)
	\end{bmatrix}} \in \mathbb{R}^{N \times N}.
\end{equation}
Next, according to the input matrix $\mathbf{S}_{\mathcal{MC}}$ in \eqref{nnm}, the output matrix becomes 
\begin{equation} \label{120001}
	\mathbf{Y}_{\mathcal{MC}} = \mathbf{X}_{\mathcal{MC}}^T \mathbf{X}_{\mathcal{MC}} \mathbf{S}_{\mathcal{MC}}=
	{\begin{bmatrix} 
			\widetilde{\mathbf{R}}_{1}\\
			\widetilde{\mathbf{R}}_{2}\\
			\vdots\\
			\widetilde{\mathbf{R}}_{N_{T}}
	\end{bmatrix}} \in \mathbb{R}^{NN_T \times N_T},
\end{equation}
where $\widetilde{\mathbf{R}}_{i}= \sum_{m^{\prime}=1}^{M} \widetilde{\mathbf{R}}_{i,m^{\prime}},~i=1,...,N_T,$ and
\begin{equation} \label{120010}
	\widetilde{\mathbf{R}}_{i,m^{\prime}} = {a_{\mathcal{M}}}\hspace{2pt}
	{\begin{bmatrix} 
			r_{i1,m^{\prime}}(0),& r_{i2,m^{\prime}}(0),& ..., & r_{iN_T,m^{\prime}}(0)\\
			\sqrt{2}\hspace{2pt}r_{i1,m^{\prime}}(1),& \sqrt{2}\hspace{2pt}r_{i2,m^{\prime}}(1),& ..., &\sqrt{2}\hspace{2pt}r_{iN_T,m^{\prime}}(1)\\
			\vdots& \vdots&  & \vdots\\
			\sqrt{2}\hspace{2pt}r_{i1,m^{\prime}}(N-1),& \sqrt{2}\hspace{2pt}r_{i2,m^{\prime}}(N-1),& ..., &\sqrt{2}\hspace{2pt}r_{iN_T,m^{\prime}}(N-1)
	\end{bmatrix}} \in \mathbb{R}^{N \times N_T}.
\end{equation}
Therefore, using the output matrix in \eqref{120001} and considering \eqref{key24525}, the loss function in \eqref{keyl} is given by $\mathcal{L}_{\mathcal{MC}} = \mathbf{a}^T_{\mathcal{MC}} \hspace{1pt} \widetilde{\mathbf{r}}_{\mathcal{MC}}$. Also, the loss function of the MIMO radar in \eqref{13} can be obtained using a similar procedure as in \eqref{1200}-\eqref{120010} considering $M=1$.
\bibliographystyle{IEEETran}
\bibliography{myreff}

\begin{thebibliography}{10}
\providecommand{\url}[1]{#1}
\csname url@samestyle\endcsname
\providecommand{\newblock}{\relax}
\providecommand{\bibinfo}[2]{#2}
\providecommand{\BIBentrySTDinterwordspacing}{\spaceskip=0pt\relax}
\providecommand{\BIBentryALTinterwordstretchfactor}{4}
\providecommand{\BIBentryALTinterwordspacing}{\spaceskip=\fontdimen2\font plus
\BIBentryALTinterwordstretchfactor\fontdimen3\font minus
  \fontdimen4\font\relax}
\providecommand{\BIBforeignlanguage}[2]{{%
\expandafter\ifx\csname l@#1\endcsname\relax
\typeout{** WARNING: IEEEtran.bst: No hyphenation pattern has been}%
\typeout{** loaded for the language `#1'. Using the pattern for}%
\typeout{** the default language instead.}%
\else
\language=\csname l@#1\endcsname
\fi
#2}}
\providecommand{\BIBdecl}{\relax}
\BIBdecl

\bibitem{stoicabook}
H.~He, J.~Li, and P.~Stoica, \emph{Waveform design for active sensing systems:
  a computational approach}.\hskip 1em plus 0.5em minus 0.4em\relax Cambridge
  University Press, 2012.

\bibitem{gini2012waveform}
F.~Gini, A.~De~Maio, and L.~Patton, \emph{Waveform design and diversity for
  advanced radar systems}.\hskip 1em plus 0.5em minus 0.4em\relax Institution
  of engineering and technology London, UK, 2012.

\bibitem{aubry2015optimizing}
A.~Aubry, A.~De~Maio, and M.~M. Naghsh, ``Optimizing radar waveform and
  {Doppler} filter bank via generalized fractional programming,'' \emph{IEEE
  Journal of Selected Topics in Signal Processing}, vol.~9, no.~8, pp.
  1387--1399, 2015.

\bibitem{scol}
\BIBentryALTinterwordspacing
M.~Skolnik, \emph{Radar Handbook, Third Edition}, ser. Electronics electrical
  engineering.\hskip 1em plus 0.5em minus 0.4em\relax McGraw-Hill Education,
  2008. [Online]. Available:
  \url{https://books.google.com/books?id=76uF2Xebm-gC}
\BIBentrySTDinterwordspacing

\bibitem{bark16}
M.~Friese, ``Polyphase barker sequences up to length 36,'' \emph{IEEE
  Transactions on Information Theory}, vol.~42, no.~4, pp. 1248--1250, 1996.

\bibitem{bark17}
P.~Borwein and R.~Ferguson, ``Polyphase sequences with low autocorrelation,''
  \emph{IEEE Transactions on Information Theory}, vol.~51, no.~4, pp.
  1564--1567, 2005.

\bibitem{bark18}
C.~Nunn, ``Constrained optimization applied to pulse compression codes and
  filters,'' in \emph{IEEE International Radar Conference, 2005.}\hskip 1em
  plus 0.5em minus 0.4em\relax IEEE, 2005, pp. 190--194.

\bibitem{bark19}
C.~J. Nunn and G.~E. Coxson, ``Polyphase pulse compression codes with optimal
  peak and integrated sidelobes,'' \emph{IEEE Transactions on Aerospace and
  Electronic Systems}, vol.~45, no.~2, pp. 775--781, 2009.

\bibitem{CAN}
P.~Stoica, H.~He, and J.~Li, ``New algorithms for designing unimodular
  sequences with good correlation properties,'' \emph{IEEE Transactions on
  Signal Processing}, vol.~57, no.~4, pp. 1415--1425, 2009.

\bibitem{lin2019efficient}
R.~Lin, M.~Soltanalian, B.~Tang, and J.~Li, ``Efficient design of binary
  sequences with low autocorrelation sidelobes,'' \emph{IEEE Transactions on
  Signal Processing}, vol.~67, no.~24, pp. 6397--6410, 2019.

\bibitem{palom}
J.~Song, P.~Babu, and D.~P. Palomar, ``Sequence design to minimize the weighted
  integrated and peak sidelobe levels,'' \emph{IEEE Transactions on Signal
  Processing}, vol.~64, no.~8, pp. 2051--2064, 2015.

\bibitem{CD}
M.~A. Kerahroodi, A.~Aubry, A.~De~Maio, M.~M. Naghsh, and M.~Modarres-Hashemi,
  ``A coordinate-descent framework to design low {PSL/ISL} sequences,''
  \emph{IEEE Transactions on Signal Processing}, vol.~65, no.~22, pp.
  5942--5956, 2017.

\bibitem{he2009designing}
H.~He, P.~Stoica, and J.~Li, ``Designing unimodular sequence sets with good
  correlations; including an application to {MIMO} radar,'' \emph{IEEE
  Transactions on Signal Processing}, vol.~57, no.~11, pp. 4391--4405, 2009.

\bibitem{song2016sequence}
J.~Song, P.~Babu, and D.~P. Palomar, ``Sequence set design with good
  correlation properties via majorization-minimization,'' \emph{IEEE
  Transactions on Signal Processing}, vol.~64, no.~11, pp. 2866--2879, 2016.

\bibitem{li2016design}
Y.~Li, S.~A. Vorobyov, and Z.~He, ``Design of multiple unimodular waveforms
  with low auto-and cross-correlations for radar via
  majorization-minimization,'' in \emph{2016 24th European Signal Processing
  Conference (EUSIPCO)}.\hskip 1em plus 0.5em minus 0.4em\relax IEEE, 2016, pp.
  2235--2239.

\bibitem{li2017fast}
Y.~Li and S.~A. Vorobyov, ``Fast algorithms for designing unimodular waveform
  (s) with good correlation properties,'' \emph{IEEE Transactions on Signal
  Processing}, vol.~66, no.~5, pp. 1197--1212, 2017.

\bibitem{cui2017constant}
G.~Cui, X.~Yu, M.~Piezzo, and L.~Kong, ``Constant modulus sequence set design
  with good correlation properties,'' \emph{Signal Processing}, vol. 139, pp.
  75--85, 2017.

\bibitem{tang2014construction}
J.~Tang, N.~Zhang, Z.~Ma, and B.~Tang, ``Construction of {Doppler} resilient
  complete complementary code in {MIMO} radar,'' \emph{IEEE Transactions on
  Signal Processing}, vol.~62, no.~18, pp. 4704--4712, 2014.

\bibitem{wang2021designing}
J.~Wang and Y.~Wang, ``Designing unimodular sequences with optimized
  auto/cross-correlation properties via consensus-{ADMM}/{PDMM} approaches,''
  \emph{IEEE Transactions on Signal Processing}, vol.~69, pp. 2987--2999, 2021.

\bibitem{welch}
M.~Soltanalian, M.~M. Naghsh, and P.~Stoica, ``On meeting the peak correlation
  bounds,'' \emph{IEEE Transactions on Signal Processing}, vol.~62, no.~5, pp.
  1210--1220, 2014.

\bibitem{alaee2019designing}
M.~Alaee-Kerahroodi, M.~Modarres-Hashemi, and M.~M. Naghsh, ``Designing sets of
  binary sequences for {MIMO} radar systems,'' \emph{IEEE Transactions on
  Signal Processing}, vol.~67, no.~13, pp. 3347--3360, 2019.

\bibitem{soltanalian2012computational}
M.~Soltanalian and P.~Stoica, ``Computational design of sequences with good
  correlation properties,'' \emph{IEEE Transactions on Signal processing},
  vol.~60, no.~5, pp. 2180--2193, 2012.

\bibitem{soltanalian2013fast}
M.~Soltanalian, M.~M. Naghsh, and P.~Stoica, ``A fast algorithm for designing
  complementary sets of sequences,'' \emph{Signal Processing}, vol.~93, no.~7,
  pp. 2096--2102, 2013.

\bibitem{wang2021quasi}
J.~Wang, P.~Fan, Z.~Zhou, and Y.~Yang, ``Quasi-orthogonal z-complementary pairs
  and their applications in fully polarimetric radar systems,'' \emph{IEEE
  Transactions on Information Theory}, 2021.

\bibitem{wu2019fast}
Z.-J. Wu, T.-L. Xu, Z.-Q. Zhou, and C.-X. Wang, ``Fast algorithms for designing
  complementary sets of sequences under multiple constraints,'' \emph{IEEE
  Access}, vol.~7, pp. 50\,041--50\,051, 2019.

\bibitem{chai2021deep}
J.~Chai, H.~Zeng, A.~Li, and E.~W. Ngai, ``Deep learning in computer vision: A
  critical review of emerging techniques and application scenarios,''
  \emph{Machine Learning with Applications}, vol.~6, p. 100134, 2021.

\bibitem{Redmark}
M.~Ahmadi, A.~Norouzi, N.~Karimi, S.~Samavi, and A.~Emami, ``Redmark: Framework
  for residual diffusion watermarking based on deep networks,'' \emph{Expert
  Systems with Applications}, p. 113157, 2019.

\bibitem{khobahi2020model}
S.~Khobahi and M.~Soltanalian, ``Model-based deep learning for one-bit
  compressive sensing,'' \emph{IEEE Transactions on Signal Processing},
  vol.~68, pp. 5292--5307, 2020.

\bibitem{zeng2021simultaneous}
Y.~Zeng, X.~Xu, S.~Jin, and R.~Zhang, ``Simultaneous navigation and radio
  mapping for cellular-connected uav with deep reinforcement learning,''
  \emph{IEEE Transactions on Wireless Communications}, 2021.

\bibitem{sharma2019distributed}
M.~K. Sharma, A.~Zappone, M.~Assaad, M.~Debbah, and S.~Vassilaras,
  ``Distributed power control for large energy harvesting networks: A
  multi-agent deep reinforcement learning approach,'' \emph{IEEE Transactions
  on Cognitive Communications and Networking}, vol.~5, no.~4, pp. 1140--1154,
  2019.

\bibitem{d2019uplink}
C.~D'Andrea, A.~Zappone, S.~Buzzi, and M.~Debbah, ``Uplink power control in
  cell-free massive {MIMO} via deep learning,'' in \emph{2019 IEEE 8th
  International Workshop on Computational Advances in Multi-Sensor Adaptive
  Processing (CAMSAP)}.\hskip 1em plus 0.5em minus 0.4em\relax IEEE, 2019, pp.
  554--558.

\bibitem{rezaei2020learning}
O.~Rezaei, M.~Ahmadi, and M.~M. Naghsh, ``A learning approach to design binary
  sequences with good correlation properties,'' in \emph{2020 IEEE Radar
  Conference (RadarConf20)}.\hskip 1em plus 0.5em minus 0.4em\relax IEEE, 2020,
  pp. 1--6.

\bibitem{hinton1994autoencoders}
G.~E. Hinton and R.~S. Zemel, ``Autoencoders, minimum description length and
  {Helmholtz} free energy,'' in \emph{Advances in neural information processing
  systems}, 1994, pp. 3--10.

\bibitem{li2018random}
P.~Li and P.-M. Nguyen, ``On random deep weight-tied autoencoders: {Exact}
  asymptotic analysis, phase transitions, and implications to training,'' in
  \emph{International Conference on Learning Representations}, 2018.

\bibitem{courbariaux2016binarized}
M.~Courbariaux, I.~Hubara, D.~Soudry, R.~El-Yaniv, and Y.~Bengio, ``Binarized
  neural networks: Training deep neural networks with weights and activations
  constrained to+ 1 or-1,'' \emph{arXiv preprint arXiv:1602.02830}, 2016.

\bibitem{tensor}
M.~Abadi, P.~Barham, J.~Chen, Z.~Chen, A.~Davis, J.~Dean, M.~Devin,
  S.~Ghemawat, G.~Irving, M.~Isard \emph{et~al.}, ``Tensorflow: A system for
  large-scale machine learning,'' in \emph{12th $\{$USENIX$\}$ Symposium on
  Operating Systems Design and Implementation ($\{$OSDI$\}$ 16)}, 2016, pp.
  265--283.

\bibitem{adam}
D.~P. Kingma and J.~Ba, ``{Adam}: A method for stochastic optimization,'' in
  \emph{ICLR}, 2015.

\bibitem{goodfellow2016deep}
I.~Goodfellow, Y.~Bengio, and A.~Courville, \emph{Deep learning}.\hskip 1em
  plus 0.5em minus 0.4em\relax MIT press, 2016.

\bibitem{nunn2008best}
C.~J. Nunn and G.~E. Coxson, ``Best-known autocorrelation peak sidelobe levels
  for binary codes of length 71 to 105,'' \emph{IEEE transactions on Aerospace
  and Electronic Systems}, vol.~44, no.~1, pp. 392--395, 2008.

\end{thebibliography}
\end{document}